\documentclass[fleqn,usenatbib]{mnras}
\usepackage{newtxtext,newtxmath}
\usepackage{xcolor}
\pagecolor{white}
\usepackage[T1]{fontenc}
\usepackage{graphicx}
\usepackage{amsmath}
\usepackage{subcaption}
\usepackage{hyperref,xurl}
\usepackage[normalem]{ulem}
\title[Spatio-spectral-temporal Modelling of Young PWNe]{Spatio-spectral-temporal Modelling of Two Young Pulsar Wind Nebulae}
\author[A, Kundu et al.]{
A. Kundu,$^{1,2,3,4}$\thanks{E-mail: anukundu@umbc.edu}
Jagdish C. Joshi,$^{5,6}$
C. Venter,$^{1, 7}$
N. E. Engelbrecht,$^{1}$
W. Zhang,$^{8}$
Diego F. Torres,$^{8,9,10}$
\newauthor
I. Sushch,$^{11,12,13,1,14}$
Shuta J. Tanaka$^{15,16}$
\\
$^{1}$Centre for Space Research, North-West University, Private Bag X6001, Potchefstroom 2520, South Africa\\
$^{2}$Center for Space Sciences and Technology, University of Maryland, Baltimore County, Baltimore, MD 21250\\
$^{3}$Astrophysics Science Division, NASA Goddard Space Flight Center, Greenbelt, MD 20771\\
$^{4}$Center for Research and Exploration in Space Science and Technology, NASA/GSFC, Greenbelt, MD 20771\\
$^{5}$Aryabhatta Research Institute of Observational Sciences (ARIES), Manora peak, Beluwakhan, Uttarakhand 263002, India\\
$^{6}$Centre for Astro-Particle Physics (CAPP) and Department of Physics, University of Johannesburg, PO Box 524, Auckland Park 2006, South Africa\\
$^{7}$ National Institute for Theoretical and Computational Sciences (NITheCS), Stellenbosch, South Africa \\
$^{8}$ Institute of Space Sciences (ICE, CSIC), Campus UAB, Carrer de Magrans s/n, E-08193 Barcelona, Spain \\
$^{9}$ Instituci\'o Catalana de Recerca i Estudis Avançats (ICREA), E-08010 Barcelona, Spain \\
$^{10}$ Institut d’Estudis Espacials de Catalunya (IEEC), 08034 Barcelona, Spain\\
$^{11}$ Centro de Investigaciones Energ\'eticas, Medioambientales y Tecnol\'ogicas (CIEMAT), E-28040 Madrid, Spain\\
$^{12}$ Gran Sasso Science Institute,Via F.Crispi 7, 67100 L’Aquila, Italy\\
$^{13}$ INFN-Laboratori Nazionali del Gran Sasso, Via G. Acitelli 22, Assergi (AQ), Italy \\
$^{14}$ Astronomical Observatory of Ivan Franko National University of Lviv, Kyryla i Methodia 8, 79005 Lviv, Ukraine\\
$^{15}$ Department of Physical Sciences, Aoyama Gakuin University, 5-10-1 Fuchinobe, Sagamihara, Kanagawa 252-5258, Japan \\
$^{16}$ Graduate School of Engineering, Osaka University, 2-1 Yamadaoka, Suita, Osaka 565-0871, Japan
}
\date{Accepted XXX. Received YYY; in original form ZZZ}
\pubyear{2024}
\begin{document}
\label{firstpage}
\pagerange{\pageref{firstpage}--\pageref{lastpage}}
\maketitle
\begin{abstract}
Recent observations of a few young pulsar wind nebulae (PWNe) have revealed their morphologies in some detail. Given the availability of spatio-spectral-temporal data, we use our multi-zone (1D) leptonic emission code to model the PWNe associated with G29.7$-$0.3 (Kes~75) and G21.5$-$0.9 (G21.5) and obtain (by-eye) constraints on additional model parameters compared to spectral-only modelling. Kes~75 is a Galactic composite supernova remnant (SNR) with an embedded pulsar, PSR~J1846$-$0258. X-ray studies reveal rapid expansion of Kes~75 over the past two decades. PWN G21.5 is also a composite SNR, powered by PSR~J1833$-$1034. For Kes~75, we study a sudden plasma bulk speed increase that may be due to the magnetar-like outbursts of the central pulsar. An increase of a few percent in this speed does not result in any significant change in the model outputs. For G21.5, we investigate different diffusion coefficients and pulsar spin-down braking indices. We can reproduce the broadband spectra and X-ray surface brightness profiles for both PWNe, and the expansion rate, flux over different epochs, and X-ray photon index vs epoch and central radius for Kes~75 quite well. The latter three features are also investigated for G21.5. Despite obtaining reasonable fits overall, some discrepancies remain, pointing to further model revision. We find similar values to overlapping parameters between our 1D code and those of an independent 0D dynamical code ({\sc TIDE}). Future work will incorporate spatial data from various energy wavebands to improve model constraints.
\end{abstract}
\section{Introduction}
Pulsars are compact, rapidly rotating, highly magnetised objects. They inject a relativistic stream of charged particles known as the pulsar wind into their surroundings. These charged particles ($e^{\pm}$ and possibly nuclei and ions) in combination with the local magnetic field generate an ambient structure termed the pulsar wind nebula (PWN; see, e.g., \citealt{Gaensler06}, and references therein), which is visible across the electromagnetic spectrum. The multi-wavelength radiation is produced via the synchrotron radiation (SR) process (radio to gamma rays, up to the MeV-GeV range) and via synchrotron self-Compton (SSC) or external-Compton (EC) mechanisms in the very-high energy (VHE; 100~GeV $\le E_{\gamma} < $ 100~TeV) and ultrahigh energy (UHE; $E_{\gamma
} \ge 100$~TeV) regimes \citep[e.g.,][]{1996ApJ253D,2006AnA621D, 2007MNRAS625F,2010ApJ1248T,Martin2012, 2013Ap7634T,2018AnA609A110Z,van2020MNRAS3091V,2023MNRAS5858J}. While the majority of PWN models (including this work) invoke a leptonic origin for the broadband emission spectrum, in recent years the tail of the TeV-PeV gamma-ray spectrum in the Crab PWN has been linked to the presence of cosmic-ray protons beyond $10^{15}$ eV in this object \citep{2021ApJ221L,2022ApJ7P}.

PWNe are inferred to evolve quite rapidly with time, as deduced by studying objects of different ages. The bulk of the evolution apparently happens within the span of tens of thousands of years. It is all the more novel to directly observe temporal variability on many different scales; for example, the expansion of a particular PWN over the time scale of decades, as in the case of Kes~75 \citep{Reynolds2018} or to observe the dynamic morphological changes in PWN elements such as the corkscrew jet of Vela \citep{Kargaltsev03}.
The general evolution of a PWN is characterised by three main phases coupled with the surrounding supernova remnant (SNR; see, e.g., \citealt{SlanePWN2017} for a review). In the initial phase, a young PWN expands inside the freely expanding SNR ejecta. Early analytical studies were done by \citet{Reynolds1984,ChevalierFransson1992}, and later, numerical simulations for cases ranging from classical to relativistic hydrodynamical systems, were performed by \citet{vanderSwaluw1998, Blondin2001,vanderSwaluw2001,Bucciantini2003,Bucciantini2004}. The next stage involves an unsteady phase for the PWN due to the collision with the SNR reverse shock \citep{McKee1974} which happens for PWNe older than $>1.5$~kyr \citep{vanderSwaluw2001}. The reverberations from the reverse shock result in contraction and expansion of the PWN \citep{vanderSwaluw1998,Blondin2001,vanderSwaluw2001,Torres2018,Bandiera2023a,Bandiera2023}. In the final phase of the evolution, the PWN has left or survived the host SNR, and its interaction with the interstellar medium results in a bow-shock structure for the PWN (see, e.g., \citealt{Wilkin1996} in context of generic bow-shocks; \citet{Olmi2019} and references therein for numerical simulation studies).

Most theoretical work has been done on the first and simplest PWN phase, and mostly within the context of single-zone (0D) models. This evolutionary phase will also be the focus of the current paper. However, in order to improve our model constraints, we will jointly fit spectral, temporal, and spatial data using a multi-zone (1D) model, for two particular young objects: G29.7$-$0.3 (Kes~75) and G21.5$-$0.9 (G21.5). The reasons for this choice are as follows. In the case of Kes~75, in addition to the recently analysed archival spatial data \citep{Hu2022}, there are intriguing temporal X-ray data that indicate the rapid expansion of the PWN over the past two decades \citep{Reynolds2018}. The embedded pulsar also exhibits mysterious magnetar-like bursts that may lead to an increase in the bulk flow speed of the surrounding plasma, with the possibility of related observational signatures. In the case of G21.5, we can use the broadband radiation spectrum and surface brightness (SB) profile to constrain parameters such as the diffusion coefficient and $B$-field spatial behaviour.

In this article, we will use a 1D leptonic emission code that calculates relativistic particle injection, transport, and emission as the particles traverse the PWN \citep{CvR2018MNRAS_G09}. We also use the 0D code {\sc TIDEfit}, see \cite{Martin2022} and references therein, to cross check our models. Below, we give an observational review for each of our sources (Section~\ref{sec:ObsReview}). A brief description of our 1D model and its implementation is provided in Section~\ref{sec:model}, with a detailed parameter study and results showing a simultaneous representation of multiple data for both sources presented in Section~\ref{sec:param} and Section~\ref{sec:res}, respectively. The discussions of our model results are given in Section~\ref{sec:disc} and  conclusions follow in Section~\ref{sec:con}.

\section{Observational Review of Sources}\label{sec:ObsReview}

\subsection{Kes 75}\label{sec:ObsReviewKes75}
Kes~75 \citep{Kesteven1968} is a prototypical composite SNR that contains a PWN powered by pulsar PSR~J1846$-$0258 (hereafter J1846). It is also one of the youngest SNRs in the Galaxy. \citet{Becker1984} first argued that this is a ``composite SNR'' containing both a shell and a PWN radio component, the latter having a radio index $\alpha\sim0.0$ ($S_\nu\propto \nu^\alpha$) and a highly polarised, centrally-peaked radio distribution, upon imaging this source with the Very Large Array (VLA) at 2, 6, and 20~cm. \citet{Salter1989} added data at 84 GHz and concluded there may be a spectral steepening between $15 - 84$~GHz ($\alpha\sim-0.3$). Its distance was estimated to be $d=5.8\pm0.5$~kpc \citep{Verbiest12}, but recently \citet{2018AJ204R} updated the estimate to $d=5.6\pm0.3$~kpc. 

Kes~75 has been detected at VHEs by the High Energy Stereoscopic System (H.E.S.S.), with a flux of 2.4 $\times$ 10$^{-12}$ erg cm$^{-2}$ s$^{-1}$ and a photon spectral index of $\Gamma \sim 2.3$ \citep{Terrier2008}. At these energies, the source manifests as a point source, and it is not clear what the relative flux contribution of the SNR shell versus that of the PWN might be (a similar observational situation is found in the detection by \textit{INTEGRAL}; cf.\ \citealt{McBride08}). However, the size of this point-like TeV source is $0.010\pm0.013^\circ$, which is much smaller than the diameter of the SNR shell of $3^\prime$ but comparable to the PWN's diameter of $40^{\prime\prime}$ in the radio band \citep{HESSgps2018, Ng08}. Furthermore, assuming a distance of 6 kpc for this source, the estimated PWN spectral energy distribution (SED) and the derived conversion efficiency of the spin-down luminosity of the pulsar to the $\gamma$-ray emissions are consistent with the origin of a PWN \citep[see, e.g.,][]{Torres2014, Tanaka_Takahara_2011}. Moreover, this TeV source is classified as a PWN in the TeVCat catalogue \citep{2008ICRC823D}. Therefore, we also consider this VHE source to be the TeV counterpart of PWN Kes~75.

In the GeV band, \citet{StraalEtAl2023} detected flux above 100~MeV which they considered to be a combination of both pulsar magnetospheric and PWN emission. They concluded that the emission in the range 100~MeV $-$ 2~GeV should originate from the pulsar’s magnetosphere, and they model the emission above 10~GeV as originating from the PWN. They regarded the point source with a log-parabola spectrum, 4FGL J1846.9$-$0247c, to be the counterpart of Kes~75. This has been the case up to and including the Data Release 3 of the Fourth \textit{Fermi} Point Source Catalog \citep[4FGL-DR3;][]{FermiDR2_2020,2023Fermi4FGL_DR4}. However, with the release of 4FGL-DR4 \citep{2023Fermi4FGL_DR4,2023FermiDR4} and subsequent GeV PWN studies, a new faint point-like source, 4FGL J1846.4$-$0258, is now added as the counterpart to Kes~75 instead (private communication, J.\ Eagle). Also, since the GeV emission from Kes~75 is faint, and 4FGL J1846.9$-$0247c is at a distance of $ >0.2^{\circ} $ from Kes~75, it can further be argued that the latter source is not associated with the Kes~75 PWN. In what follows, we indicate the published \textit{Fermi} Large Area Telescope (LAT) data from \citet{StraalEtAl2023} and compare the SED prediction from our PWN model to these older data. Since our model does not include pulsed emission from the pulsar, we show their data below 2~GeV (interpreted to be from the pulsar) for representation purposes, but we do not model those data.

In the X-ray regime, the \textit{Chandra} X-ray Space Telescope was, however, able to resolve the morphology of the source as well as infer its expansion from 2006 to 2016, tentatively detecting no discernible proper motion of the pulsar \citep{Reynolds2018}. \citet{Ng08} found a jet-torus PWN structure in the X-rays, with the absence of an observable counterjet, yielding a lower limit on the flow speed of $0.4c$. On the other hand, \citet{Reynolds2018} found a northern and southern jet in the keV band, and observed great changes in their morphologies on a timescale of a decade,  attributable to the short-term pulsar activity (See their Fig.~3). Absence of shell emission in the east points to a density gradient of the medium into which Kes~75 is expanding. Recently, \cite{Hu2022} analysed archival \textit{Chandra} data for various sources (spanning $2000$ to $2016$ for Kes~75). They fitted spatial features of several sources using a pure diffusion model, suggesting the dominance of diffusion in the particle transport in the majority of these sources. To explain the discrepancy in consistency of parameters for a couple of their targets, they proposed the need for a spatially-dependent magnetic field, which is something we consider in this paper.

Kes~75's embedded pulsar, J1846, was discovered as a rotation-powered pulsar in 2000 \citep{Gotthelf2000} in the X-ray band, but no pulsations were detected in the radio band \citep{Archibald08} and also not significantly (4.2$\sigma$) in the $30 - 100$~MeV gamma-ray band \citep{Kuiper2018} nor in the $>100$~MeV band \citep{StraalEtAl2023}, due to a lack of statistics. It is one of the most energetic pulsars known, with a period of $\sim 326$ ms and a spin-down luminosity of 8.1$\times 10^{36}$ erg s$^{-1}$. For a period derivative of 7.1$\times 10^{-12}$ s s$^{-1}$ \citep{Gotthelf2000}, it has a small characteristic age of $\tau_{\rm c} \equiv P/2\dot{P}\sim 728$ yr (for the canonical value of the braking index, $n=3$). Its inferred magnetic field of $5\times10^{13}$~G places it on the verge of the magnetar population. Indeed, in 2006, this pulsar exhibited magnetar-like short X-ray outbursts \citep{Gavriil2008, Kumar2008}, with a subsequent softening of the X-ray spectrum \citep{Gavriil2008} and a change in the pre-burst braking index value from $n= 2.65 \pm 0.01$ \citep{Livingstone2006} to $n=2.16 \pm 0.13$ \citep{Livingstone2011}. \citet{Archibald15} found an index $n=2.19\pm0.03$ over a seven-year period following the outburst (i.e., from 2006 to 2013), with evidence for a glitch that occurred between 2005 and 2008. After 14 years of quiescence a pulsar reactivation (outburst in August 2020) was reported by \citet{Krimm2020,Blumer2021}. \citet{HuEtal2023} detected a large spin-up glitch in \textit{NICER} (Neutron star Interior Composition ExploreR) monitoring data and an increase in pulsed X-ray flux, similar to the behaviour of the 2006 outburst. They reported a braking index of $n=2.7 \pm 0.2$ before this outburst, indicating that between 2006 and 2020, the spin-down relaxed to the original evolution before the 2006 outburst. The braking index value post-2020 outburst has not been estimated yet due to a lack of data. 

\subsection{G21.5} G21.5 has been known as a radio source in our Galaxy from the 1970s. It was later observed as an extended X-ray and radio source by the \textit{Einstein} Observatory and radio instruments \citep{1970AnAS....1..319A, becker1981ApJ23B}. It is a composite SNR, where the central region is plasma-filled and powered by a pulsar (PSR J1833$-$1034), and it also has an expanding surrounding shell of material powered by the supernova blast wave. The period of the pulsar is $P = 61.9$~ms and its derivative is $\dot{P} = 2 \times 10^{-13} $s s$^{-1}$ \citep{camilo2006ApJ637456C}. Its current spin-down power is $\simeq 3.4 \times 10^{37}$ erg s$^{-1}$, ranking it as the fifth highest $\dot{E}_{\rm rot}$ pulsar in our Galaxy \citep{2005AJ....129.1993M}. The characteristic age of the pulsar is $\tau_{\rm c} = 4.9$ kyr \citep{gupta2005CSc853G}. The revised source distance is $d=4.4 \pm 0.2 $ \citep{2018AJ204R} compared to earlier estimations of $\sim 4.7$ kpc \citep{becker1981ApJ23B,tian2008MNRAS391L4T}. 

The radio spectrum of G21.5 is similar to that of two other young PWNe: Crab and 3C~58. It is uniform over the circular projected image of the PWN of $\sim 1^\prime$ in extent \citep{2008MNRAS861411B}. Using the expansion speed of the nebula, its age is estimated\footnote{If one assumes $\tau_{\rm exp} = P/(n-1)\dot{P}$, one finds $n\sim1.4$. This is similar to the estimation made using the {\it Giant Metrewave Radio Telescope} of $n = 1.857$ \citep{roy2012MNRAS213R}.} to be $\tau_{\rm exp}=870^{+200}_{-150}$ years \citep{2008MNRAS861411B}. 

Infrared (IR) observations of this object were performed using the Very Large Telescope (VLT), the Canada-France-Hawaii Telescope, and the \textit{Spitzer} Space Telescope by \citet{2012AnA12Z}. Significant linear polarisation was found, indicating a non-thermal origin of the IR emission, which helps to constrain the synchrotron part of the spectrum. However, their observations are from a 2$^{\prime\prime}$ region around the central pulsar and also the flux levels were lower compared to the IR observations by \citet{gallan1999ESASP313G}. In our modelling, we have considered the IR observations based on \citet{gallan1999ESASP313G}, that also indicate that G21.5's PWN  has a morphology very similar to that of the Crab.

In the $0.1-10$ keV band, \textit{Chandra} X-ray observations revealed a non-thermal emitting region of size of $40^{\prime\prime}$ in which the X-ray spectrum becomes steeper with radial distance from the pulsar \citep{Matheson2005}. \textit{NuSTAR} also observed this object within a radius of $165^{\prime\prime}$, and reported that the spectra in the energy range of $3-45$~keV are better explained by a broken power-law with a break at $9.7 \pm 1.3 $~keV \citep{nynka2014ApJ_78972N}. The spectral indices below and above the break energy are $1.996 \pm 0.013$ and $2.093 \pm 0.013$, respectively. However, \textit{Hitomi} analysis selected a spatial extent of $140^{\prime\prime}$. Their observations in the energy range $0.8-80$~keV revealed a break at $7.1 \pm 0.3$~keV, characterised by a lower spectral index value of $1.74 \pm 0.02$ before the break energy. Beyond the break, the index value increased to $ 2.14 \pm 0.01$, with the emission predominantly originating from the PWN. The exact cause of the lower value of the break energy remains unknown and warrants further investigation \citep{hitomi2018PASJ7038A}. Recently, \citet{Hu2022} used $0.5-8$ keV \textit{Chandra} X-ray archival observations (spanning $1999-2014$) to derive a spectral index of $\sim 1.4$ in the innermost region, steepening to $\sim 2.2$ at a radius of $35^{\prime\prime}$. This steepening might be connected with the injection of electrons from the central pulsar and SR cooling as they traverse the PWN magnetic field \citep{slane2000ApJ533L29S}. This source is useful to cross-calibrate instruments, for example X-ray satellites, due to its lower brightness compared to the Crab PWN \citep{tsuj2011AA25A25T}. 

\textit{INTEGRAL} observations suggest that there is a relatively small
contribution ($<20\%$) from the pulsar in the $0-200$ keV energy range, and most of it is due to a PWN of size $40^{\prime\prime}$ \citep{de2009MNRAS93527D}. On the other hand, \textit{Fermi} LAT observations reported no detection of GeV radiation from this object \citep{GeV_l_2011ApJ35A}. TeV gamma-ray observations of this region identified the source HESS~J1833$-$105 \citep{Djannati2008}. This object was also observed by H.E.S.S.\ during their Galactic Plane Survey and the emission properties were consistent with the previous observations \citep{HESSgps2018}. The H.E.S.S.\ source spectral shape $(\Gamma = 2.1 \pm 0.2)$ is very similar to the \textit{INTEGRAL} hard X-ray spectral index of $\Gamma_{X} = 2.2 \pm 0.1$ \citep{de2009MNRAS93527D}.

\section{Model Assumptions and Technical Details}\label{sec:model}

\subsection{Theoretical framework}
The leptonic PWN model we use in this paper is basically the same as described in \citet{2018arXiv180200216V}, \citet{CvR2018MNRAS_G09}, and \cite{CarloPhD2020}; the only update is adding an increase in the bulk flow speed $V_0$ after some time $t_0$ in the case of Kes~75 that may be due to the magnetar-like flares that deposit additional energy into the system. \citet{Martin20} suggest that such bursts may lead to increased pair formation and hence an injection of additional energetic leptons into the PWN, or boost the mean strength of the nebular magnetic field. Such perturbations may yield increases in the flux in some radiative bands, depending on model parameters. Here, we investigate the scenario where the bulk flow is suddenly increased by a few percent, given the energy deposition via the bursts (See Appendix~\ref{app:BurstLikeEnergyInjection}).
We furthermore exploit new model outputs and fit them to new data (expansion rate, time-dependent X-ray flux; Section~\ref{sec:outputs}), in addition to previous model outputs.
In this section, we therefore summarise only the main elements of the spatio-temporal model. More details may be found in \citep{CvR2018MNRAS_G09}.

As a first step, we solve the following Fokker-Planck-type transport equation:
\begin{equation}
\begin{split}
\frac{\partial N_{\rm{e}}}{\partial t} =& -\mathbf{V} \cdot (\nabla N_{\rm{e}}) +  \kappa \nabla^2 N_{\rm{e}} + \frac{1}{3}(\nabla \cdot \mathbf{V})\left( \left[\frac{\partial N_{\rm{e}}}{\partial \ln E_{\rm{e}}} \right] - 2N_{\rm{e}} \right)   \\
&+ \frac{\partial }{\partial E}(\dot{E}_{\rm{e,rad}}N_{\rm{e}}) +  Q(\mathbf{r},E_{\rm{e}},t),
\end{split}
\label{eq:transportFIN}
\end{equation} 
where $N_{\rm{e}}(\mathbf{r},E_{\rm{e}},t)$ is the number of leptons per unit energy and volume, \textbf{V} is the bulk velocity of leptons, $\kappa$ is the spatially-independent diffusion coefficient, $E_{\rm{e}}$ is the particle energy, $\dot{E}_{\rm{e,rad}}$ is the total, i.e., the sum of the SR and inverse Compton (IC) scattering, energy loss rates, and $Q$ is the particle injection spectrum. Here $r$ is the radial dimension (assuming spherical symmetry) and $t$ is the time since the PWN's birth. We solve this transport equation in time and energy, including one spatial dimension. 

The particle injection spectrum is assumed to be a broken power-law:
\begin{equation}
 Q(E_{\rm{e}},t) = \left\{\begin{matrix}
Q_0(t)\left(\frac{E_{\rm{e}}}{E_{\rm{b}}}\right)^{-\alpha_1} \qquad E_{\rm{e,min}} \leq E_{\rm{e}}<E_{\rm{b}}\\ 
Q_0(t)\left(\frac{E_{\rm{e}}}{E_{\rm{b}}}\right)^{-\alpha_2} \qquad E_{\rm{b}} < E_{\rm{e}} \leq E_{\rm{e,max}}
\end{matrix}\right.,
\label{brokenpowerlaw}
\end{equation}
where $Q_0(t)$ is the time-dependent normalisation constant that is determined by equating the first moment of the injection spectrum to a constant fraction $\eta$ of the time-dependent pulsar spin-down luminosity $L(t)$, $E_{\rm{b}}$ is the injection spectral break energy, $E_{\rm{e,min}}$ ($E_{\rm{e,max}}$) is the minimum (maximum) energy of the injected leptons, and $\alpha_1$ and $\alpha_2$ are the spectral indices. To limit the number of free parameters in this model, we assume that $\alpha_1$ and $\alpha_2$ are time-independent. The choice of the injection spectrum as a broken power-law is motivated observationally and has been invoked in phenomenological models, where the value of $\alpha_1$ typically varies in the range $1.0-1.8$, and after the break the value of $\alpha_2$ is found to lie in the range $2.0 - 3.1$ \citep{Tanaka_Takahara_2011,2013Ap7634T,Vorster2013_3,2018AnA609A110Z}. Since we are injecting the particle spectrum at the inner boundary of our radial grid, we can formally write $Q(\mathbf{r},E_{\rm{e}},t) = Q(E_{\rm{e}},t) \delta(r - r_{\rm min})$, where $r_{\rm min}$ is taken to be the PWN shock radius.

For the low-energy part of the broken power-law distribution, \citet{Tanaka2017} found that the low-energy particles can originate from stochastic acceleration of relic particles or of particles injected at the interface of the supernova ejecta \citep{Tanaka2023, ShutaWataru2024}. Although the injection point of the low-energy particles is still a matter of contention (e.g., \citealt{AharonianAtoyan1996, AmatoEtAl2000}), the assumption that the low-energy particles are also injected from the PWN shock radius does not affect our results, since the low-energy particles are responsible for the radio emission of the spatially-integrated SED.

The energy losses in the model are due to radiative and adiabatic energy losses. The calculations for radiative losses (SR and IC scattering) are similar to the globular cluster modelling of \citet{Kopp2013} and the expressions can be found in Section~$2.3$ of \citet{CvR2018MNRAS_G09}. 

The adiabatic losses caused by the bulk motion of the particles in the expanding PWN are given by the $\frac{1}{3}(\nabla \cdot \mathbf{V})$ term in Equation~(\ref{eq:transportFIN}).
Since this is a radiative model, not a magnetohydrodynamic (MHD) one, we parameterise the bulk flow speed profile of the leptons as well as the magnetic field's spatial and temporal profile. The first is parameterised as $V(r) = V_0\left(r/r_0\right)^{\alpha_{\rm{V}}}$, where $V_0$ is the bulk flow velocity\footnote{As in \citet{Kennel1984a}, we assume that $V_0$ is a constant. This means that the velocity profile only changes with space, and not with time. It may be interesting to consider the impact of $V_0(t)$ in future, but this will introduce more free parameters to the model, that would need to be constrained by, e.g., MHD modelling.} at the termination shock at $r_0 = r_{\rm min}$, and $\alpha_{\rm{V}}$ is the bulk-flow index parameter. We parameterise the latter as $B(r,t) = B_{\rm{age}}\left(r/r_0\right)^{\alpha_{\rm{B}}}\left(t/t_{\rm{age}}\right)^{\beta_{\rm{B}}},$ where $B_{\rm{age}}$ is the present-day magnetic field at $r = r_0$ and $t = t_{\rm{age}}$ ($t_{\rm{age}}$ is the age of PWN). With \citet{Kennel1984a}, we assume that the magnetic field is toroidal (azimuthal) and the bulk flow is purely radial (further discussion of this assumption is provided in \citealt{CvR2018MNRAS_G09}). To limit the number of free parameters for this work, we assume a fixed $\beta_B = -1$ (the variation for $\beta_B$ will be explored in our follow-up works). In the free expansion stage of the PWNe, a range of values of $\beta_B$ ($-1.15$ to $-1.45$) have been used \citep{Vorster2013_3}, while in other works, this parameter was fixed to $-1.3$ \citep{Reynolds1984, 2010ApJ1248T}. Varying $\beta_B$ in our model  between this range did not substantially change the model outputs.

\begin{figure}
  \centering
  \includegraphics[width=1.0\linewidth]{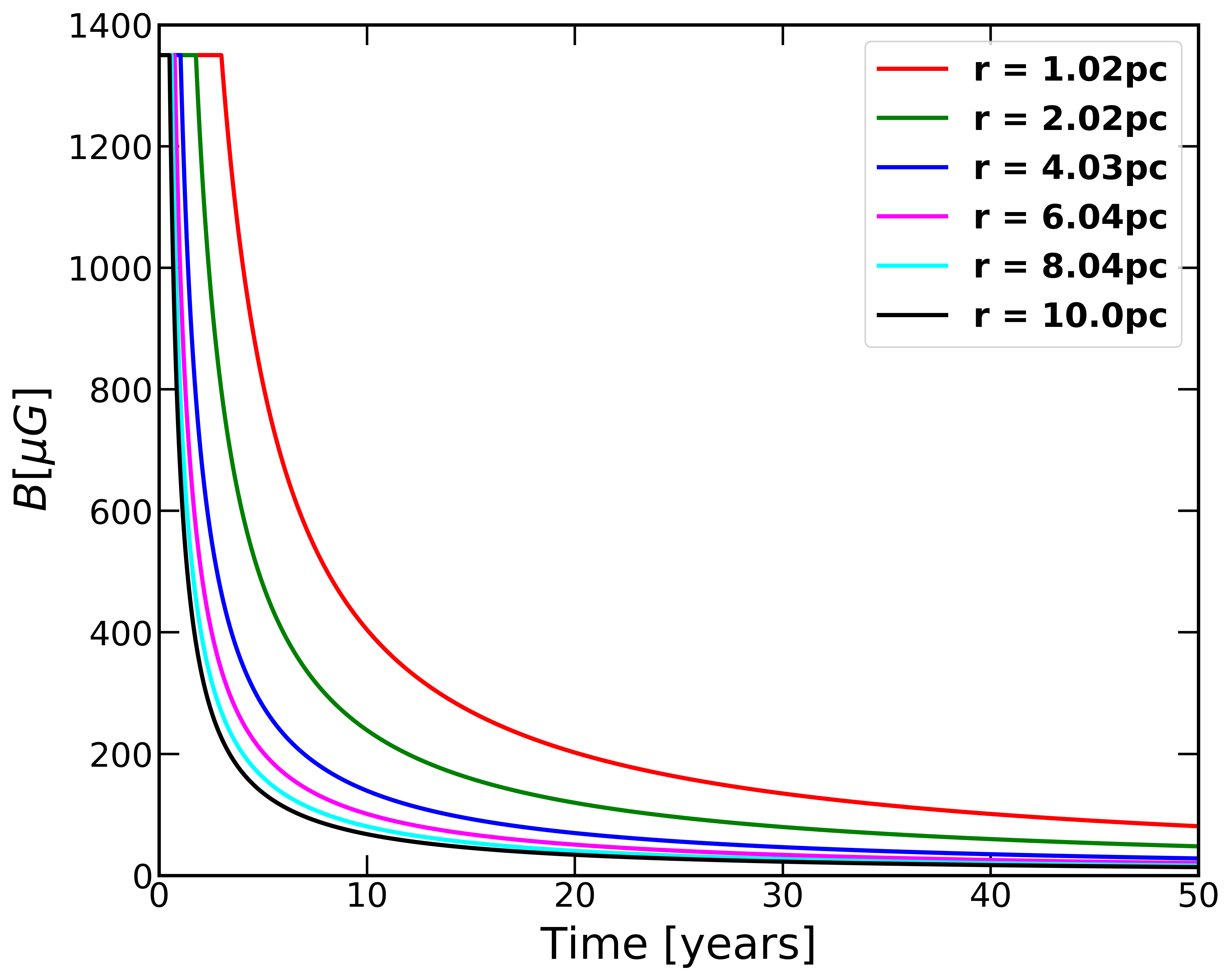}
\caption{Parameterised nebular magnetic field for a present-day field of $B_{\rm age} =135 \mu G$ ($\alpha_{\rm{B}} = -0.8$, and $\beta_{\rm{B}} = -1.0$) for the first $50$-year lifespan of Kes~75 over different radial distances in parsec (pc). }
\label{fig:magfield_plot} 
\end{figure}

Since this implies that $B(r,t)$ becomes infinite at $t=0$ (i.e., for negative values of $\beta_{\rm B}$) in this approximate parameterised expression, we limit the maximum magnetic field to $10~B_{\rm age}$. To test the robustness of this prescription, we also tested the value $B_{\rm max} = 30~B_{\rm age}$ and did not find significant changes to our model outputs, given our assumed value for current age, $t_{\rm age}$ (because the $B$-field is only very large for a very short time and for a small region in space). Raising this maximum value increases the computational time significantly while affecting the results only marginally, hence we consider this to be a reasonable limit, since we do not expect assumptions for $t_{\rm age}$ to change too much. The magnetic field variation for different radial distances over the first $50$ years of Kes~75's lifespan is shown in Fig.~\ref{fig:magfield_plot}.

We furthermore assume that, since the nebular plasma is a good conductor, we can apply the ideal MHD equations to describe the PWN wind such that the magnetic field is frozen into the outflowing plasma. In this case, Ohm's law becomes
\begin{equation}
    \mathbf{E} = -\frac{\mathbf{v}}{c}\times\mathbf{B}.
\end{equation}
By combining this with Faraday's law, we find \citep[e.g.,][]{Ferreira_deJager2008}
\begin{equation}
    \frac{\partial\mathbf{B}}{\partial t} = \nabla\times(\mathbf{v}\times\mathbf{B}).\label{eq:Faraday}
\end{equation}
To link the radial profiles of the magnetic field and bulk motion of the particles, we assume with \cite{Schock2010,Holler12,Fang2017,Fang2019} that the temporal change in the magnetic field is slow enough that we can set $\partial \mathbf{B}/\partial t \simeq 0$ in the above equation, even for a time-dependent prescription of $B$. (This assumption holds exactly for steady-state models such as those of \citealt{Kennel1984a,Vorster2013}; it may not be fully applicable for the first decade of the evolution, see Fig.~\ref{fig:magfield_plot}, although this represents only a small fraction of the total age.) From this follows 
\begin{equation}
VBr=V_0 B_{\rm age} r_0={\rm constant},
\end{equation}
which for our simplified parametric specifications of the magnetic field and bulk flow implies that
\begin{equation}\label{eq:alpValpB}
 \alpha_V + \alpha_B = -1.
\end{equation}
This relation is used to reduce the number of free parameters by one and to simplify our exploration. 

The diffusion is assumed to be either Bohm-type, 
\begin{equation}
    \kappa(E_{\rm{e}}) = \kappa_{\rm B} \frac{E_{\rm{e}}}{B(r,t)},
    \label{eqn:BohmDiffusion}
\end{equation}
with $\kappa_{\rm B} = c/3e$ and $e$ denoting the elementary charge, or a power-law type (e.g., Kolmogorov-type),
\begin{equation}
\kappa(E_{\rm{e}}) = \kappa_{\rm X} \left(\frac{E_{\rm{e}}}{E_0}\right)^\delta, 
\label{eqn:KolmogorovDiffusion}
\end{equation}
where $\kappa_{\rm X}$ and $\delta$ are the diffusion coefficient normalisation and exponent respectively, similar to what has been used in Galactic cosmic-ray studies. We present a motivation of the choices for $\delta$ below in Section~\ref{sec-dif}, whereas $\kappa_{\rm X}$ is left as a free parameter to be constrained by the spatial and spectral data.

\subsection{Diffusion}\label{sec-dif}
Although a considerable amount of theoretical work \cite[see, e.g.,][]{S02,Shalchi09,Shalchi20,EngelbrechtEA22} has been done in the context of the heliosphere, very few studies of particle diffusion in PWNe have been performed \cite[see, however,][]{Tang2012,Vorster2013,PorthEA16,OlmiEA16,Olmi2019,Zhu2023,LuEA23}. This renders studies of the relative influence of diffusion as a transport mechanism in these environments difficult. These difficulties are further exacerbated by the lack of observational constraints pertaining to both the large-scale behaviour of PWN plasma quantities as well as the nature of the plasma turbulence. The latter information is essential when modelling the diffusion coefficients of energetic charged particles. 

However, some information as to the energy dependencies of these coefficients can be gleaned from existing theories, by making assumptions as to the nature of the turbulence, as well as the large-scale PWN magnetic field. Assuming the pulsar wind magnetic field (PWMF) is essentially frozen into the outflow of charged particles when $\sigma$ (the ratio of electromagnetic energy density to that of particles) is low (i.e., a high plasma-$\beta$ scenario), and a radial outflow of said particles, the PWMF can be assumed to spatially scale as a divergence-free \citet{Parker1958} magnetic field, given by

\begin{equation}
    \mathbf{B} (r,\theta) = B_{0s}\left(\frac{r_{0}}{r}\right)^{2} \left( \hat{\mathbf{u}}_{r}-\tan{\Psi \hat{\mathbf{u}}_{\phi}} \right)
\end{equation}

\begin{equation}
    \qquad \tan \Psi=\frac{\Omega (r-r_{0}) \sin \theta}{V_{0}},
\end{equation}
where the Parker field is assumed to emanate from a source surface located at a distance $r=r_{0}$ (with a value there of $B_{0s}$), nominally where $\sigma$ drops below 1. Furthermore, $V_{0}$ denotes the pulsar wind speed, $\Omega$ is the rotation rate of the pulsar wind at $r_0$, and $\Psi$ is the winding angle between the magnetic field and the radial direction. As a 1D approach is being taken in this paper, as a first approach we can assume that the co-latitude  $\theta=90^{\circ}$ so that, given the high rotation rate and pulsar wind speed, the above reduces to the familiar  $r^{-1}$ radial dependence for the magnetic field magnitude (beyond the pulsar light cylinder, where the corotation speed equals the speed of light). Given the self-consistent constraint of Equation~(\ref{eq:alpValpB}), this would mean that $\alpha_B =-1$ and $\alpha_{\rm V} =0$. This, of course, is only valid prior to any later shock in the plasma. For instance, in the heliosphere the stellar wind drops below supersonic speeds at $\sim85$~AU generating the heliospheric termination shock, after which the Parker field would not adequately describe the heliospheric magnetic field \cite[see, e.g.,][and references therein]{KleimannEA22}. Note that the presence of analogous termination shocks in PWNe are expected from theory and observation \cite[see, e.g.,][]{Lyub,KirkEA09,Slane2016,Cerutti20}. Assumptions about the PWN field structure leads to a constraint for parameter studies, particularly for $\delta$, which governs the energy dependence of the diffusion coefficient in Equation~(\ref{eqn:KolmogorovDiffusion}). For a pulsar, such a Parkerian field would be essentially azimuthal, with the implication that the radial diffusion coefficient, which is what is technically employed in our 1D transport model, would be dominated by the diffusion coefficient perpendicular to the magnetic field. 

It should be noted, however, that the perpendicular diffusion coefficient would be expected to be considerably smaller than the diffusion coefficient parallel to the PWMF \cite[see, e.g.,][]{EngelbrechtEA22,HerbstEA22}. The strong background PWMF would imply the presence of strongly anisotropic transverse 2D turbulence, based on MHD simulations performed by \citet{Shebalin83}, which could in principle lead to effective cross-field pitch-angle scattering. In such a scenario, a reasonable model for the perpendicular diffusion coefficient would be the nonlinear guiding centre (NLGC) theory of \citet{MattEA03}. Cosmic-ray modulation studies employing diffusion coefficients derived from this theory have successfully reproduced spacecraft observations of Galactic cosmic-ray protons and antiprotons \cite[e.g.][]{EngelbrechtMoloto21}, and said diffusion coefficients are in good agreement with numerical test-particle simulations in synthetic turbulence \cite[e.g.,][]{MinnieEA07}. Although more complex expressions for electron diffusion coefficients have been proposed and employed in modulation studies \cite[see, e.g.,][]{Engelbrecht19,DempersEngelbrecht20}, the NLGC expression has the benefits of being relatively tractable as well as successful in reproducing heliospheric observations.

The NLGC theory, however, requires at the very least an estimate for the form of the 2D magnetic turbulence power spectrum, which is obviously currently impossible to ascertain from observations for PWN contexts. MHD simulations can provide useful insights, such as those of \citet{PorthEA16}, who analyse turbulence generated via their MHD model and report a spectrum with a wavenumber-independent energy-containing range, and a Kolmogorov inertial range. The NLGC diffusion coefficient also needs, at the very least, an estimate of the energy/rigidity dependence of the parallel mean free path. A reasonable approach, based on the assumption that highly energetic particles would only be resonantly scattered by turbulent fluctuations with characteristic length scales comparable to the particle Larmor radii, and that such fluctuations would be found within the wavenumber-independent energy-containing range of the turbulence power spectrum, would be to assume that the parallel mean free path scales as $R^2$, where $R$ denotes the particle's rigidity, following what is expected from magnetostatic quasilinear theory \citep{Jokipii66,TeufelSchlickeiser03}. The NLGC theory predicts a perpendicular mean free path that scales as $\lambda_\perp\sim \lambda_{\parallel}^{1/3}$, with $\lambda_{\parallel}$ the parallel mean free path, leading to a perpendicular diffusion coefficient that scales, at very high energies, as $R^{2/3}$, implying that $\delta =2/3$. Should, however, the particles in question be resonantly scattered by fluctuations in the inertial range, whether the spectral index of this be the Kolmogorov value ($-5/3$) or the Iroshnikov-Kraichnan value ($-5/2$), the parallel mean free path would scale as $R^{1/3}$ or $R^{1/2}$, respectively \cite[see, e.g.,][]{CaballeroEA19}. This would then imply an NLGC perpendicular mean free path that scales as $R^{1/9}$ or $R^{1/6}$, implying that $\delta = 1/9$ or $\delta = 1/6$. In what follows, we investigate the effect of assuming different values of $\delta$, and also consider $\kappa_{\rm X}$ to be free, to find optimal by-eye fits to the data. See Table~\ref{tbl:Kes75} and~\ref{tbl:G21p5} for the best qualitative fit values of these parameters.

\subsection{Implementation}\label{sec:TheCode}
Our model is spherically-symmetric, spatially multi-zonal, and time-dependent, and is implemented in a code that calculates the transport and radiation of particles as they traverse a PWN. We use a spatial grid of concentric spherical shells of increasing radius, centred on the pulsar, which we call zones. The code spans three dimensions: space/radius, time, and energy. The radial binning is linear and is determined by $\delta_{\rm r} = \left( r_{\rm max} - r_{\rm min} \right) / \left( N - 1\right)$, where $\delta_{\rm r}$ is the radial step, $r_{\rm min}$ and $r_{\rm max}$ are the chosen minimum and maximum values of $0.02$ and $10.0$~pc, respectively, and $N$ is the total number of radial bins. The results are a bit sensitive to the choice of $r_{\rm min}$. If the inner boundary is chosen to be too distant, we cannot reproduce some of the spatial data at lower radii. Therefore, we tested different values of $r_{\rm min}$ and found that $0.02$ is most suitable. We choose $r_{\rm max} = 10.0$~pc, which is much larger than the observed radii of both our sources ($\sim 0.8$~pc), to ensure that the solution is not prematurely cut off due to an imposed boundary condition\footnote{While we assume that the PWN is in a freely-expanding stage, we effectively ignore the outer SNR conditions; there is therefore no hydrodynamic boundary in our model.}. We tested that our spectral results are not too dependent on this choice by checking for convergence within acceptable limits for larger values of this parameter. The energy bins are calculated on a logarithmic scale, using $\delta_{\rm E} = \ln\left( E_{\rm max}/E_{\rm min} \right) / \left( M - 1\right)$, where $\delta_{\rm E}$ is the logarithmic energy step, and $E_{\rm max}$ and $E_{\rm min}$ are the maximum and minimum lepton energy values (different for particles and also different for the various radiation mechanisms), and $M$ is the total number of energy bins. For a complete description of the binning procedure, refer to Appendix A.2 in \cite{CarloPhD2020}. We tested for convergence of our outputs for both the spatial and energy dimensions for a different number of bins. We found that beyond a combination of $N = 180$ and $M = 200$, there are not any appreciable differences in the SED and hence, we fix our number of bins at these values.

The simulation begins at $t = 0$ representing the time of the birth of the PWN. The simulation is allowed to go up to the $t_{\rm age}$. The transport of the population of particles inside the PWNe is solved numerically using Equation~(\ref{eq:transportFIN}) over the entire age of the PWN \citep{2018arXiv180200216V}. The pulsar is allowed to spin down according to the usual braking formula \citep[e.g.,][]{Pacini1973} $L(t) = L_0(1 + t/\tau_0)^{-\gamma},$ with $\tau_0\equiv P_0/(n-1)\dot{P}_0 = P/(n-1)\dot{P}-t_{\rm age}$ the characteristic age at birth. The canonical value of $n=3$ (implying $\gamma= (n+1)/(n-1)=2$) represents a dipolar magnetic field, and measurements yield typical values between $n\sim2-3$ (see, for e.g., \citealt{Gaensler06}), although this value maybe be greater, or vastly greater when measured on small time scales (\citealt{2016ArchibaldbrakingindexGT3, 2016Eksi, 2023LiBiao, 2017TongKau} or \citealt{Parthasarathy2020MNRAS, Lower2021, Johnston2017}, respectively). Such large braking indices may be connected, among other things, to changes in the pulsar moment of inertia, magnetic field decay, glitches, or the alignment of the magnetic and spin axes with time.

We lastly perform a line-of-sight (LOS) calculation \citep{CvR2018MNRAS_G09} to project radiation from multiple spherical shells onto the plane of the sky, which enables us to estimate the SB and flux as a function of radial distance.

\subsection{The output plots}\label{sec:outputs}

By solving Equation~\eqref{eq:transportFIN} numerically, we are able to predict the following PWN properties in a spherically-symmetric scenario:

\begin{table*}
\begin{tabular}{|lllll|}
\hline
\textbf{Output plot} & \textbf{Energy range} & \textbf{References} & \textbf{Energy range} & \textbf{References} \\ 
\hline
 & \textbf{Kes~75} &  & \textbf{G21.5} & \\
\hline
SED	&  Radio & \cite{Salter1989} & Radio to X-ray &  \cite{2020ApJ90432H} \\ 
&   & \cite{BockGaensler2005} \\ 
 & X-ray & \cite{Gotthelf_2021} \\
 &  TeV & \cite{HESSgps2018} & TeV & \cite{HESSgps2018}\\ 
 &  \textit{Fermi}  & \cite{StraalEtAl2023}\textsuperscript{*} & \textit{Fermi} upper limits & \cite{GeV_l_2011ApJ35A}\\ 
 Integral flux over different epochs & $1 - 8$~keV & \cite{Reynolds2018} & $3 - 8$~keV & \cite{2020ApJ90432H} \\
 SB profile & $0.5 - 8$~keV & \cite{Hu2022}  & $0.5 - 8$~keV & \cite{Hu2022}\\
 Expansion over different epochs & $0.7 - 8$~keV & \cite{Reynolds2018} & -- & --\\
 X-ray photon index versus distance & $0.5 - 8$~keV & \cite{Hu2022} & $0.5 - 8$~keV & \cite{Hu2022}\\
 X-ray photon index versus epoch & $1 - 8$~keV & \cite{Reynolds2018} & $3 - 8$~keV & \cite{2020ApJ90432H}\\
X-ray photon index versus energy &  --&--  & $0.3 - 45$~keV & \cite{2020ApJ90432H}\\
\hline 
\end{tabular}
\caption{A summary of data per energy range we used, corresponding with different model output plots. \textsuperscript{*}As described in Section~\ref{sec:ObsReviewKes75}, we differentiate the \textit{Fermi} data from \protect\cite{StraalEtAl2023} using dashed and solid lines to highlight the proposed contribution of the pulsar vs the PWN \protect\citep[as considered by][]{StraalEtAl2023}, respectively.}
\label{tbl:Data_energyrange_ref}
\end{table*}

\begin{itemize}
    \item \textit{Broadband SED}: The SED shows the full radiation spectrum for the PWN, integrated over the spherical radial coordinate (or equally, over the cylindrical LOS-projected radius), spanning from radio, through X-rays to the TeV energy range, and including both an SR and an IC component, but neglecting the pulsed GeV component expected from the central pulsar. 
    \item \textit{Integral flux for different epochs:} The integrated flux over a particular energy band is then calculated at different epochs (snapshots in time during the PWN evolution), but again integrated over all radii. In the case of Kes~75, this will be compared with the results in Table~3 of \citet{Reynolds2018}, where the flux was spatially integrated over the whole PWN and the pulsar was excluded. It should be noted that the observed decrease in X-ray flux was not spatially uniform, with most of the decline associated with the dimming of the northern knot, thus violating our spherically symmetric model that we apply as a first approximation here. We therefore emphasise that we do not attempt to model a significant flux decrease limited to a particular region, as our model cannot resolve such a small spatial scale. But we can compare changes in the total flux from our model distributed over the whole PWN with that given in their paper.
    \item \textit{SB profiles}: LOS calculations are then performed by finding the intersection volume of radiating spheres and cylindrical\footnote{These should actually trace out cones, but given the vast distance to the PWN compared to its size, a cylindrical approximation is well motivated.} lines of sight, thereby projecting the 3D PWN as a 2D image on the plane of the sky. This yields the predicted SB profile (as a function of projected radius $\rho$) at any snapshot in time, for a particular energy band. Details may be found in Section~2.9 of \citet{CvR2018MNRAS_G09}. A significant update from previous works using this model is that instead of calculating the SB profile (and some other features) for a single energy value (the midpoint value of the relevant energy band), we integrate quantities over the energy band as specified by the observations (provided in Table~\ref{tbl:Data_energyrange_ref}).
    \item \textit{PWN expansion for different epochs}: \citet{Reynolds2018} calculated an average expansion of Kes~75 by taking account of changes in the physical image (i.e., spatial) and SB (i.e., flux) over time, with the expansion centred on the pulsar. They focused on the relatively faint northwestern rim that is far removed from the jets where significant morphological and flux variations were expected. Combining three independent expansion rate measurements over different epochs, they finally found an average expansion rate of $\sim0.249\% \pm 0.023\%$~yr$^{-1}$. This corresponds to a PWN expansion velocity of $\sim1000$~km\,s$^{-1}$, depending on the distance. To match these observations, we calculate the cumulative flux contained within a particular projected radius (specifically, the integral of $dN_\gamma/dE_\gamma$ over $0.7-8$~keV, normalised to the total flux found by integrating over all radii) and then use interpolation to find the $\rho_{68}$, which is the radius containing $68\%$ of this integrated flux\footnote{We fit the integrated flux profile using a Stretched Exponential Decay function of the form $a {\rm e}^{(-r/b)^{c}}$ to ensure smoothness of the profile for interpolation.}. We use $\rho_{68}$ as the effective PWN size at a particular time, and compute the relative expansion of this radius over several different epochs. We compared our outputs to the expansion rates at different epochs (spanning $\sim7$, $10$, and $16$ years), as listed in Table~2 of \citet{Reynolds2018}.
    \item \textit{X-ray photon index versus distance}: The SR component of the emitted SED for each projected annulus on the sky (i.e., after the LOS calculation) is used to calculate the X-ray photon index as a function of $\rho$ by fitting a power-law function to the model SED in the X-ray energy band for several annuli centred on the pulsar. This is done over different epochs, and for different $\rho$ (lines of sight).
    \item \textit{X-ray photon index versus epoch}: We calculate the photon index as before, but for the SED integrated over all radii, taking a snapshot at a particular epoch.
    \item \textit{X-ray photon index versus energy}: We calculate the photon index as before but it is averaged over energy for specific energy intervals to be able to compare the X-ray part of our SED to the X-ray energy data from \cite{2020ApJ90432H} in the case of G21.5.
\end{itemize}

Depending on data availability for each PWN, many of these predictions are simultaneously explored. Whether a 1D model parameter set represents data well or not, is judged by eye (i.e., not via a statistical prescription, but qualitative fits only) and the preferred model parameters are thus determined. We summarise the energy ranges for the data and their references in Table~\ref{tbl:Data_energyrange_ref}.

\section{Parameter study}\label{sec:param}
Below, we calculate time scale versus energy plots for relevant cases to study the dominant energy loss mechanism at a particular point in space, time, and energy. The synchrotron cooling timescale, $\tau_{\rm SR}$, is given by \citep{1970RvMP237B}: 
\begin{equation}
\tau_{\rm SR} \simeq 1.3 \times 10^{3}~{\rm yr} \left(\frac{B(r,t)}{100~\mu G}\right)^{-2} \left(\frac{E_{\rm e}}{1~{\rm TeV}}\right)^{-1},
\label{eq:tausyn}
\end{equation}
with $E_{\rm e}$ the particle energy.
The IC losses are calculated numerically using the total Klein-Nishina cross section \citep{RybickiLightman1979}, with the associated time scale being $\tau_{\rm IC} \equiv E_{\rm e}/\dot{E}_{\rm e,IC}$. Due to the diffusion process, we have the Bohm time scale
\begin{equation}
\tau_{\rm diff}^{\rm Bohm} \simeq 3.6 \times 10^{6}~{\rm yr} \left(\frac{B(r,t)}{100 \mu {\rm G}}\right) \left(\frac{E_{\rm e}}{1 ~{\rm TeV}}\right)^{-1} \left(\frac{R_{\rm PWN}(t)}{2 {\rm pc}}\right)^2,
\end{equation}
where $R_{\rm PWN}$ is the radius of the PWN (here a fiducial value to indicate its effect on the diffusion time scales, and not used in the current model context). Similarly, we also investigate the power-law nature of diffusion coefficients (for more details, see Section~\ref{sub_diff_Kes75} and Section~\ref{sub_diff_G21p5}) and the corresponding time scale here for a given choice in $\kappa_{\rm X}$ is
\begin{equation}
\tau_{\rm diff}^{\rm \delta} \simeq 2 \times 10^{5}~{\rm yr} \left(\frac{\kappa_{\rm X}}{6\times10^{24}~{\rm cm}^{2}\,{\rm s}^{-1}}\right) \left(\frac{E_{\rm e}}{1 ~{\rm TeV}}\right)^{-\delta} \left(\frac{R_{\rm PWN}(t)}{2 {\rm pc}}\right)^2.
\end{equation}
Some theoretical background for different scenarios for the parameter $\delta$ is discussed in Subsection~\ref{sec-dif}. We treat $\kappa_{\rm X}$ as a free parameter, taking note of the Galactic value of this normalisation as a reference value \citep[e.g.,][see Section~\ref{sec:G21diff}]{Vla2012ApJ68V}. Theoretically, this quantity would depend on turbulence levels and related parameters. As these are currently unknown, the present approach is taken. 

As a comparison, we calculate a typical time scale to traverse a particular zone, given the bulk flow speed of $V(r)$:
\begin{equation}
    \tau_{\rm esc} \approx \frac{\delta_{\rm r}}{V(r)} \approx 160~{\rm yr}\left(\frac{r_{\rm max}}{10~{\rm pc}}\right)\left(\frac{179}{N-1}\right)\left(\frac{10^8~{\rm cm}\,{\rm s}^{-1}}{V_0}\right).
\end{equation}

\begin{figure}
    \centering
\includegraphics[width=0.8\linewidth]{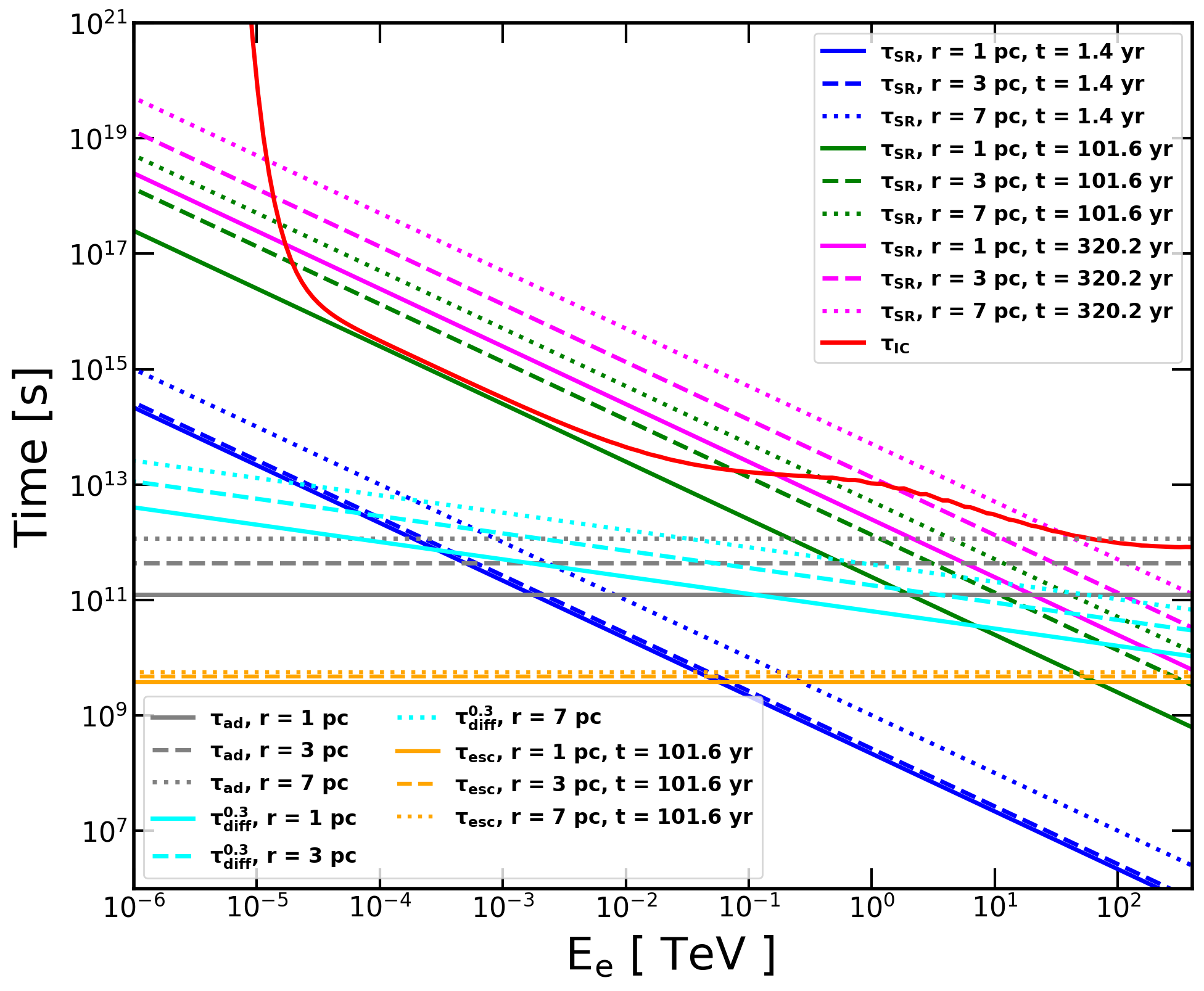}
    \caption{Time scale variation for PWN Kes~75 at three epochs, $1.4$, $101.6$, and $320.2$ yr for the different mechanisms as indicated in the legend. The radial distances are highlighted by different line types: solid, dashed, and dotted for $\sim 1$, $\sim 3$, and $\sim7$ pc. The model parameters are as given in Table~\ref{tbl:Kes75}.}
\label{fig:Kes75BestFit_Timescales}
\end{figure}

We have also computed the adiabatic loss time scales using $\tau_{\rm ad} = 3/(\nabla.\mathbf{V}) $ \citep[e.g.,][]{2013ApJ773139V}. 
The corresponding cooling time scales are shown in Fig.~\ref{fig:Kes75BestFit_Timescales} and Fig.~\ref{fig:Tscaleg21p5} for Kes~75 and for G21.5, respectively.

\subsection{Kes 75}
In this section, we show the influence of several free parameters in our model, corresponding to the present-day magnetic field, bulk-flow motion, diffusion, and braking index on various model outputs for Kes~75, and explain the physical reasons for the corresponding variations. Apart from the parameter under investigation, all the model parameters are fixed as listed in Table~\ref{tbl:Kes75}.

\subsubsection{Loss time scales}
We note from Fig.~\ref{fig:Kes75BestFit_Timescales} that IC losses are not that important. At low energies, 
particle escape from a zone due to the bulk motion dominates, while at higher energies (and small radii and early times), SR dominates. The SR timescales become longer with both time and distance, given the drop in the $B$-field with time and radius. The combined effects of these timescales lead to various observational signatures, as they determine the particle transport in the PWN.

\begin{figure*}
  \begin{subfigure}[t]{0.5\linewidth}
    \centering
    \includegraphics[width=0.8\linewidth]{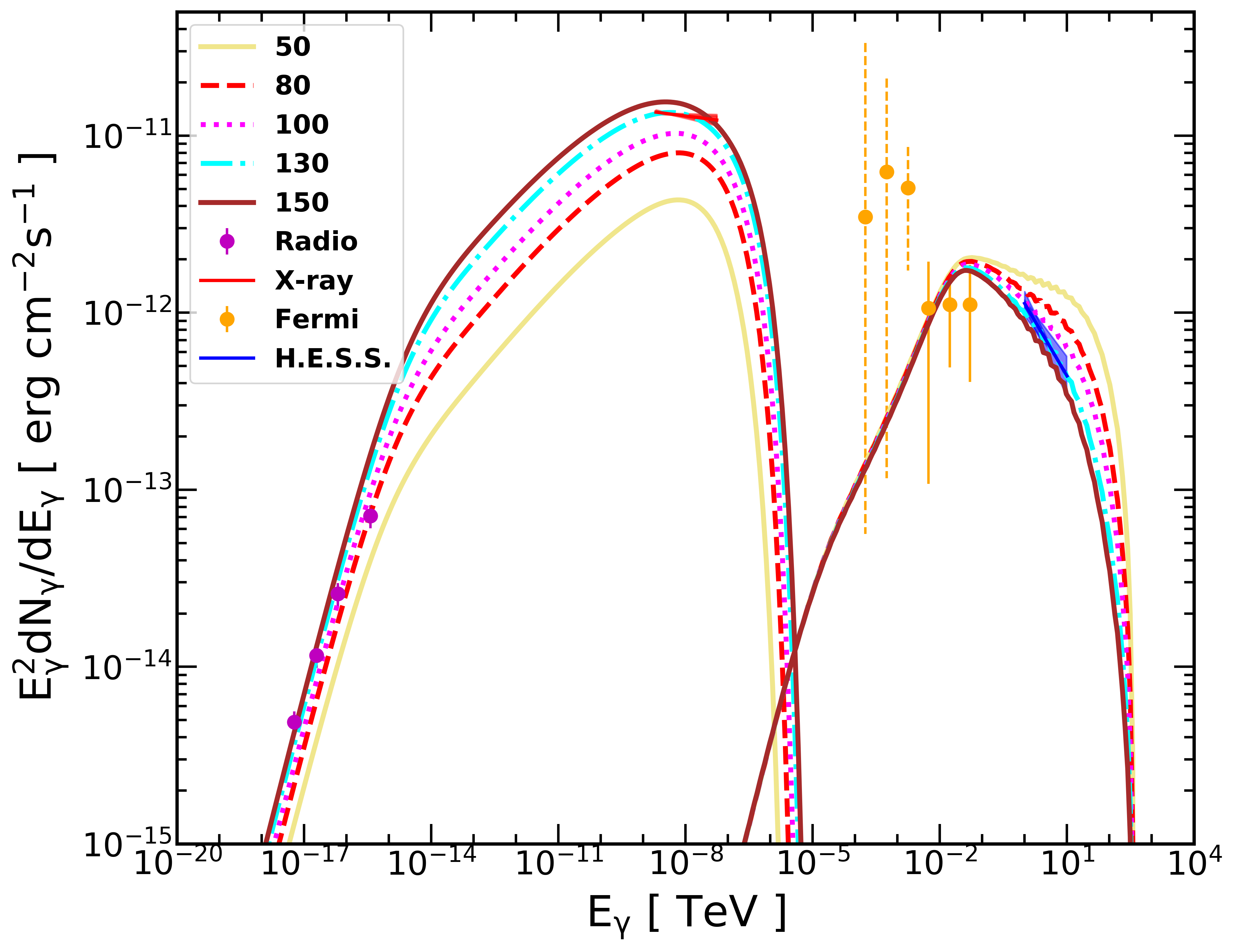}
    \caption{Variation in $B_{\rm{age}}$ (in $\mu$G).} 
    \label{fig:Kes75_mf_SED}
  \end{subfigure}
  \begin{subfigure}[t]{0.5\linewidth}
    \centering
    \includegraphics[width=0.8\linewidth]{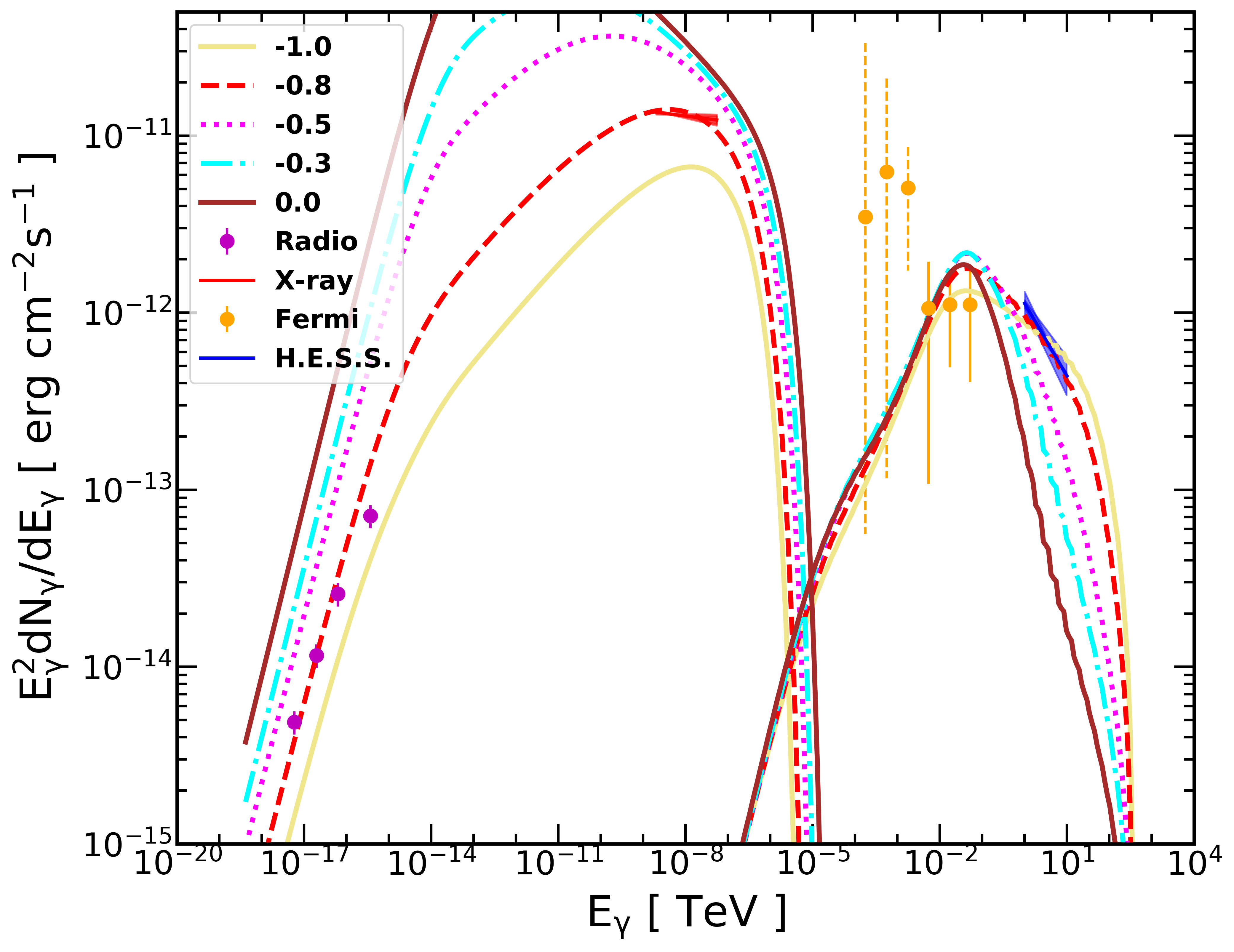}
    \caption{Variation in $\alpha_{\rm B}$.}
    \label{fig:Kes75_aB_SED}
  \end{subfigure} 
  \caption{The predicted SED for Kes~75, showing the effect of varying (a) the present-day magnetic field strength, $B_{\rm{age}}$, ($\mu$G) and (b) the radial dependence of the field via the parameter $\alpha_{\rm B}$ (for a fixed $B_{\rm{age}}=135\,\mu$G). All the other parameters are set to the preferred values given in Table~\ref{tbl:Kes75}.}
  \label{Kes75magfield} 
\end{figure*}

\subsubsection{Magnetic field}
\label{Kes75:Bfield}
The magnetic field prescription is one of the most influential parameters determining the shape of the SED. The parameters $B_{\rm{age}}$, $\alpha_{\rm B}$, and $\beta_{\rm B}$ collectively determine the peak location of the SED components. In Fig.~\ref{fig:Kes75_mf_SED}, the effects of considering different $B_{\rm{age}}$ are shown, where the values are varied from $50$ to $150~\mu$G, represented by different colours as shown in the figure key. Since the SR loss rate $\dot{E}_{\rm{e,SR}} \propto E_{e}^{2} B^{2} $, for a particular energy the SR losses (and thus SR flux) are higher for a higher value of $B_{\rm{age}}$ (since $B(r,t) \propto B_{\rm{age}}$). Also, it can be clearly seen that since for a lower magnetic field there are more surviving high-energy particles as opposed to the case for a higher $B_{\rm{age}}$ value, the tail of the IC spectrum is higher for lower magnetic fields.

In Fig.~\ref{fig:Kes75_aB_SED}, fixing $B_{\rm{age}}$ to be $135 \mu G$, we show variations in the SED for different $\alpha_{\rm B}$ values. Since we impose the condition $\alpha_V + \alpha_B = -1$, the variation in $\alpha_{\rm B}$ is linked to changes in the bulk flow's spatial profile. However, we see the SR spectrum going from higher to lower values in its peak as we increase the negative $\alpha_B$ value because of the relation $B \propto r^{\alpha_B}$. For example, moving from ${\alpha_B} = 0.0$ to $-0.5$ to $-1.0$, magnetic field scales as a constant to a $1/\sqrt{r}$ and to a $1/r$ profile, and for the same radial distances and age, it means that the magnetic field is smaller for more negative ${\alpha_B}$. Hence, the trend becomes similar to scaling for a higher or lower magnetic field value as discussed for Fig.~\ref{fig:Kes75_mf_SED} above. However, the effect is not as linear, because the linked change in  $\alpha_V$ also affects the bulk-flow motion and thus the adiabatic losses. One also notes that the SR cutoff moves to higher energies as the $B$-field is increased (in both panels of the Figure), since the critical SR frequency where the spectrum peaks $\nu_{\rm SR}\propto E^2_{\rm e}B$.

\begin{figure}
    \centering
    \includegraphics[width=0.8\linewidth]{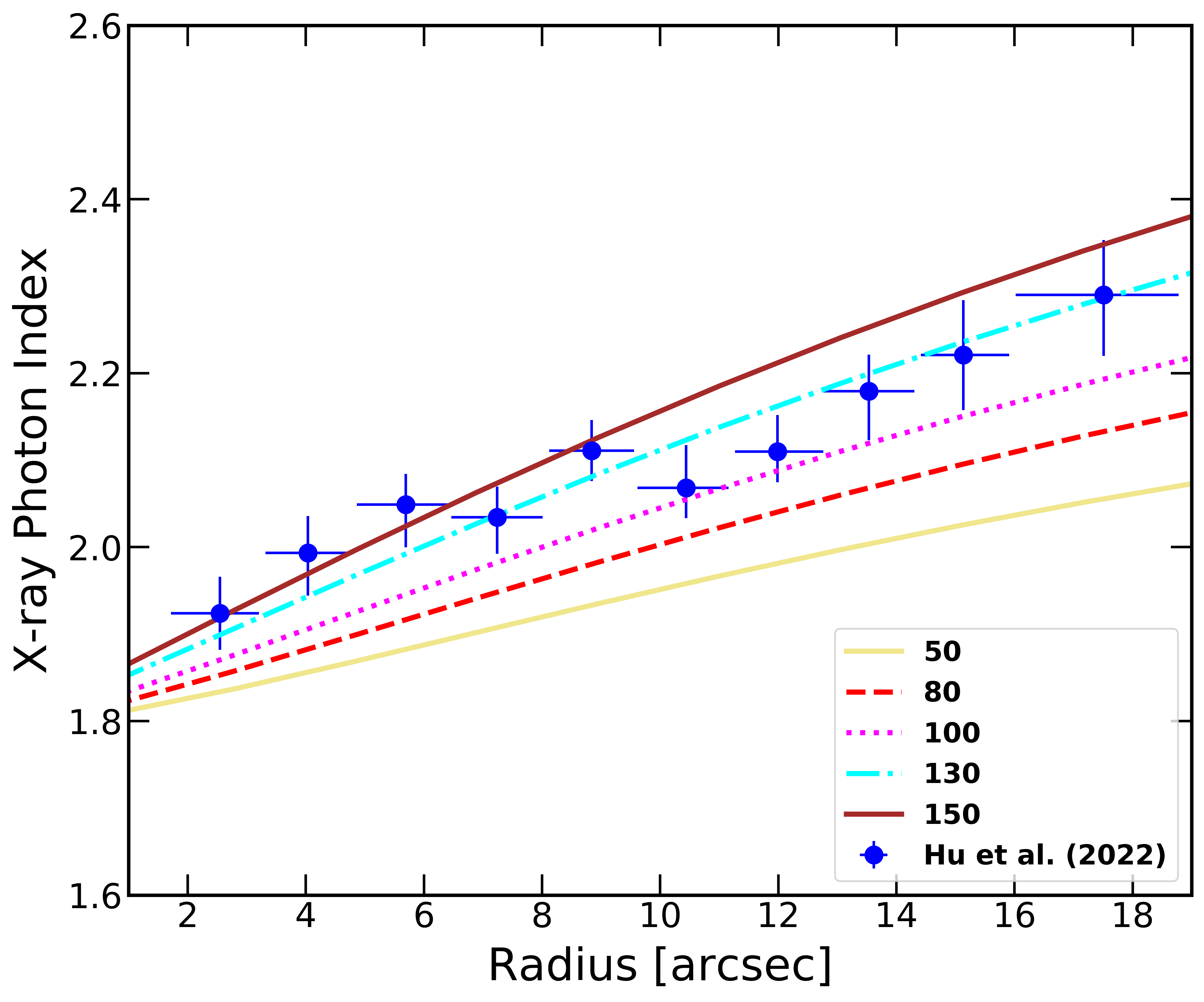} 
    \caption{The variation in X-ray photon index over projected radial distance for Kes~75. The assumed values for $B_{\rm{age}}$ ($ \mu G$) are noted in the key.} 
    \label{fig:Kes75_mf_vsr_Index} 
\end{figure}

We also show the effect of varying $B_{\rm{age}}$ on the X-ray photon index over radial distance in Fig.~\ref{fig:Kes75_mf_vsr_Index}. Since the SR mechanism is responsible for the X-ray emission, we fit a power-law to the SED for every $B_{\rm{age}}$ in the energy range $0.5-8$~keV and derive a spectral index that we can  match with the X-ray photon index data taken from \cite{Hu2022}. The calculated spectral index is shown in the figure, and as expected, the trend for different magnetic field values is similar to the SED in Fig.~\ref{fig:Kes75_mf_SED} - but narrowed down to the X-ray energy range. For a higher $B_{\rm{age}}$, we have increased SR losses and therefore we see a softer spectrum with a correspondingly larger spectral index and vice versa. We find that a $B_{\rm{age}}$ in the range of $130 \pm 10 \mu G$ reasonably reflects the photon index profile.

\subsubsection{Bulk flow}\label{Kes75:Bflow}
\begin{figure}
    \centering
    \includegraphics[width=0.8\linewidth]{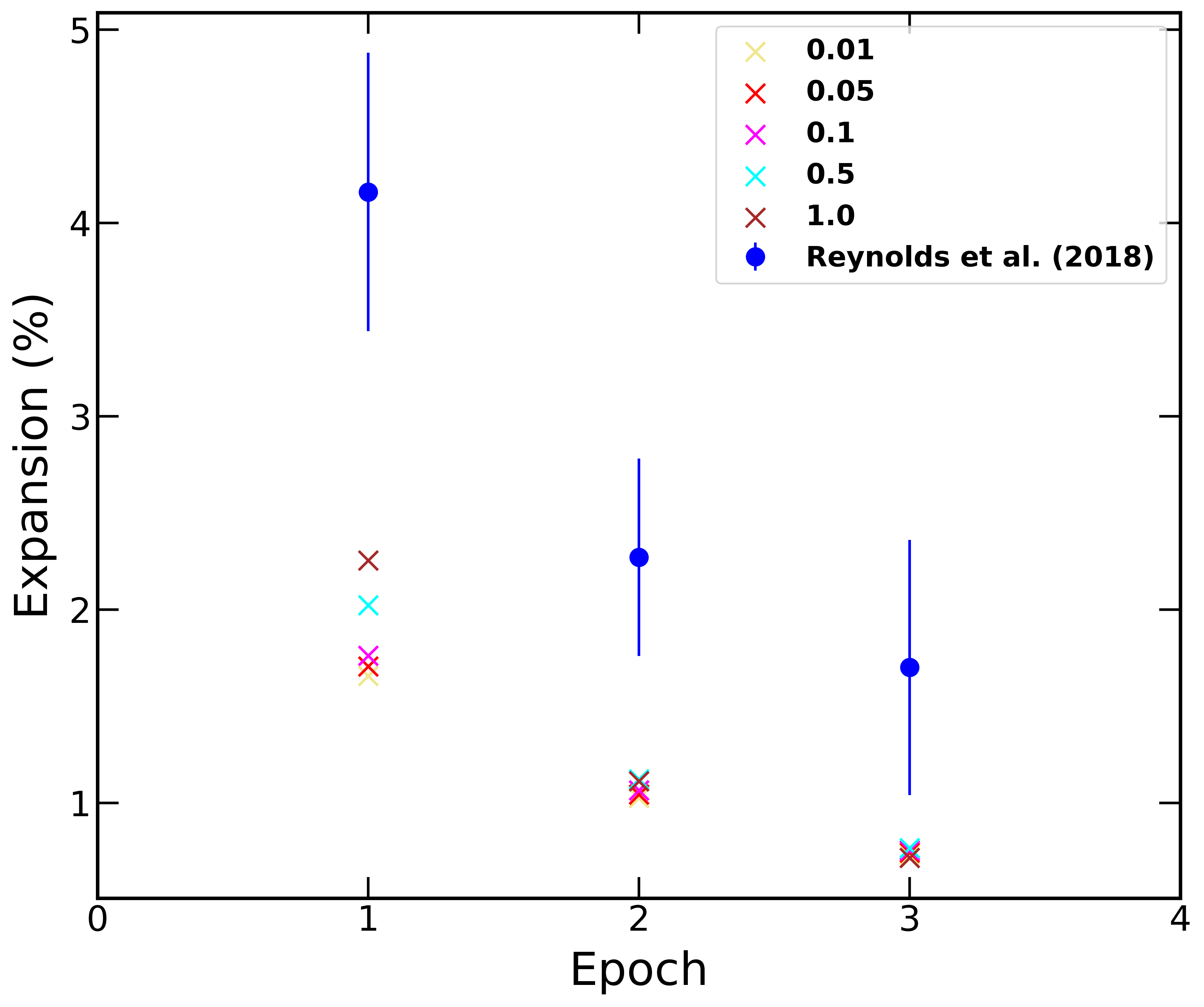}
    \caption{Expansion of  Kes~75 for different fractional increases $dV_0/V_0$ (as noted in the key) due to the burst-like injection of energy, introduced in the last $50$ years of its lifetime. Epochs $1$, $2$, and $3$ correspond to the intervals $2000-2016$, $2006-2016$, and $2009-2016$, respectively.}
    \label{fig:Kes75ExpansionBulkFlowBurstPercent}
\end{figure}

To model the effect of the magnetar-like outbursts linked with the glitching behaviour of J1846, we invoke a sudden injection of energy in Kes~75 in the last $50$ years of its lifetime, and link this to a sudden increase in the bulk speed. In Appendix~\ref{app:BurstLikeEnergyInjection}, we describe this burst-like injection of energy into the PWN in some detail, working out limits on the fractional increase in the bulk flow speed, $dV_0/V_0$, given some constraints on the fractional change in pulsar spin-down luminosity. We show that a change in $dV_0/V_0$ of up to a few percent is reasonably justified. In Fig.~\ref{fig:Kes75ExpansionBulkFlowBurstPercent} we show the effect on the expansion of Kes~75 for different burst-like injection scenarios. The fractional increase in $dV_0/V_0$ is shown in the legend of the figure, varying it from $1\%$ to $100\%$, where the latter means doubling the bulk flow suddenly. One sees that a higher bulk speed generally leads to a higher expansion rate, as expected. However, there is not much difference in the expansion for only a few percent increase in the bulk speed. However, for the initial epochs, there is a noticeable difference even for increments larger than $10\%$. We also varied the time of injection of this energy, ranging from $300$ to $5$~yr before the present age of the PWN, but it does not result in any considerable change in model outputs. We therefore infer that unless there is a substantial change in the fractional bulk speed, we cannot hope to see significant changes in the expansion of Kes~75.

\subsubsection{Diffusion}\label{sub_diff_Kes75}

\begin{figure*}
\begin{subfigure}[b]{0.5\linewidth}
\centering
\includegraphics[width=0.8\linewidth]{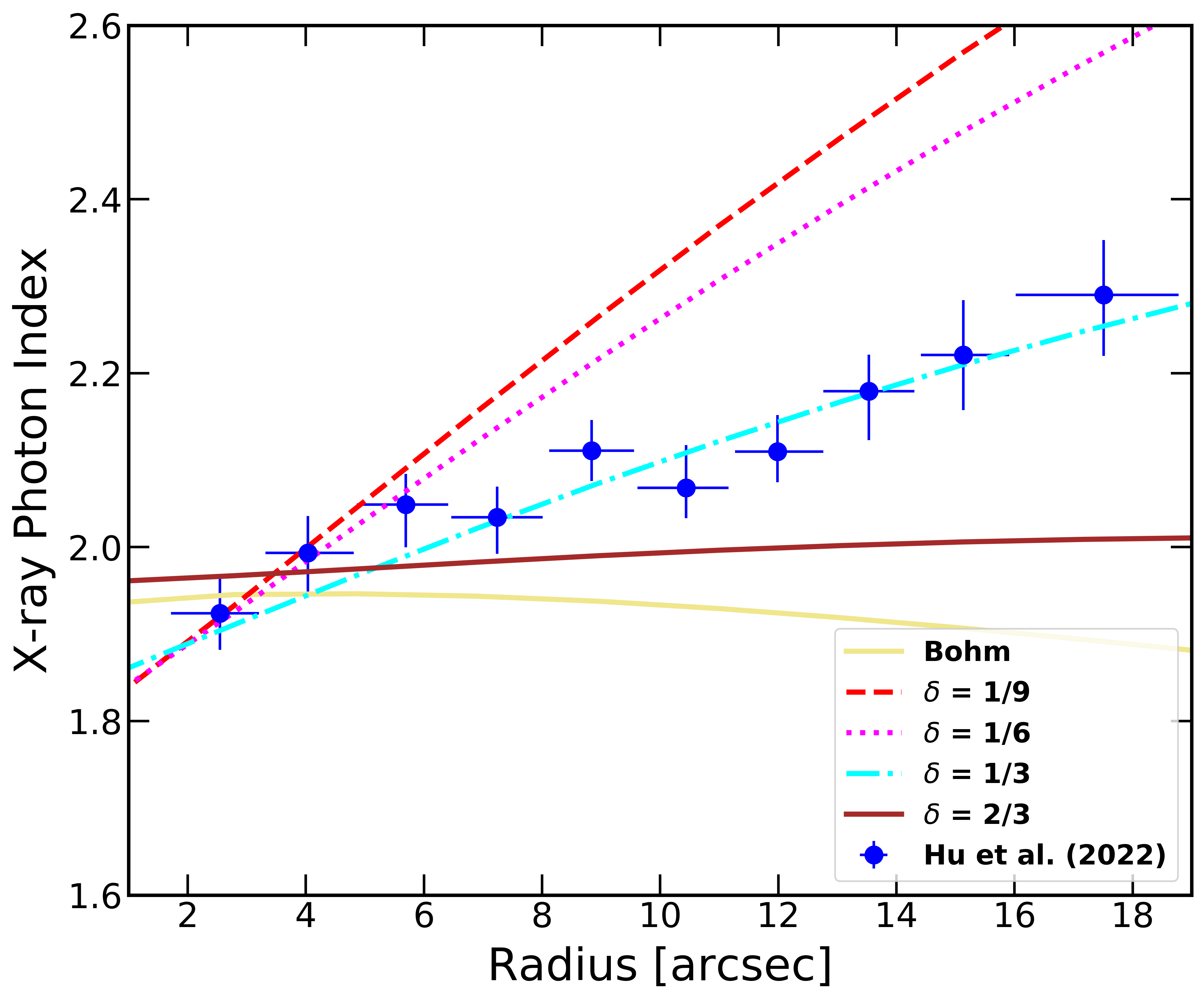}
\caption{X-ray photon index versus projected radial distance.} 
\label{fig:Kes75Diffusion:a} 
\vspace{4ex}
\end{subfigure}
\begin{subfigure}[b]{0.5\linewidth}
\centering
\includegraphics[width=0.8\linewidth]{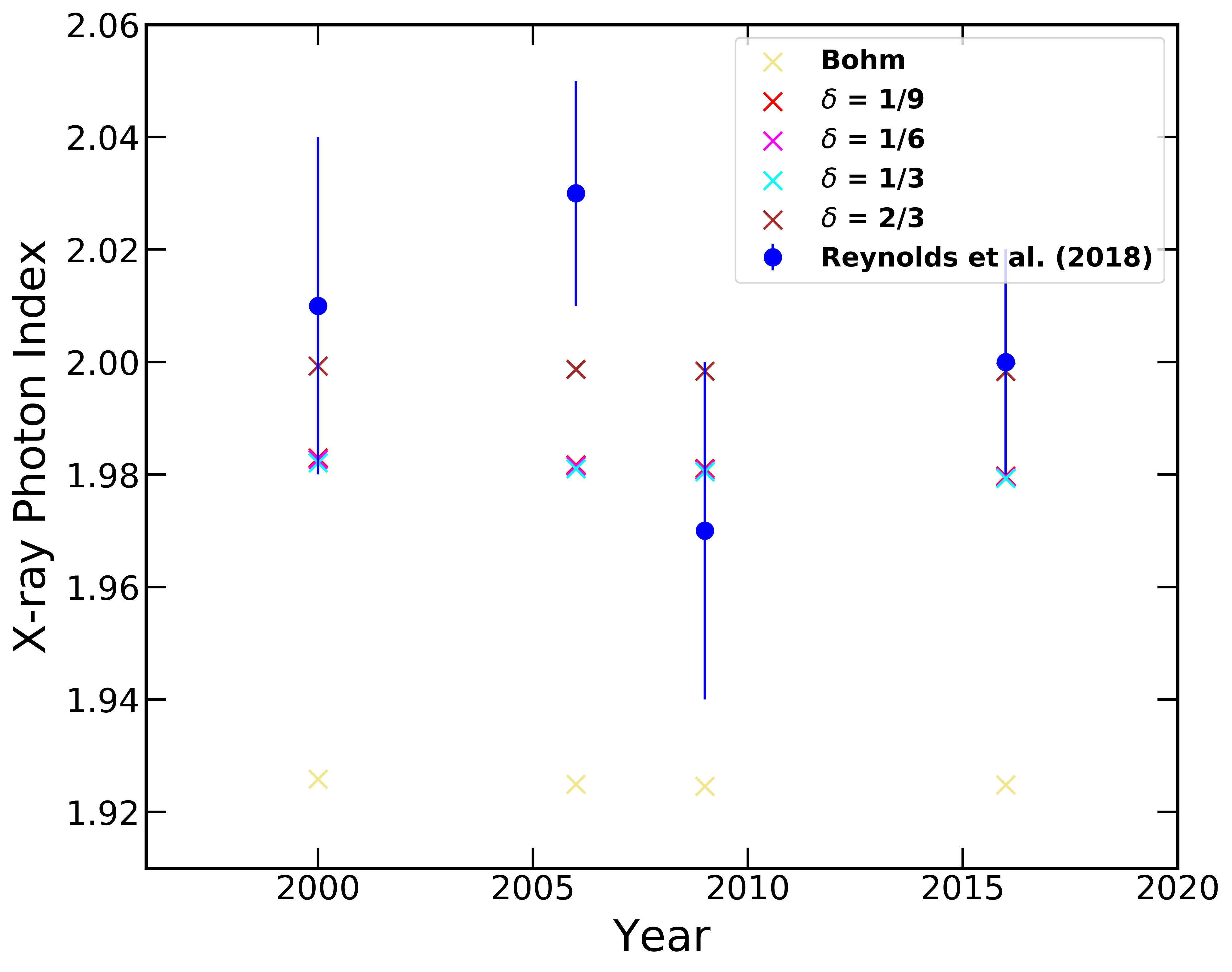} 
\caption{X-ray photon index versus epoch.} 
\label{fig:Kes75Diffusion:b} 
\vspace{4ex}
\end{subfigure} 
\begin{subfigure}[t]{0.5\linewidth}
\centering
\includegraphics[width=0.8\linewidth]{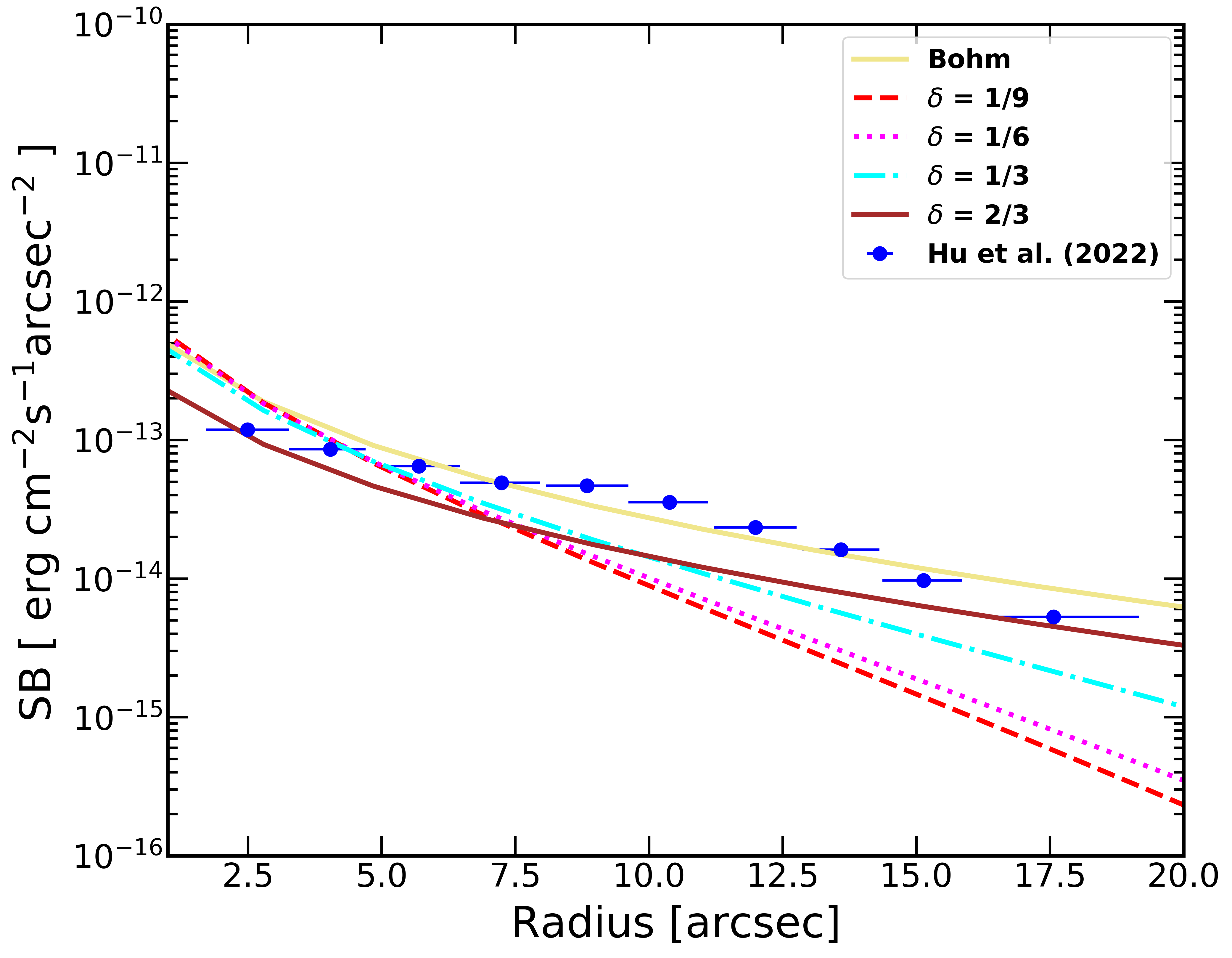} 
\caption{SB profile for different diffusion exponents $\delta$.} 
\label{fig:Kes75Diffusion:c} 
\end{subfigure} 
\hspace{-2ex}
\begin{subfigure}[t]{0.5\linewidth}
\centering
\includegraphics[width=0.8\linewidth]{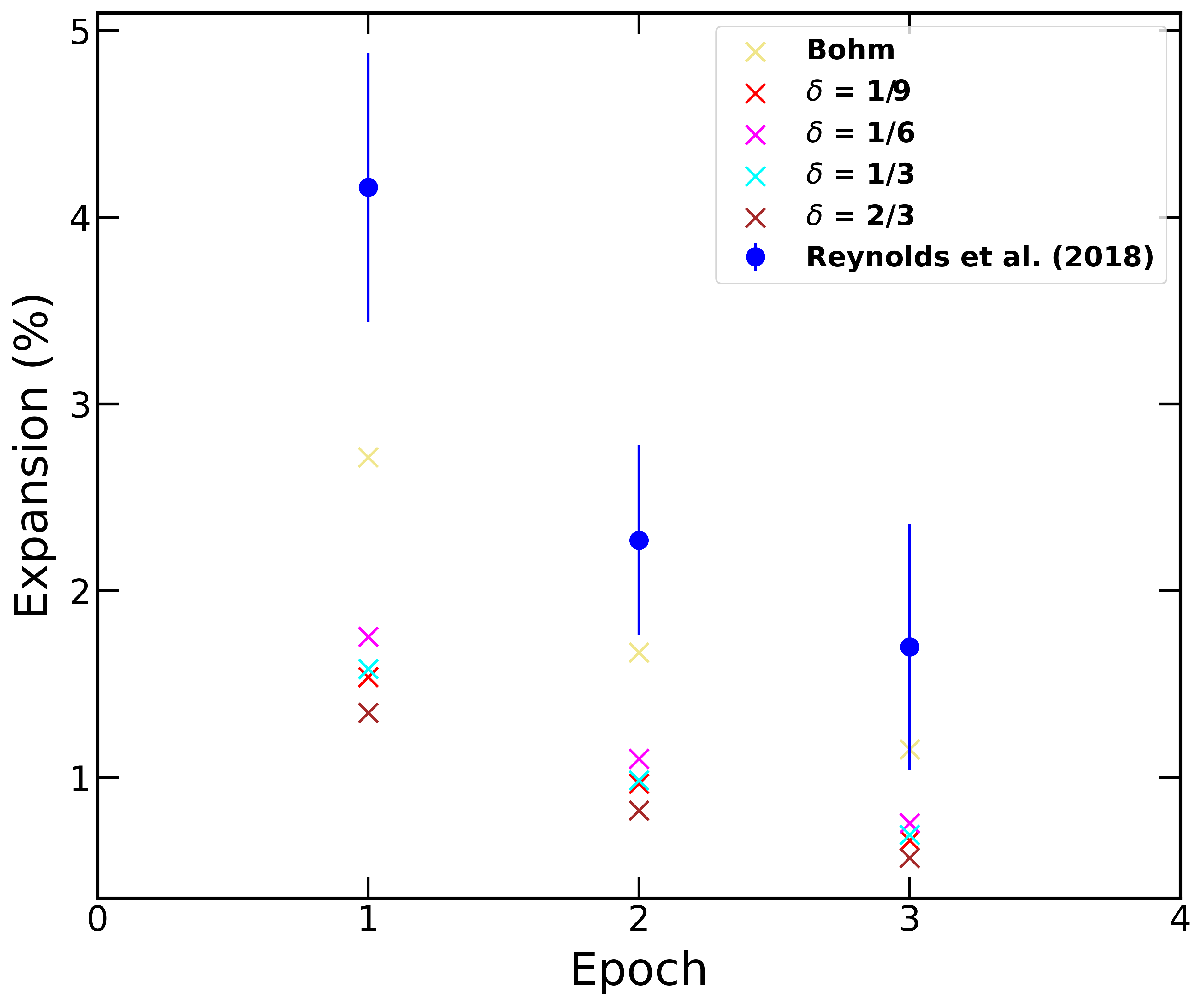} 
\caption{Expansion rate versus epoch for different diffusion exponents $\delta$.}
\label{fig:Kes75Diffusion:d} 
\end{subfigure}
\caption{The effects of considering Bohm diffusion and different diffusion exponents $\delta$ for a power-law-type diffusion, are presented for various features of Kes~75. All values are given in the key. (Epochs in (d) are same as mentioned in the caption of Fig.~\ref{fig:Kes75ExpansionBulkFlowBurstPercent}).}
\label{fig:Kes75Diffusion} 
\end{figure*}

\begin{figure}
    \centering
    \includegraphics[width=0.8\linewidth]{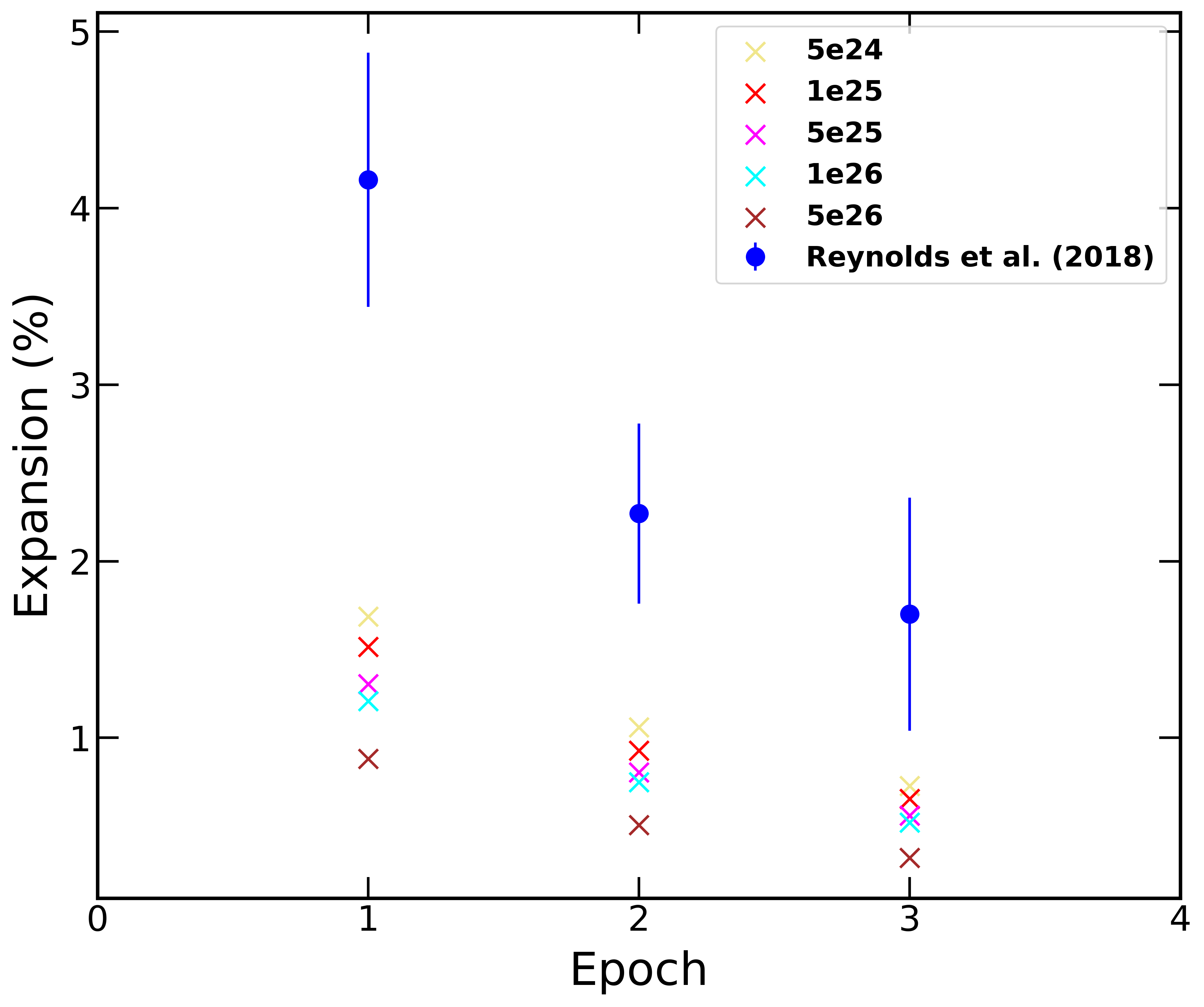}
    \caption{The variation of the power-law-type diffusion coefficient normalisation factor $\kappa_{\rm{X}}$ on the expansion versus epoch for Kes~75 (values in units of ${\rm cm}^2\,{\rm s}^{-1}$ are given in the key; epochs are same as mentioned in the caption of Fig.~\ref{fig:Kes75ExpansionBulkFlowBurstPercent}).}\label{fig:Kes75_kappa_scale_exp_new}
\end{figure}

We either use a Bohm-like diffusion coefficient (Equation~\ref{eqn:BohmDiffusion}), or a power-law-type coefficient (Equation~\ref{eqn:KolmogorovDiffusion}), with the free parameters being the exponent $\delta$ and coefficient normalisation factor $\kappa_{\rm{X}}$. In Fig.~\ref{fig:Kes75Diffusion}, the effects of considering different diffusion types and free parameters are presented for spatial and temporal features. Different values for the index $\delta$ (which is unity in the Bohm case) impact on the slopes of the diffusion loss time scales (Fig.~\ref{fig:Kes75BestFit_Timescales}), leading to a change in the spectrum of particles in a particular shell (given the interplay of SR losses), thus impacting the resulting X-ray spectrum (slope and flux). 
When $\delta = 1/3, 2/3$ we find that our modelled X-ray photon index (both for radial distances (Fig.~\ref{fig:Kes75Diffusion:a}) and epochs (Fig.~\ref{fig:Kes75Diffusion:b}) aligns with the data, but the model SB profile and expansion rate versus epoch do not (Fig.~\ref{fig:Kes75Diffusion:c} and Fig.~\ref{fig:Kes75Diffusion:d}). On the other hand, it is apparent that Bohm-like diffusion is highly preferred to produce the SB profile and expansion rate, but it is not preferred for producing various spectral characteristics. This highlights the importance of jointly fitting several data sets to break any model parameter degeneracies. 

To constrain the free parameter $\kappa_{\rm{X}}$, we study the expansion rate of Kes~75 over different epochs as shown in Fig.~\ref{fig:Kes75_kappa_scale_exp_new}. We see significant improvement as we move from values of order of $10^{26}$ to $10^{24}$~cm$^2$\,s$^{-1}$, showing that a lower coefficient value (leading to a longer diffusion time, and thus more dominant SR losses at high energies) is preferred to explain the expansion of the nebula. This is lower than the Galactic value \citep[e.g.,][]{Vla2012ApJ68V}, but consistent with the findings of other PWN models \citep[e.g.,][]{Abe2017Sci911A,dimar20203035D}.

\subsubsection{Braking index}\label{sec_brakingindex}
\begin{figure}
    \centering
    \includegraphics[width=0.8\linewidth]{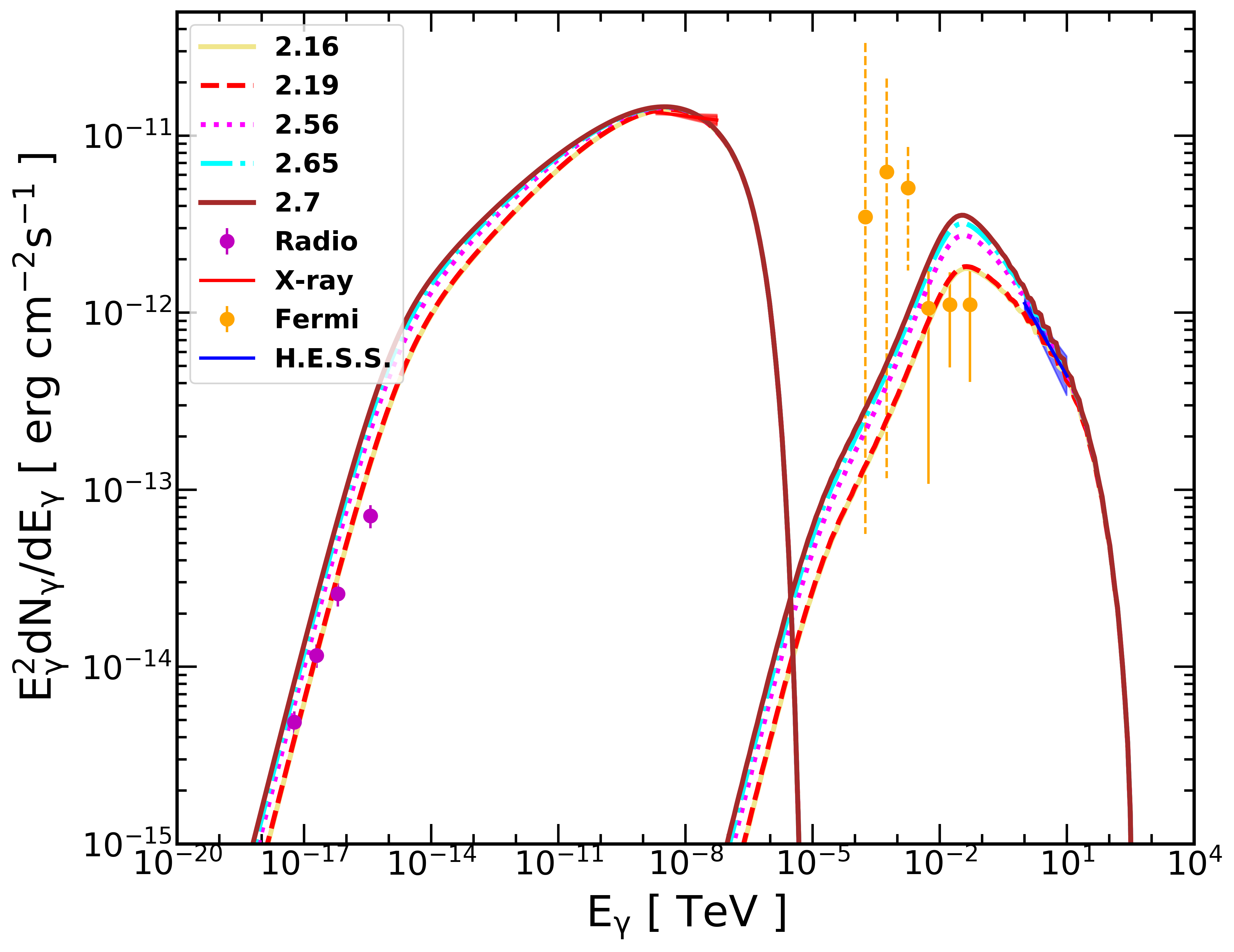}
    \caption{The impact of considering various values of $n$ (taken from the literature, see text for details) is shown on the SED for Kes~75. Values are given in the key.}
    \label{fig:Kes75_n_SED}
\end{figure}

We studied the effect of using the different reported values of $n$ over the past two decades as discussed in Section~\ref{sec:ObsReviewKes75} on the SED of Kes~75. The result is shown in Fig.~\ref{fig:Kes75_n_SED}. There are no significant differences in the spatial outputs for different values of $n$, hence we do not show them here.

\subsection{G21.5}
In the subsections below, we focus on describing the impact of varying the diffusion and braking index parameters on the model outputs. We do not show here the magnetic field variation study for G21.5, since the conclusions were found to be same as for Kes 75, discussed in Section~\ref{Kes75:Bfield}.

\subsubsection{Loss time scales}
Similar to Kes~75, we note from Fig.~\ref{fig:Tscaleg21p5} that IC and adiabatic losses do not dominate. However, the IC time scale is longer in this case, given the lower soft-photon temperature and hence photon density. At low energies, diffusion dominates, especially for smaller values of $\delta$.
Only at the highest energies (and small distances) does SR come into play at the current age. The SR time scales become longer with time and distance, given the assumed drop in the $B$-field, so SR is indeed more dominant at earlier times than later in the evolutionary history of the PWN. The normalisation factor $\kappa_{\rm X}$ influences the diffusion time scales as well, with a lower value leading to longer diffusion time scales.

\subsubsection{Diffusion}
\label{sub_diff_G21p5}

Based on the observational data, we found that the normalisation of the diffusion coefficient should be
$\kappa_{\rm X} \sim 10^{26} {\rm cm}^2\,{\rm s}^{-1}$ at $1$~TeV for this source (Fig.~\ref{fig:sedg21p5}, \ref{fig:sb_g21p5}, and\ref{fig:plot_diff_index}). This is approximately $\sim$ 100 times lower than the interstellar medium (ISM) value and also the typical value inside other PWNe \citep[e.g.,][]{Abe2017Sci911A}. 
The value of $\delta$ is constrained to a value of $1/6$ using these fits.
The SED, SB, and X-ray photon index are shown for the above value of $\kappa_{\rm X}$ and for various values of $\delta$ in Fig.~\ref{fig:sedg21p5}, \ref{fig:sb_g21p5}, and \ref{fig:plot_diff_index}. The magenta solid line is used for $n =1.857$ and other curves are for $n=3.0$. The other model parameters are listed in Table~\ref{tbl:G21p5}. We show the effect of varying  $\kappa_{\rm X}$ on the SB and X-ray photon index in Fig.~\ref{fig:diffg21_var} and \ref{fig:plot_diff_kappaX}. 

\begin{figure}
    \centering
    \includegraphics[width=0.8\linewidth]{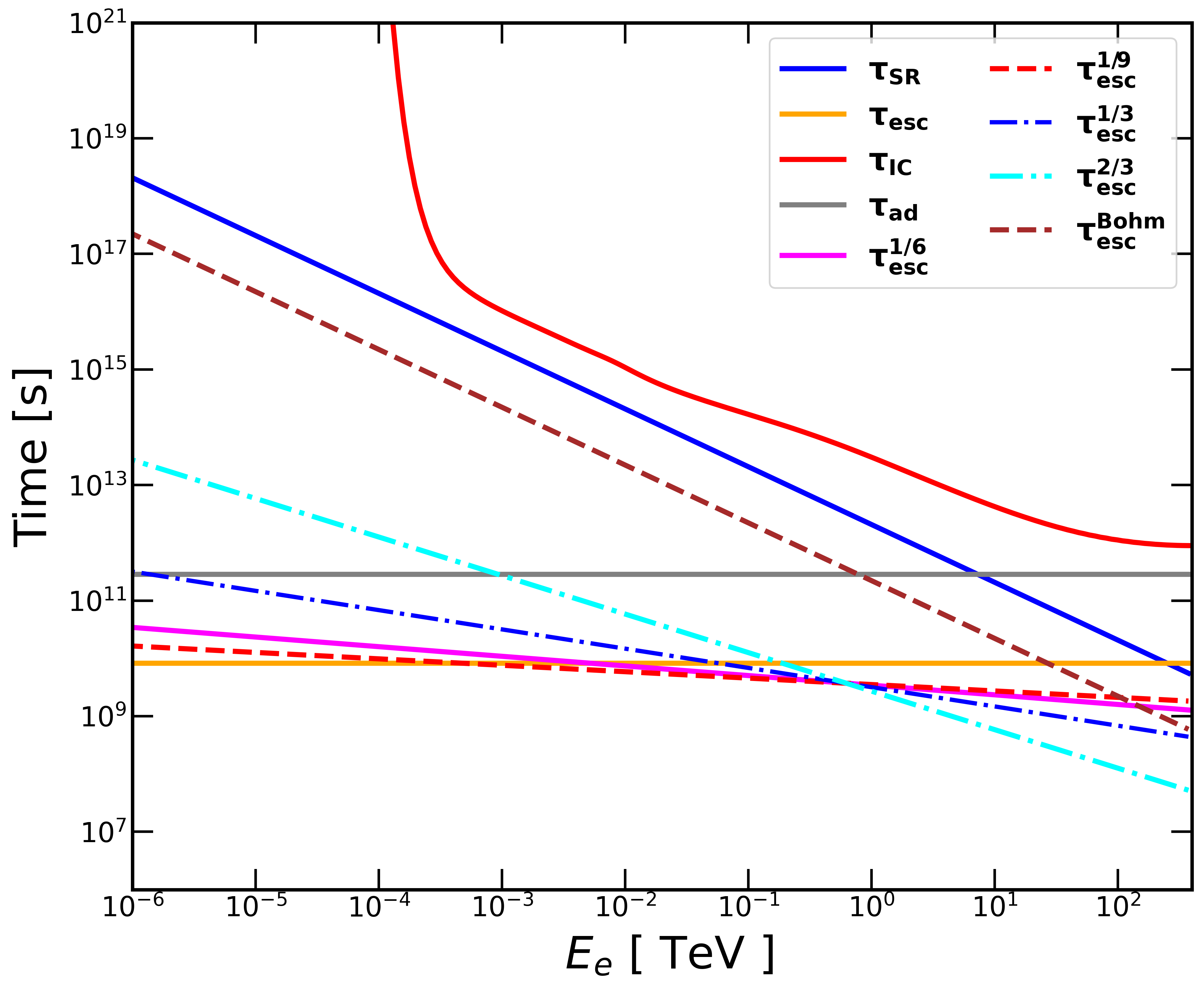}
    \caption{The variation of time scales with particle energy for G21.5 for various cooling processes at 1~pc and the current age of the PWNe. The normalisation $\kappa_{\rm X}$ for the power-law diffusion is taken as $10^{26}~{\rm cm^2s^{-1}}$ for $n=1.857$, with the relevant parameters listed in Table \ref{tbl:G21p5}.}
    \label{fig:Tscaleg21p5}
\end{figure}

Compared to $\delta$, we find that the above curves are very sensitive to the choice of $\kappa_{\rm X}$, as shown in Fig.~\ref{fig:diffg21_indexvr}. The model overpredicts the X-ray photon index values in the core region of the PWNe but there is some agreement above 20$^{\prime\prime}$. We find that for a value $\kappa_{\rm X} < 10^{26}~{\rm cm^2s^{-1}}$ the SR flux is well explained but the GeV-TeV gamma-ray flux is too low (the SED predictions are not shown here). This implies that the diffusion is slower, leading to SR to now dominate diffusion, at the expense of the IC process.
The effects become severe for the value of $\kappa_{\rm X} > 10^{26}~{\rm cm^2s^{-1}}$, when the SR flux also lowers, because diffusion then dominates. Even lower energetic particles escape the emission region before substantial SR may be produced (Fig.~\ref{fig:Tscaleg21p5}). Hence the SB and X-ray photon index data are very crucial to constrain the diffusion coefficient in PWNe modelling.

\subsubsection{Braking Index}
\begin{figure}
    \centering
    \includegraphics[width=0.8\linewidth]{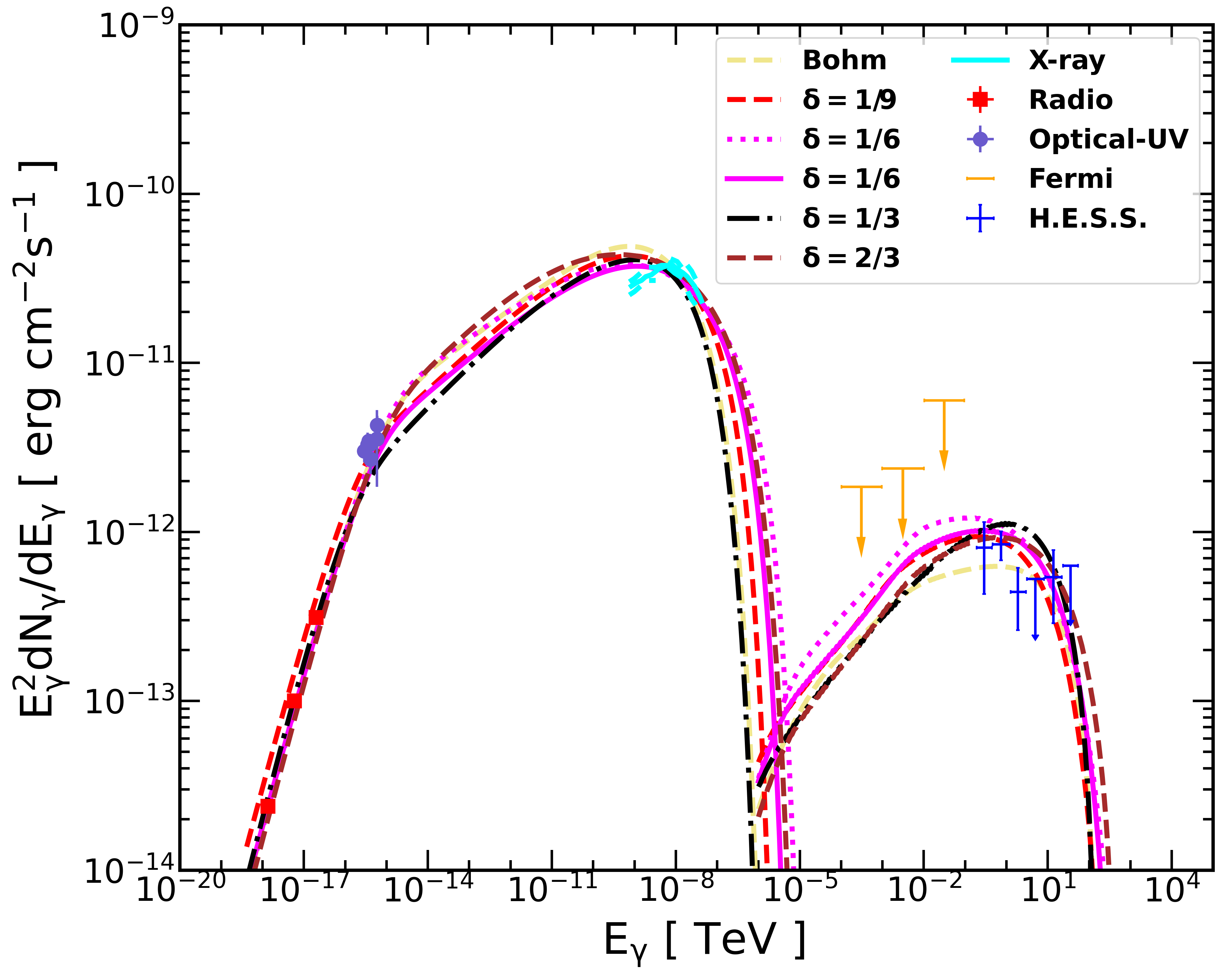}
    \caption{The SED for G21.5 for the various diffusion models considered in this work. The curves are plotted for $\kappa_{\rm X} = 10^{26}~{\rm cm}^2\,{\rm s}^{-1}$ and $n = 3.0$. We also indicate the curve for $n = 1.857$ by the magenta solid line. The data points are taken from published literature as listed in Table~\ref{tbl:Data_energyrange_ref}.}
    \label{fig:sedg21p5}
\end{figure}

\begin{figure*}
  \begin{subfigure}[t]{0.5\linewidth}
    \centering
    \includegraphics[width=0.8\linewidth]{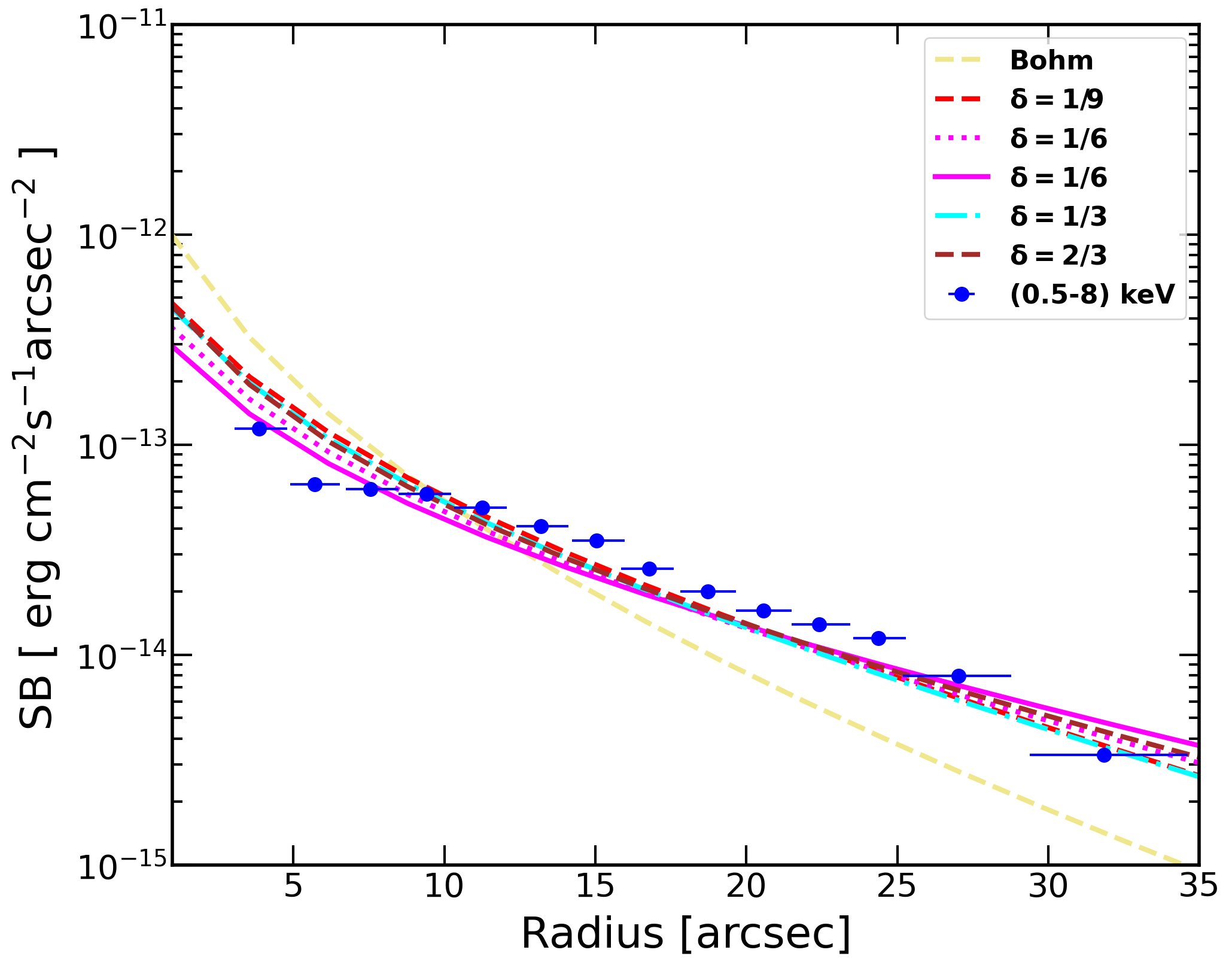}
    \caption{The SB profile for G21.5 superposed on the X-ray data. Diffusion scenarios other than the Bohm case better match the observed brightness levels. } 
    \label{fig:sb_g21p5}
  \end{subfigure}
  \hspace{-2ex}
  \begin{subfigure}[t]{0.5\linewidth}
    \centering
    \includegraphics[width=0.8\linewidth]{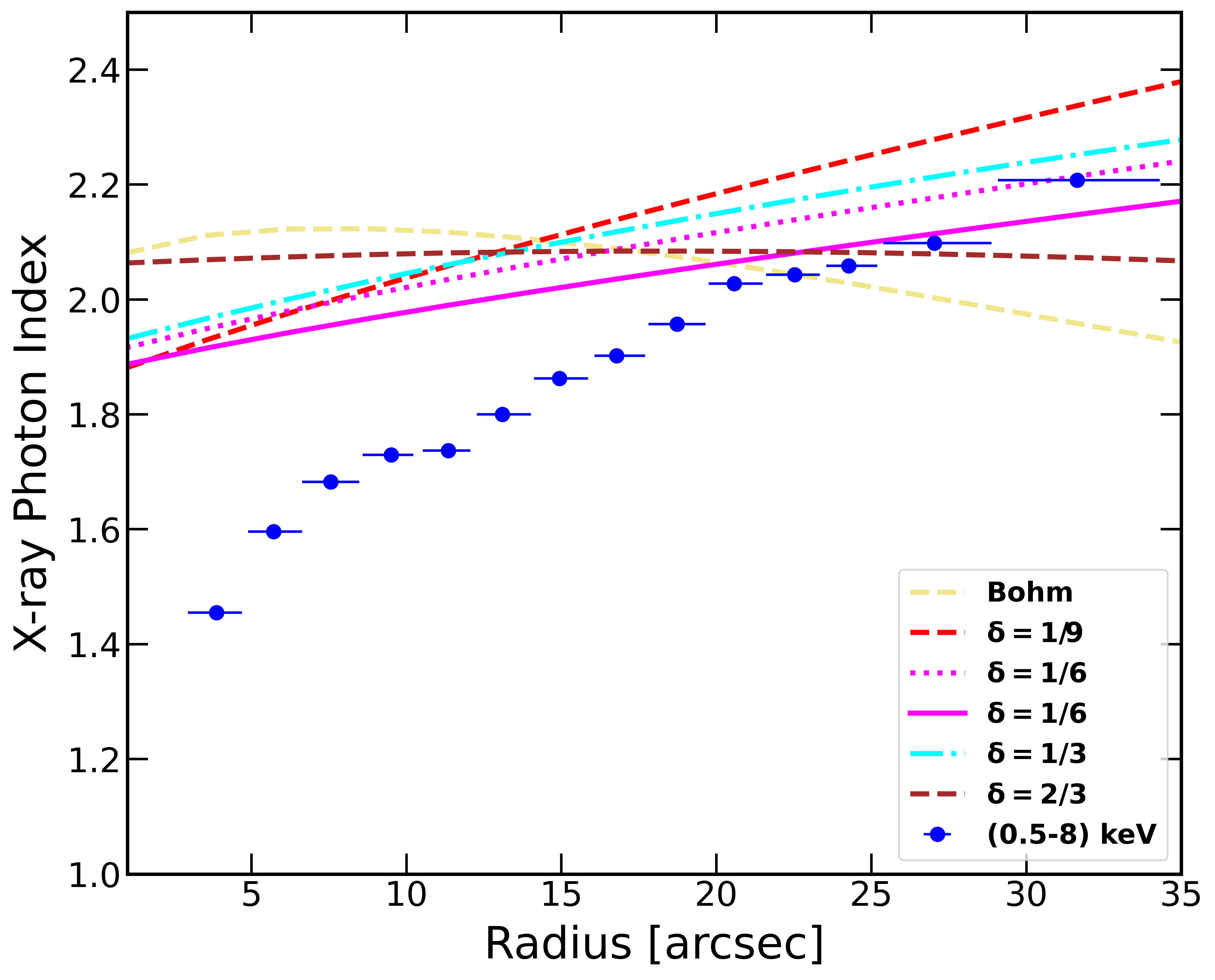}
    \caption{Predicted and observed X-ray photon index profile.}
    \label{fig:plot_diff_index}
  \end{subfigure} 
  \caption{Impact of changing $\delta$ on the SB and the photon spectral index profiles for G21.5. The data points are taken from \protect\cite{Hu2022}. The solid magenta line is for a braking index $n = 1.857$ and the remaining curves are for a value $n = 3.0$.}
  \label{fig:diffg21_delta} 
\end{figure*}

\begin{figure*}
  \begin{subfigure}[t]{0.5\linewidth}
    \centering
    \includegraphics[width=0.8\linewidth]{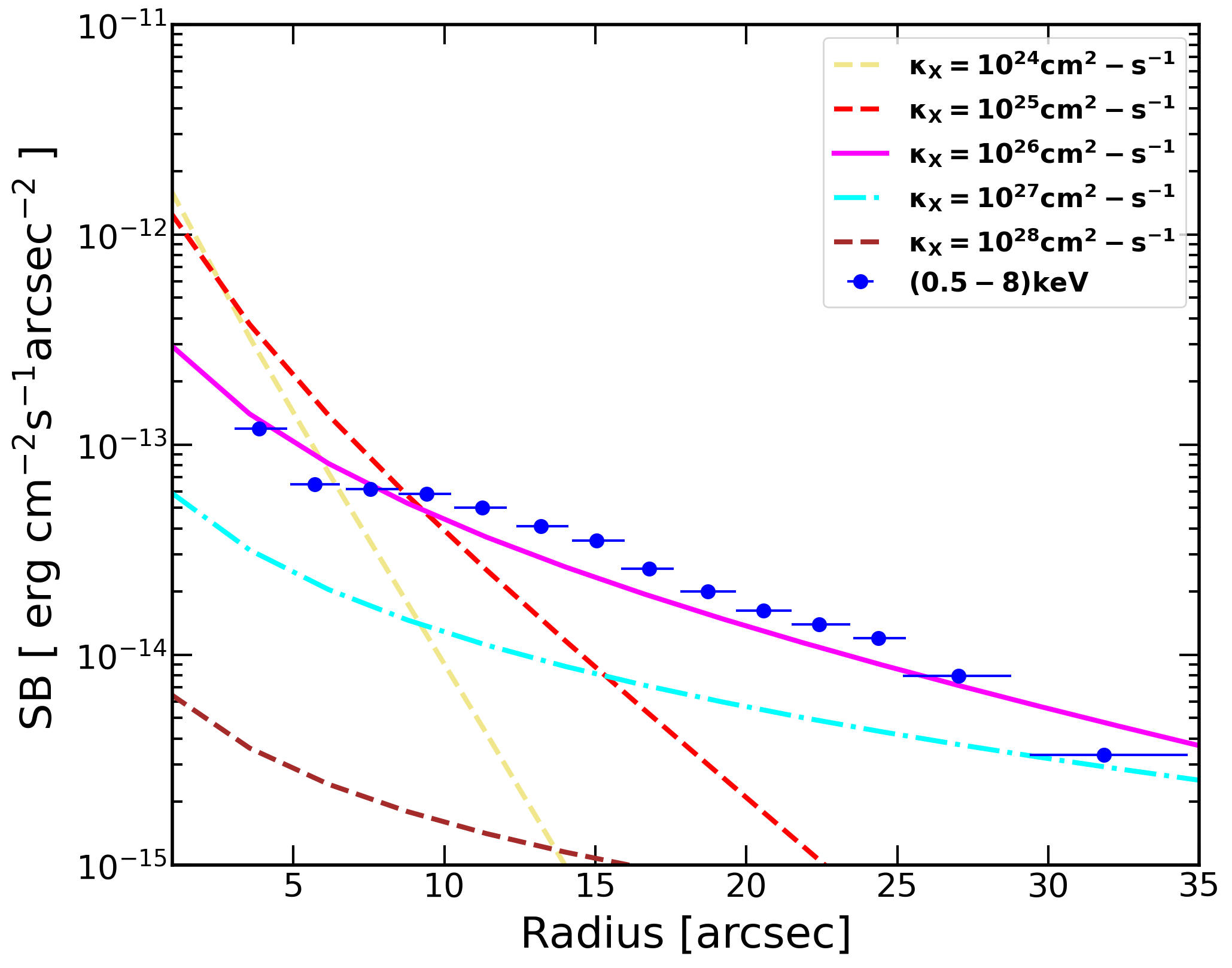}
    \caption{SB profiles for G21.5.} 
    \label{fig:diffg21_var}
  \end{subfigure}
  \hspace{-2ex}
  \begin{subfigure}[t]{0.5\linewidth}
    \centering
    \includegraphics[width=0.8\linewidth]{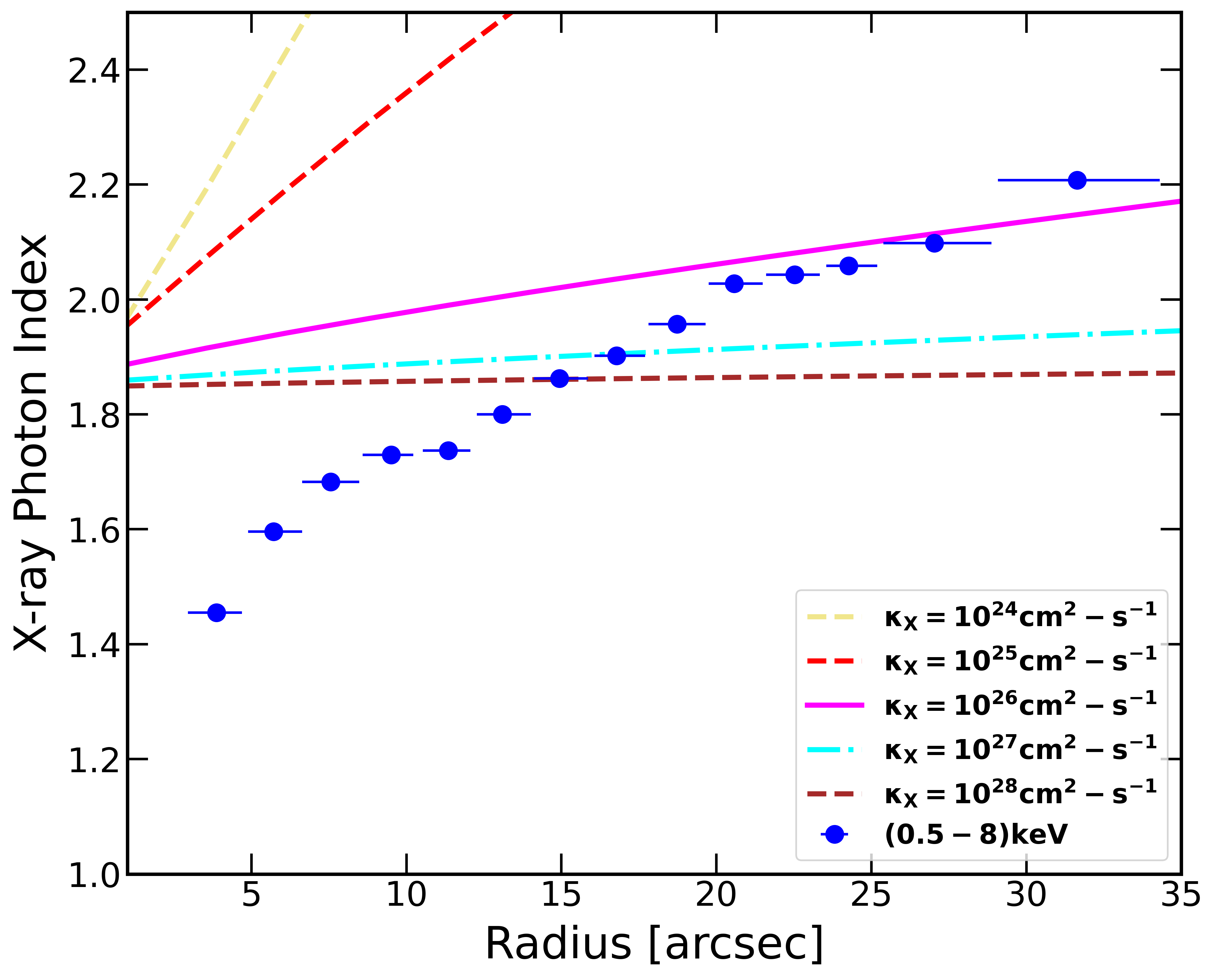}
    \caption{X-ray photon index variation for G21.5.}
    \label{fig:plot_diff_kappaX}
  \end{subfigure} 
  \caption{(a) SB fits for G21.5 used to constrain the value of $\kappa_{\rm X}$. (b) X-ray photon index vs projected radius. The curves are plotted for various values of $\kappa_{\rm X}$ and $n = 1.857$, and for a fixed $\delta = 1/6$.}
  \label{fig:diffg21_indexvr} 
\end{figure*}

The model outputs change slightly when we use $n = 1.857$ compared to a value of $n =3.0$. The lower value of $n$ will increase the spin-down time scale of the pulsar, but decrease the spin-down rate. For $n = 3.0$ we show the SED, SB, and X-ray photon index by the magenta dotted line in Fig.~\ref{fig:sedg21p5}, \ref{fig:sb_g21p5}, and \ref{fig:plot_diff_index}. In these plots, the magenta solid line is used to represent the case for $n = 1.857$. We find that the model outputs are not very sensitive to these changes in $n$.

\section{Results}\label{sec:res}
\begin{table}
\begin{tabular}{|ll|}
\hline
\textit{\textbf{Fixed (measured) parameters}} & \textbf{Kes~75} \\ 
\hline 
Pulsar period ($P$) (s)	& $0.326$ [1] \\ 
Time derivative of period ($\dot P$) (s s$^{-1}$) &  $7.1\times10^{-12}$ [2] \\
Age ($t_{\rm{age}} = \tau_{\rm c}$) (yr)  &  $728$ [3]\footnotemark \\ 
Present-day spin-down luminosity ($L_{\rm{age}}$) (erg/s) & $8.1\times10^{36}$ [1]\\ 
Braking index ($n$)  & $2.16$ [1]\\
Distance to the source (kpc) & $5.6$ [4] \\ 
\hline
\textit{\textbf{Fixed (assumed) parameters}} &  \\ 
\hline 
Index of the injected spectrum ($\alpha_1$) & $1.1$ \\ 
Index of the injected spectrum ($\alpha_2$)  & $2.4$ \\
Minimum energy of injected leptons ($E_{\rm{e,min}}$, ergs)  & $8.2\times10^{-7}$ \\
Maximum energy of injected leptons ($E_{\rm{e,max}}$) (ergs)  & $7.0\times10^{2}$ \\
Break energy ($E_{\rm{b}}$) (ergs) &$0.2$ \\
Magnetic energy conversion efficiency ($\eta$)& $0.2$ \\
Particle energy conversion efficiency ($\epsilon$)& $0.8$ \\
Sigma parameter ($\sigma$)& $0.25$ \\
Magnetic field time dependence ($\beta_{\rm B}$)& $-1.0$ \\
\textit{Soft-photon components:}&   $T$ (K), $u$ (eV\,cm$^{-3}$) \\ 
Cosmic microwave background (CMB)& $2.76$, $0.23$\\
Infrared& $30.0$, $0.3$\\
Optical& $6500$, $32$ \\
\hline
\textit{\textbf{Fitted parameters}}	&	\\ \hline 
Radial parameter of the magnetic field  ($\alpha_{\rm B}$) & $-0.8$ \\
Present-day magnetic field, $B_{\rm{age}}$ ($\mu$G)	&  $135$  \\ 
Bulk flow normalisation ($V_0$) ($10^{10}$ cm s$^{-1}$)	 &  $0.01$ \\ 
Diffusion coefficient normalisation ($\kappa_{\rm X}) (10^{24} {\rm cm}^2\,{\rm s}^{-1}$) & $8$\\
Diffusion coefficient exponent ($\delta$)    & $0.3$
\\ \hline
\end{tabular}
\caption{Preferred model parameters for PWN Kes~75. References:
[1] \protect\cite{Livingstone2011}, [2] \protect\cite{Gotthelf2000}, [3] \protect\cite{2005AJ....129.1993M}, [4] \protect\cite{2018AJ204R}.}
\label{tbl:Kes75}
\end{table}
\footnotetext{\url{https://www.atnf.csiro.au/research/pulsar/psrcat/}}

\subsection{Kes 75}
\begin{figure*}
  \begin{subfigure}[b]{0.5\linewidth}
    \centering
\includegraphics[width=0.8\linewidth]{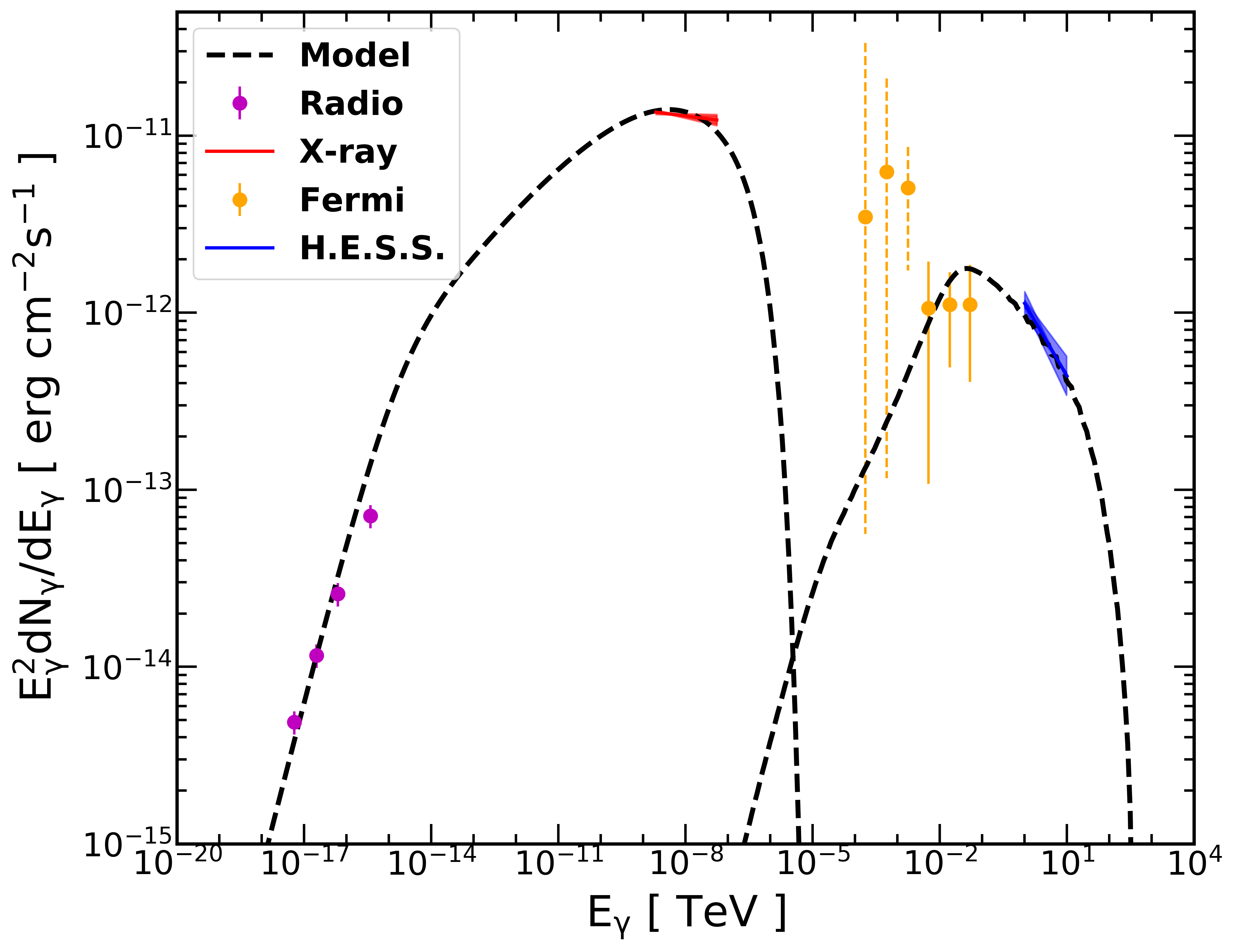}
    \caption{SED.} 
    \label{BestFit_Kes75:a} 
    \vspace{4ex}
  \end{subfigure}
  \begin{subfigure}[b]{0.5\linewidth}
    \centering
    \includegraphics[width=0.8\linewidth]{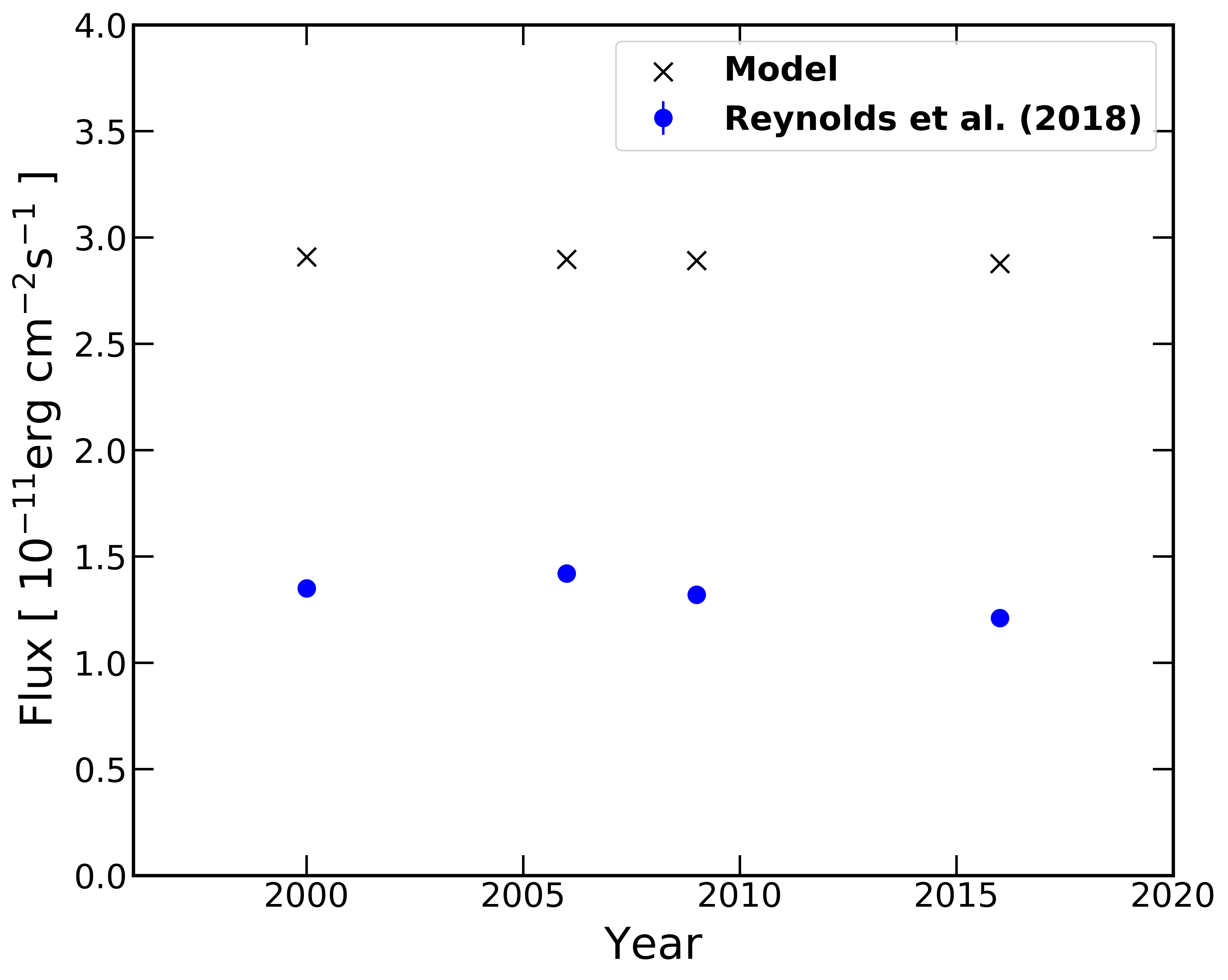} 
    \caption{Integrated flux over different epochs.} 
    \label{BestFit_Kes75:b} 
    \vspace{4ex}
  \end{subfigure}
  \begin{subfigure}[b]{0.5\linewidth}
    \centering
    \includegraphics[width=0.8\linewidth]{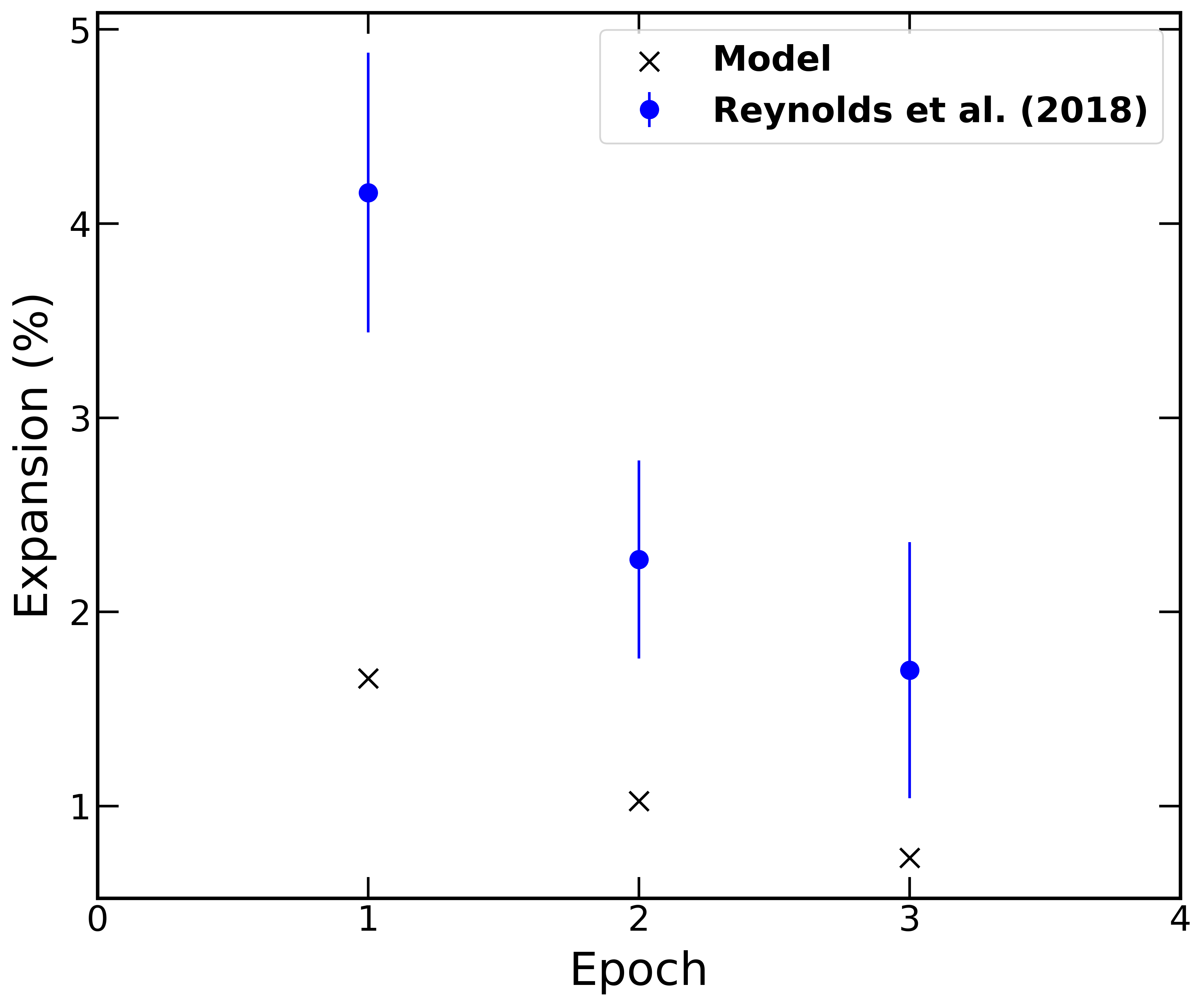} 
    \caption{Expansion rate for different epochs.}
    \label{BestFit_Kes75:c} 
  \end{subfigure}
  \begin{subfigure}[b]{0.5\linewidth}
    \centering
    \includegraphics[width=0.8\linewidth]{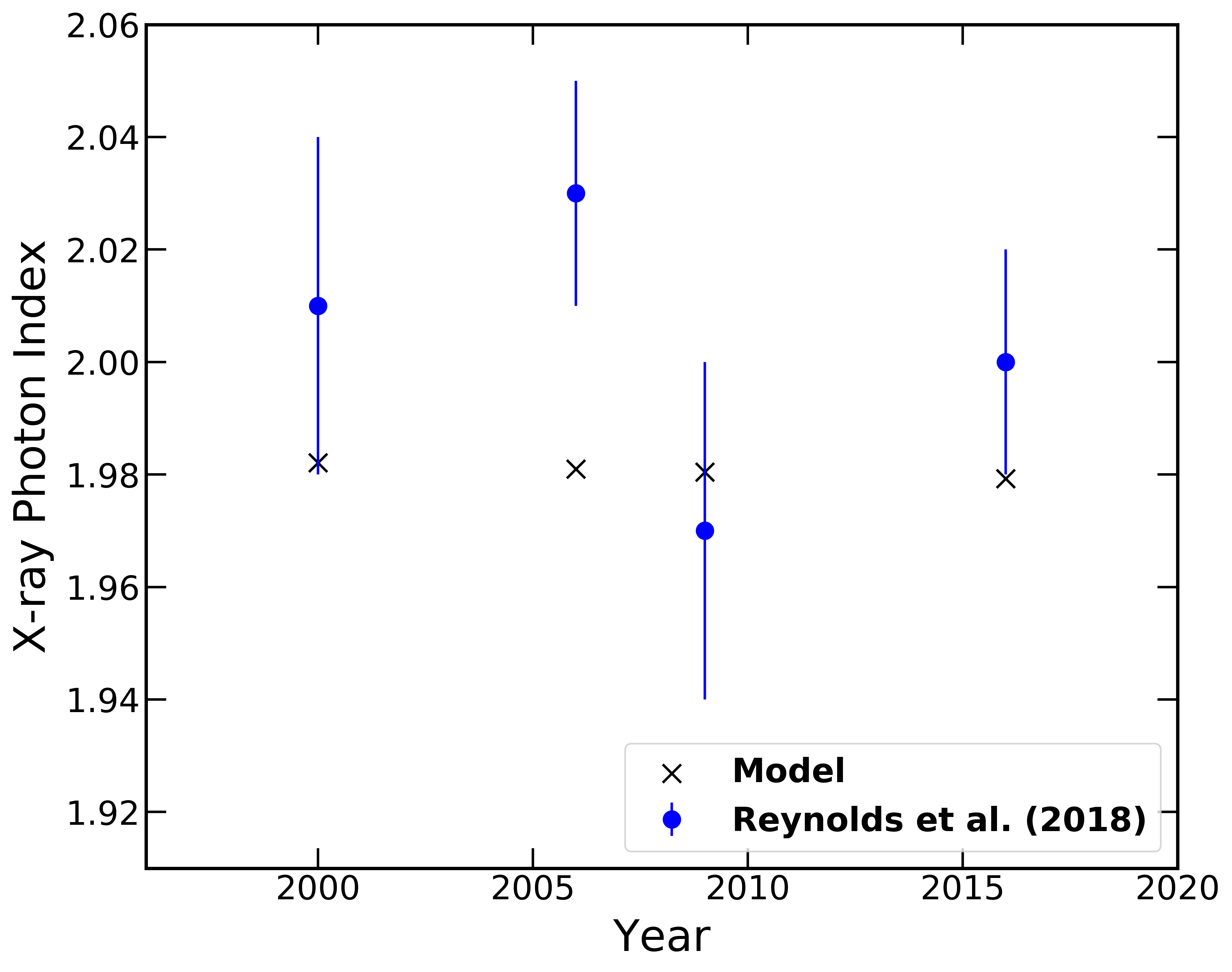} 
    \caption{X-ray photon index versus epoch.} 
    \label{BestFit_Kes75:d} 
  \end{subfigure}
      \begin{subfigure}[b]{0.5\linewidth}
    \centering
    \includegraphics[width=0.8\linewidth]{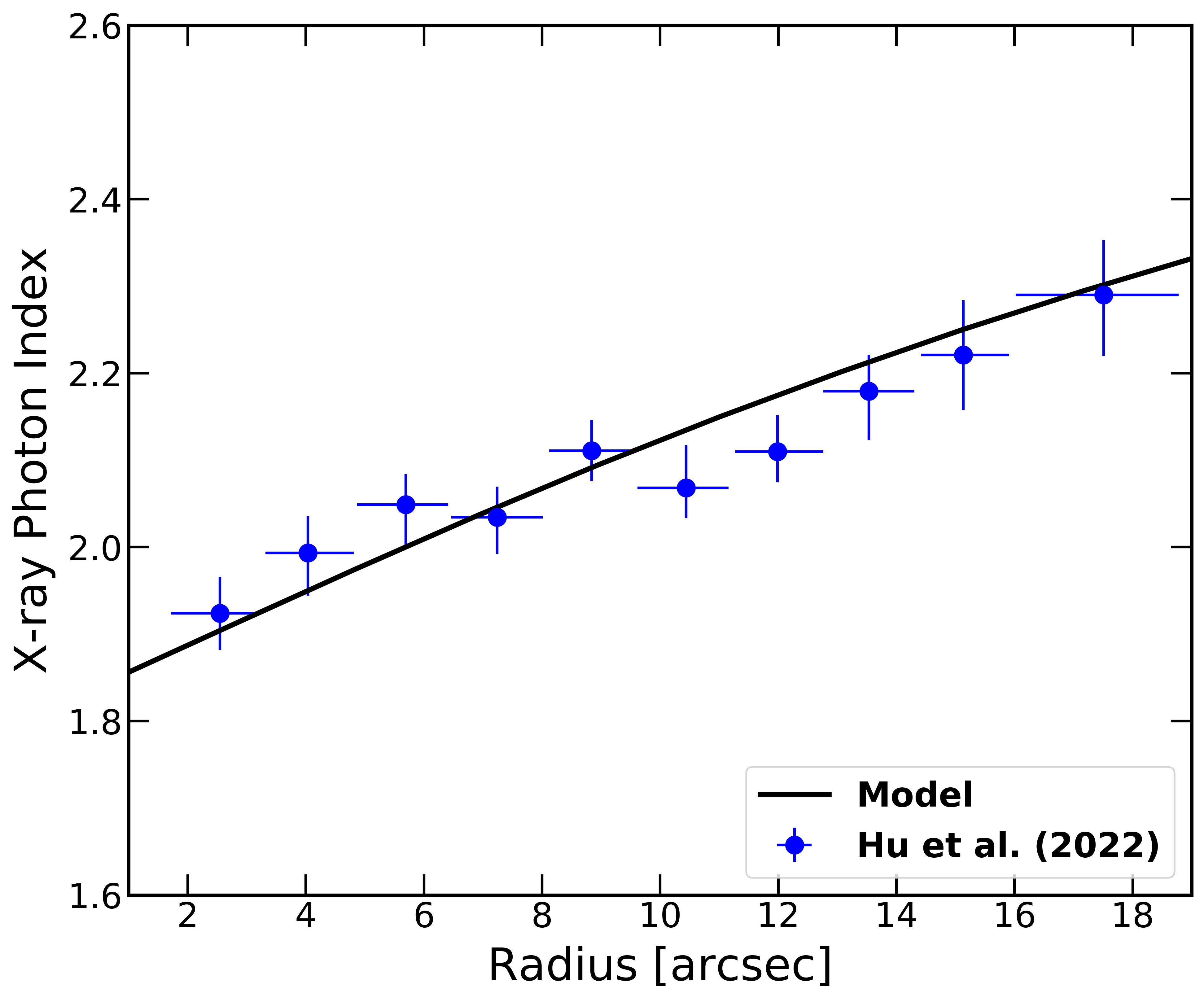} 
    \caption{X-ray photon index versus projected radius.}
    \label{BestFit_Kes75:e} 
  \end{subfigure}
  \begin{subfigure}[b]{0.5\linewidth}
    \centering
    \includegraphics[width=0.8\linewidth]{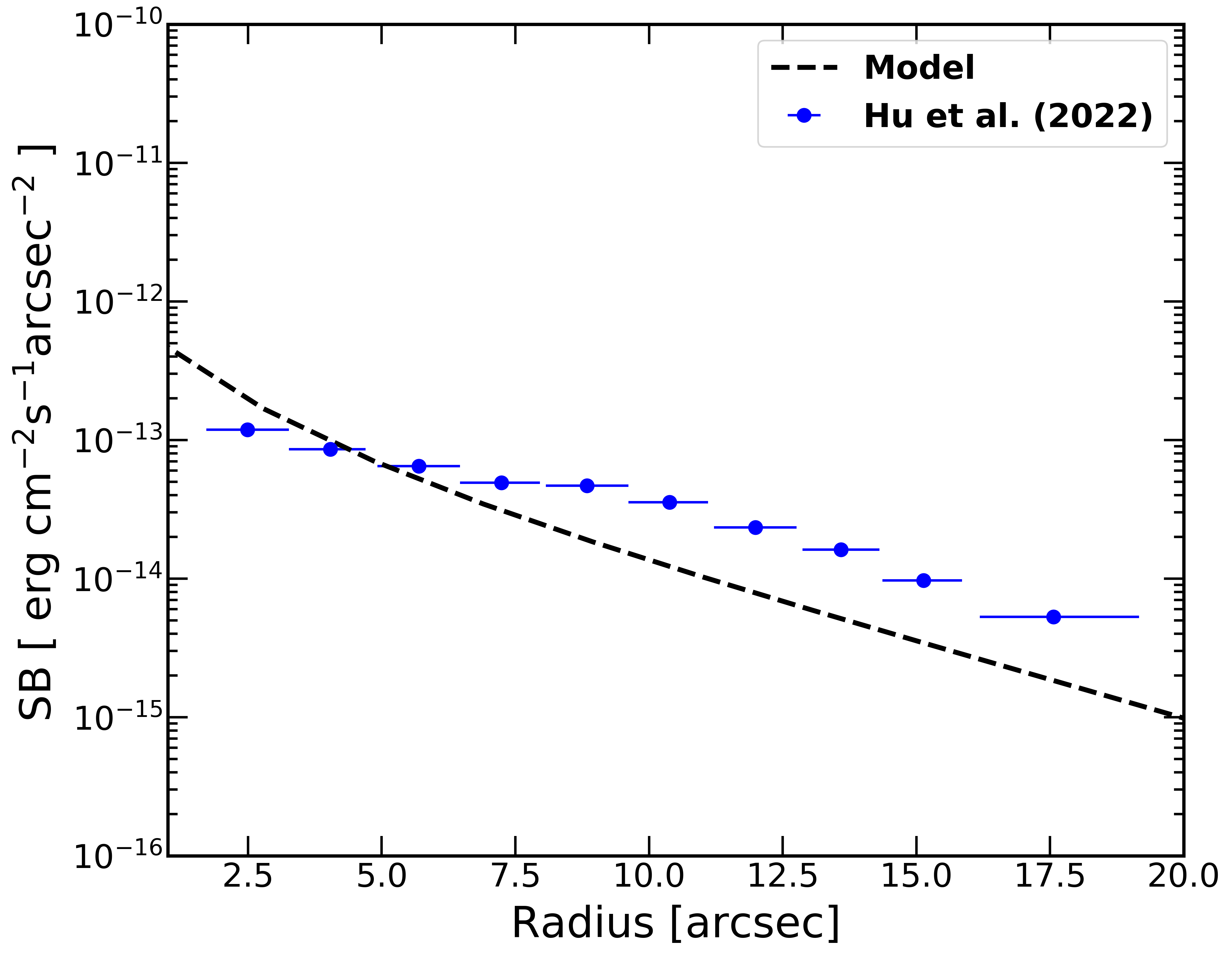} 
    \caption{SB profile versus projected radius.} 
    \label{BestFit_Kes75:f} 
  \end{subfigure} 
  \caption{Preferred model to simultaneously interpret spectral, spatial, and temporal data available for Kes~75, obtained by eye. Note that the error margins on the data in (b) are too small to be visible in the plot. Refer to Table~\ref{tbl:Data_energyrange_ref} for energy ranges and references for the data used. (Epochs in (c) are same as mentioned in the caption of Fig.~\ref{fig:Kes75ExpansionBulkFlowBurstPercent}.)}
  \label{fig:BestFit_Kes75} 
\end{figure*}

Our preferred models for Kes~75 are shown in Fig.~\ref{fig:BestFit_Kes75}. 
The corresponding model parameters are provided in Table~\ref{tbl:Kes75}.

For the SED (see Fig.~\ref{BestFit_Kes75:a}), our predicted radio emission is slightly shifted in energy (the index is slightly hard), but it explains the X-ray and TeV emission properly. \textit{Fermi} data is shown only for representation purposes (see Section~\ref{sec:ObsReviewKes75}). Some of the previous works like those of \citet{Bucciantini2011, Gotthelf_2021, StraalEtAl2023} report excellent fitting for the SED of Kes 75, and we note that our parameter values are similar to those of the latter two works. The high-energy particle index ($\alpha_{2}$) in our model is preferred to be $2.4$, which is closer to the reported values of $2.17$ and $2.3$ by \cite{Gelfand2014} and \cite{Torres2014}, respectively. We do not require a softer spectrum of  $\alpha_2\sim 3.0$ as in the works of \citet{Gotthelf_2021, StraalEtAl2023} (note that the latter study is also different from this work because of the inclusion of GeV data in their modelling). The slight difference in the quality of the model fit comes from the fact that we have fine-tuned our parameters to produce simultaneous model solutions for multiple (spatial) data features, and not just spectral data. See Section~\ref{sec:0D} below for further comparison to the best-fit parameters of a 0D code.

The predicted integrated X-ray flux and expansion rate over several epochs as shown in Fig.~\ref{BestFit_Kes75:b} and~\ref{BestFit_Kes75:c}, are too low by a factor $\sim 2$. Since repeated measurements are done to improve precision, not simply to measure temporal evolution, we also calculate the average expansion rate per year to compare with a similar observational quantity. We calculate this from our model by dividing our predicted expansion rate (in percent) by the spans of the corresponding epochs ($16, 10,$ and $7$~yr, respectively) and taking the mean of the three values. The average expansion rate we obtain is $0.104\%$ yr$^{-1}$, which is similar to the yearly expansion rates over individual epochs, and within a factor $\sim2$ of the average measured value of $0.249\% \pm 0.023\%$ yr$^{-1}$, as given in \cite{Reynolds2018}. Similarly, for the integral flux vs time, our model reflects the trend of no substantial time variation, as also seen in data, while overshooting the X-ray flux by a factor of $\sim2$.

We find good correspondence between the model and the data for the (lack of) temporal and spatial variation of the X-ray photon index, as shown in Fig.~\ref{BestFit_Kes75:d} and~\ref{BestFit_Kes75:e}. Within the dynamic range of the $y$-axis, and given the large error bars, the X-ray photon index does not seem to vary significantly, which is also reflected by the model results. Although the SB profile is underestimated by a factor $\sim5$ at some radii, we show in Fig.~\ref{fig:Kes75Diffusion:c} that it is well explained with a Bohm diffusion scenario; however, this does not explain other data sets.

\subsection{G21.5}
\label{sec:G21diff}
\begin{figure*}
  \begin{subfigure}[t]{0.5\linewidth}
    \centering
    \includegraphics[width=0.8\linewidth]{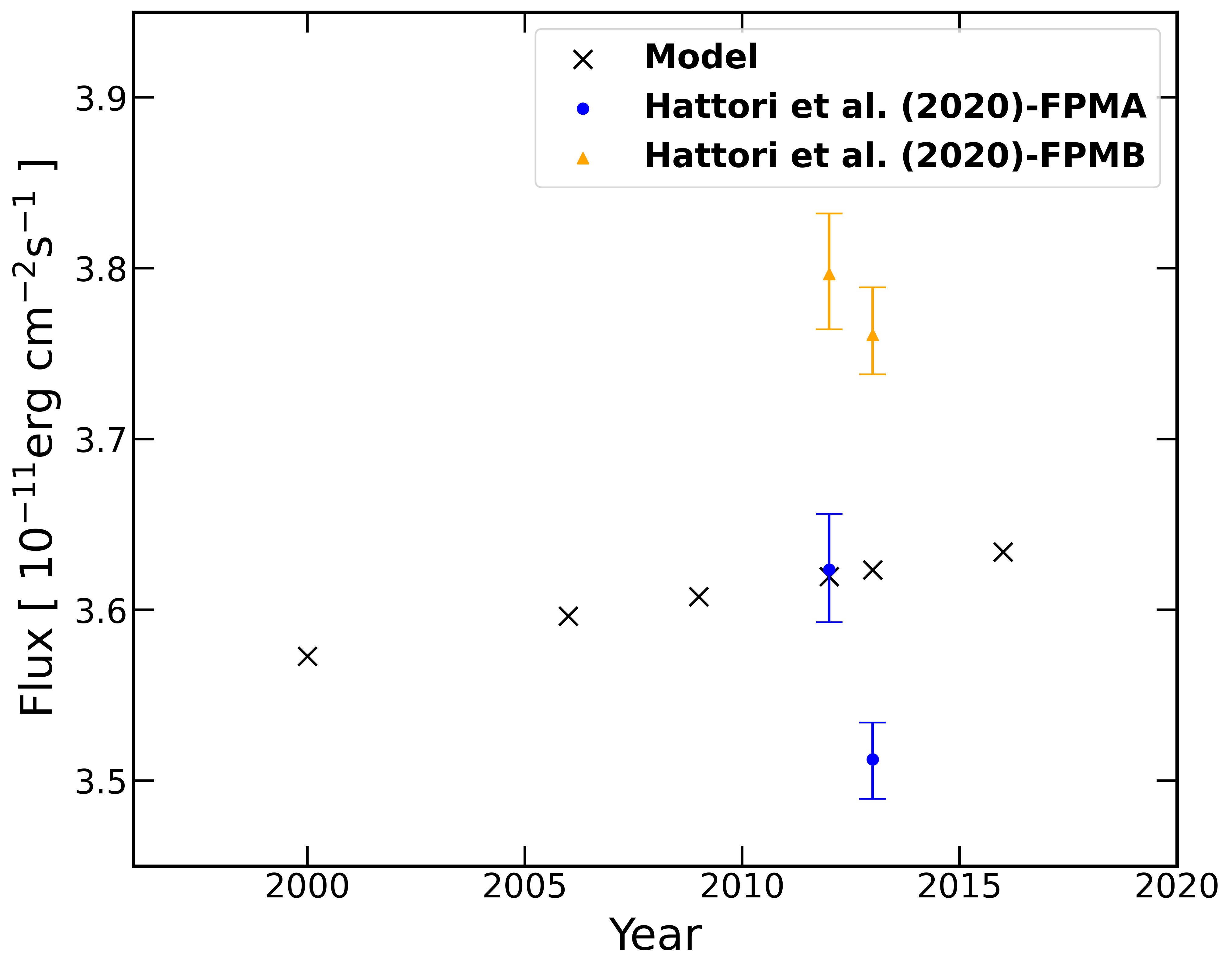}
    \caption{Integrated flux over different epochs.}
  \end{subfigure}
  \hspace{-2ex}
  \begin{subfigure}[t]{0.5\linewidth}
    \centering
    \includegraphics[width=0.8\linewidth]{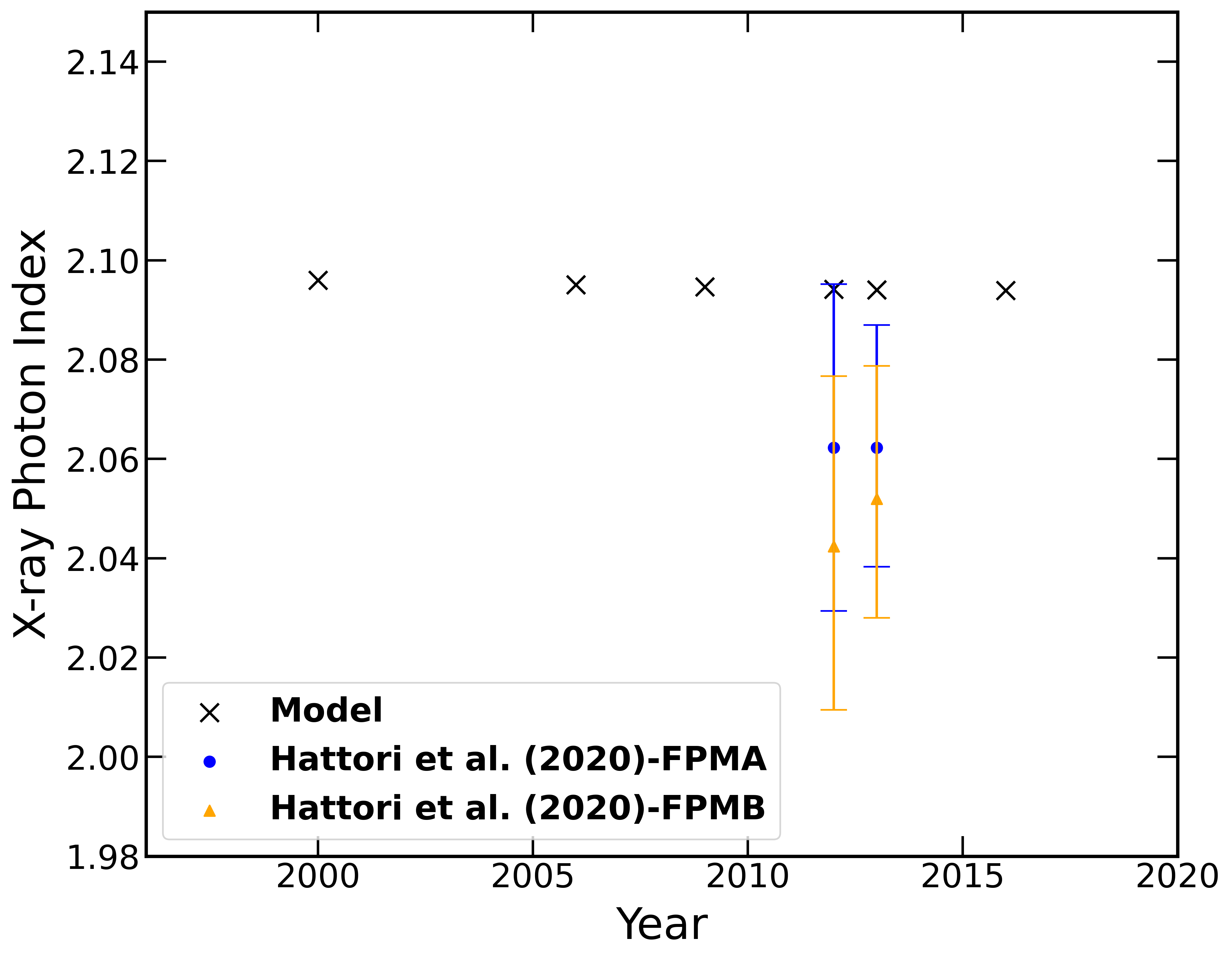}
    \caption{X-ray photon index versus epoch.}
  \end{subfigure} 
  \caption{The flux and index data for G21.5 are from \protect\cite{2020ApJ90432H}, with FPMA and FPMB indicating two co-aligned telescopes in \protect\textit{NuSTAR}. The epochs 2012 and 2013 are chosen with observation IDs 4002 and 6003, respectively, for the energy range $3-8$~keV. The model output is shown at different epochs, with the parameter values listed in Table~\ref{tbl:G21p5} and corresponding to $n = 1.857$.}
\label{fig:G21_fluxandindex} 
\end{figure*}

\begin{figure}
    \centering
    \includegraphics[width=0.8\linewidth]{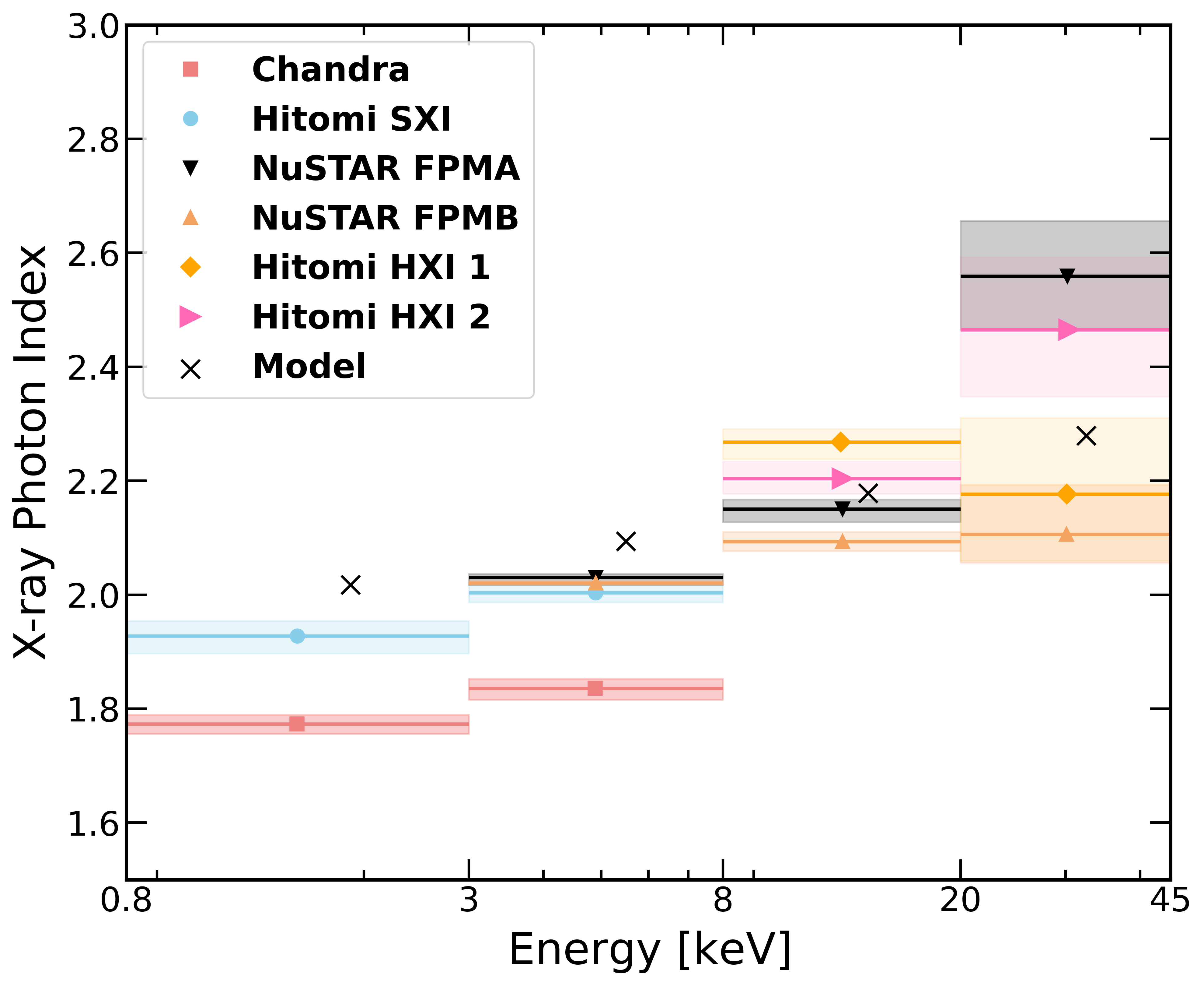}
    \caption{The index variation vs energy within the $0.8-45$~keV range is shown for G21.5. The data points are from \citet{2020ApJ90432H} based on \textit{Chandra} (0.8-8) keV, \textit{Hitomi} (Soft X-ray Imager (SXI), (0.8-8) keV and Hard X-ray Imager (HXI) having two independent instruments, HXI 1 and HXI 2 covering energy range 8 to 45 keV ) and \textit{NuSTAR} with two co-aligned telescopes, in the range of 3 to 45 keV. The model output is shown at different epochs for the parameters listed in Table \ref{tbl:G21p5}, corresponding to $n = 1.857$.}
    \label{fig:G21_Index_vs_E}
\end{figure}

In Fig.~\ref{fig:sedg21p5}, the SED for G21.5 is shown, for possible diffusion scenarios within the PWN, as discussed in subsection \ref{sec-dif}. On the basis solely of the SED fit, it is difficult to identify the most favourable diffusion coefficient inside the PWN. However, for the power-law diffusion model, the normalisation $\kappa_{\rm X}$ at 1 TeV could be constrained to $\sim 10^{26}~{\rm cm^2\,s^{-1}}$ for $\delta \le 1/6$ and it lowers by a factor of $\sim 3$ for $\delta = 1/3$ and by $\sim 10$ for $\delta = 2/3$, respectively. The normalisation $D_0$ at 1 TeV, for a value of $\delta =0.3$ in the ISM is approximately $ 2.5 \times 10^{29} {\rm cm}^2\,{\rm s}^{-1}$ \citep{Vla2012ApJ68V}. Hence we consider a diffusion model with $\kappa_{\rm X} \sim 10^{26} {\rm cm}^2\,{\rm s}^{-1}$ as a suitable choice, as also discussed in other works \citep{dimar20203035D}.

In addition to the SED, we also model the SB and X-ray photon index profile (Fig.~\ref{fig:sb_g21p5} and~\ref{fig:plot_diff_index}). The SB profile is better explained by the power-law type diffusion parameterisation compared to the Bohm diffusion one. However, no parameter combination can reproduce the X-ray photon index profile below a radius $\sim 20^{\prime\prime}$, given the very hard photon index of $\sim 1.4-2.0$ compared to that of Kes~75, which is well explained by our model. For $\delta =1/6$, we reach the observed spectral index asymptotically, as shown in Fig.~\ref{fig:plot_diff_index}. Further, the normalisation of the diffusion coefficient was also found to be $~\sim 10^{26}~ {\rm cm^2\,s^{-1}}$ by utilising the spatial variation of the SB and X-ray photon index data, as shown in Fig.~\ref{fig:diffg21_indexvr}. Our range of inferred model parameters based on the above model is listed in Table~\ref{tbl:G21p5}. Based on these model parameters, we estimated the integrated flux and X-ray photon indices over different epochs, which are shown in Fig.~\ref{fig:G21_fluxandindex}. The model outputs are shown in comparison to the available data, where two points are picked corresponding to the epochs 2012 and 2013, with observation IDs 4002 and 6003, as given in Table 4 of \cite{2020ApJ90432H}. We note that while \cite{2020ApJ90432H} do present the results of observations at different times, they do not claim that the differences between measurements in flux and photon index in the same energy ranges are due to temporal variations of these quantities, but rather the result of statistical and systematic uncertainties in these measurements that may mimic a temporal variation. As such, we simply show our model outputs vs time in order to compare with these data, without claiming temporal variations.
We do not show the intermediate data available for epochs within this one year, as we have not investigated the implications of such finer temporal details in this work. Within error bars, the average value of the \textit{NuSTAR} X-ray flux is higher in the FPMB telescope and slightly lower in FPMA telescope, and forms an envelope for the model predictions, with rather small deviations. However, the average measured X-ray photon index is lower than what is predicted by the model, although this deviation is also rather small. We also show the variation of the X-ray photon index with respect to energy in Fig.~\ref{fig:G21_Index_vs_E}. The data points are taken from \cite{2020ApJ90432H} without a pulsar blackbody component, based on \textit{Chandra}, \textit{Hitomi}, and \textit{NuSTAR} observations. As seen in the figure, our model captures the trend of the softening of the photon index with energy.

\begin{table*}
\begin{tabular}{|lll|}
\hline
 \textbf{PWN G~21.5} & $n = 3.0$ & $n =1.857$ \\ 
\hline 
\textit{\textbf{Fixed parameters}} &  &  \\
\hline 
Pulsar period ($P$) (s)	& $0.062$ [1]  \\ 
Time derivative of period ($\dot P$) (s s$^{-1}$) & $2.0\times10^{-13}$ [1] \\
Age ($t_{\rm{age}}$) (yr)                                  &  $1000$  \\
Present-day spin-down luminosity ($L_{\rm{age}}$) (erg/s) & $3.4\times10^{37}$ [2]  \\ 
Distance to the source (kpc)   & $4.4 \pm 0.2$ [3]  \\ 
\hline
\textit{\textbf{Range of Assumed Parameters}} &  \\ 
\hline 
Index of the injected spectrum ($\alpha_1$) & $1.1$  & $1.1$\\ 
Index of the injected spectrum ($\alpha_2$)  & $[2.46-2.57]$  & $2.52$\\
Minimum energy of injected leptons ($E_{\rm{e,min}}$) (ergs)  & $8.2 \times 10^{-7}$ & $8.2 \times 10^{-7}$\\
Maximum energy of injected leptons ($E_{\rm{e,max}}$) (ergs)  & [$3.0\times10^{2}, 9.0\times10^{2}]$ & $6.0\times10^{2}$\\
Break energy ($E_{\rm{b}}$) (ergs) &$[0.03, 0.06]$ & 0.05  \\
Magnetic energy conversion efficiency ($\eta$)& $0.2$ & 0.2 \\
Particle energy conversion efficiency ($\epsilon$)& $0.8$   & 0.8\\
Sigma parameter ($\sigma$)& $0.25$  & 0.25 \\
Magnetic field time dependence ($\beta_{\rm B}$)& $-1.0$ & $-1.0$ \\
\textit{Soft-photon components:}&~~~~~~~~  $T$ (K), $u$ (eV\,cm$^{-3}$)\\ 
Cosmic microwave background (CMB)& [$2.72$, $0.23$] &  [$2.72$, $0.23$]\\
Infrared& [$30$, $1.1$] &[$30$, $0.1$] \\
Optical& [$3500$, $1.0$] & [$3500$, $1.0$]\\
\hline
\textit{\textbf{Fitted parameters}}	&			\\ \hline 
Radial parameter of the magnetic field ($\alpha_{\rm B}$) & $-0.6$ & $-0.6$\\
Present-day magnetic field, $B_{\rm{age}}$ ($\mu$G)			&  $[110-140]$ & $140$ \\ 
Bulk flow normalisation, $V_0$, ($10^{10}$ cm s$^{-1}$)	   &  $0.01$&  $0.01$ \\ 
Diffusion coefficient normalisation, $\kappa_{\rm X}, (10^{24} {\rm cm}^2\,{\rm s}^{-1}$) & $[10-100]$ & 100 \\
Diffusion coefficient exponent ($\delta$)  & [$1/6-2/3$] & $1/6$\\
\hline
\end{tabular}
\caption{Model parameters used in this work for the simultaneous interpretation of the SED, surface brightness and X-ray photons index for G21.5.  References: [1]  \protect\cite{camilo2006ApJ637456C}, [2] \protect\cite{2005AJ....129.1993M}, [3] \protect\cite{2018AJ204R}.}
\label{tbl:G21p5}
\end{table*}

\subsection{Comparison between the preferred model parameters by independent 0D and 1D codes}\label{sec:0D}

\begin{table*}
\setlength\tabcolsep{4pt}
	\centering
  \scriptsize
	\caption{Summary of the inferred parameters of the 0D model using {\sc TIDE}. Lower and upper limits are presented in brackets.}
	\label{pars_0D}

	\begin{tabular}{lllll}
            \hline
	    \textbf{Parameters} & \textbf{Symbol} & \textbf{Kes~75} &\textbf{G21.5} &\textbf{Fitting range}\\
	    \hline
            \textbf{Fixed (measured) parameters}\\
            \hline
		      Characteristic age (kyr) &$\tau_{\rm c}$ &0.728 &4.85 &\\
		      Present spin-down luminosity ($10^{37}$erg s$^{-1}$) &$L_{\rm age}$ &0.81 &3.4 &\\
		      Distance (kpc) & $d$ &5.6 &4.4 &\\
                Observed PWN radius (pc) &$R_{\rm PWN}$&0.814 &0.853 &\\
            \hline
                \textbf{Fixed (assumed) parameters }\\
            \hline
                ISM density (cm$^{-3}$) &$n_{\rm ISM}$ &0.1 &- &\\
                Breaking index & $n$ &2.16 &1.857 &\\
                Initial spin-down age (kyr) &$\tau_0$ &$({2\tau_c})/({n-1}) - t_{\rm age}$ &- &\\
                Initial spin-down luminosity (erg s$^{-1}$) &$L_0$ &$L_{\rm age}(1+\frac{t_{\rm age}}{\tau_0})^{\frac{n+1}{n-1}}$ &- &\\
                Minimum energy at injection ($m_{\rm e}c^2$) &$\gamma_{\rm min}$ & 1 & - &\\
                Containment factor  &$\epsilon$ &0.5 &- &\\
                SN explosion energy (erg) &$E_{\rm SN}$ &$10^{51}$ &- &\\
                CMB temperature (K)  &$T_{\rm CMB}$ &2.725 &- &\\
		      CMB energy density (eV cm$^{-3}$)  &$U_{\rm CMB}$ &0.26 &- &\\
                FIR temperature (K)  &$T_{\rm FIR}$ &70 &- &\\
		      NIR temperature (K)  &$T_{\rm NIR}$ &3000 &- &\\
	    \hline
            \textbf{Fitted parameters} \\
            \hline
                Age (kyr) &$t_{\rm age}$ &1.05 [1.03, 1.06]  &0.82 [0.70, 0.94] &$0.3 - 1.25$ (Kes 75); $0.3 - 5.0$ (G21.5) \\
		      Break energy ($10^5 m_{\rm e}c^2$)&$\gamma_{\rm b}$ &4.07 [3.51, 4.71] &0.62 [0.47, 0.79] &$0.1 - 100$\\
		      Low-energy index &$\alpha_1$ &1.67 [1.64, 1.69] &1.00 [1.00, 1.22] &$1 - 4$\\
		      High-energy index &$\alpha_2$  &2.199 [2.197, 2.201] &2.58 [2.55, 2.61] &$1 - 4$\\
		      Ejected mass $(M_{\odot})$  &$M_{\rm ej}$ &7.54 [7.16, 7.96] &7.00 [7.00, 8.55] &$7 - 20$ \\
		      Magnetic energy fraction ($10^{-2}$)  &$\eta$ &2.64 [2.47, 2.81] &1.30 [1.02, 1.79] &$1 - 20$\\
		      FIR energy density (eV~cm$^{-3})$ &$U_{\rm FIR}$ &9.93 [5.56, 9.94] &1.68 [0.01, 4.17] &$0.01 - 10$ \\
		      NIR energy density (eV~cm$^{-3})$ &$U_{\rm NIR}$ &9.97 [0.01, 9.99] &9.97 [0.01, 9.99] &$0.01 - 10$ \\
		      PWN radius now (pc) &$R_{\rm PWN}$ &0.84 &0.84 &\\
		      Present magnetic field ($\mu$G) & $B_{\rm age}$ &29.79 &46.84 &\\
                Degrees of freedom &D.O.F. &4 &15 &\\
		      Reduced $\chi^2$ &$\chi^2$/D.O.F. &1.70 &1.39 &\\
		      Systematic uncertainty &$\sigma^\prime$ &0.01 &0.12 &$0.01 - 0.5$\\
        \hline
	\end{tabular}
\end{table*}

\begin{figure}
    \includegraphics[width=0.9\columnwidth]{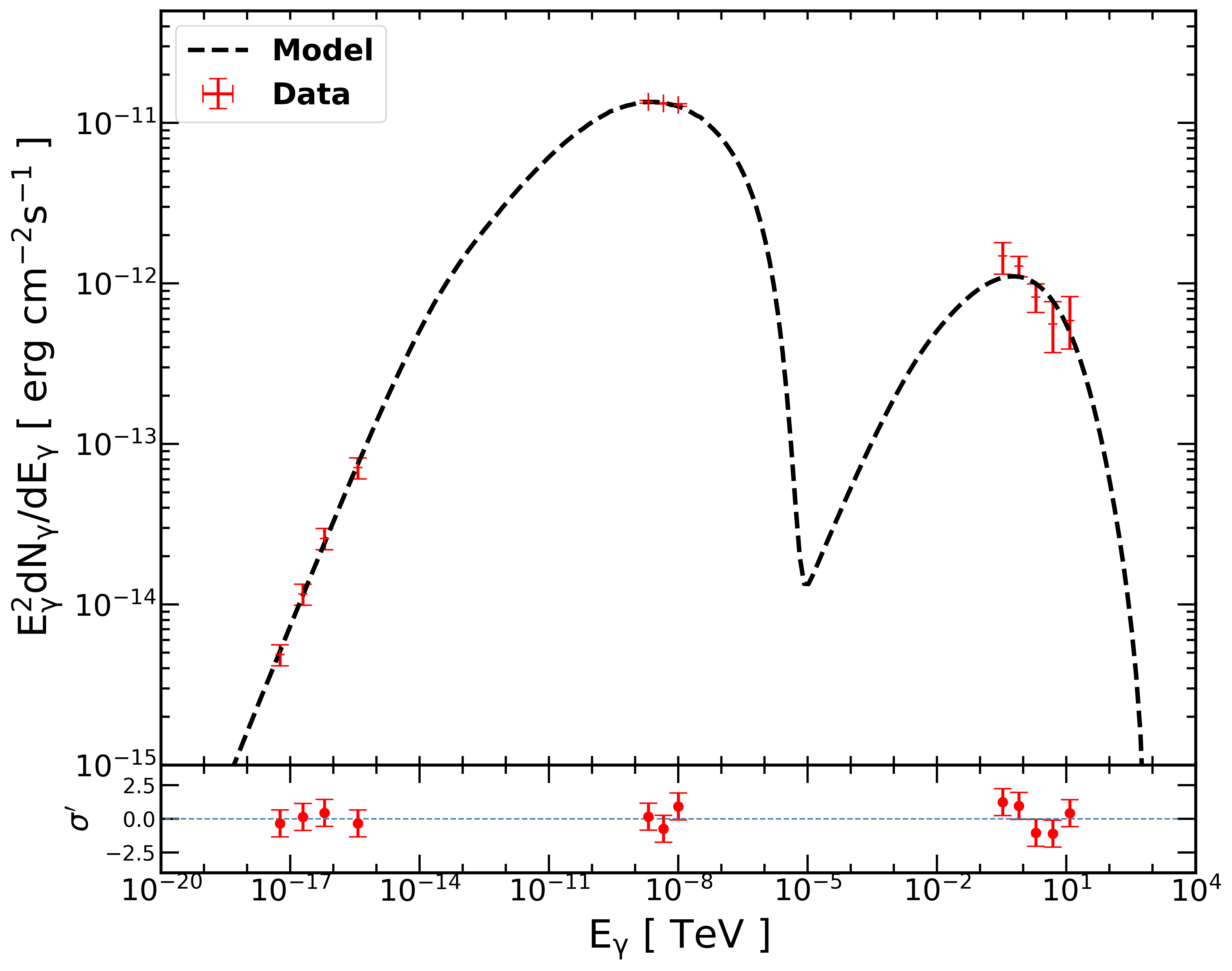}
    \includegraphics[width=0.9\columnwidth]{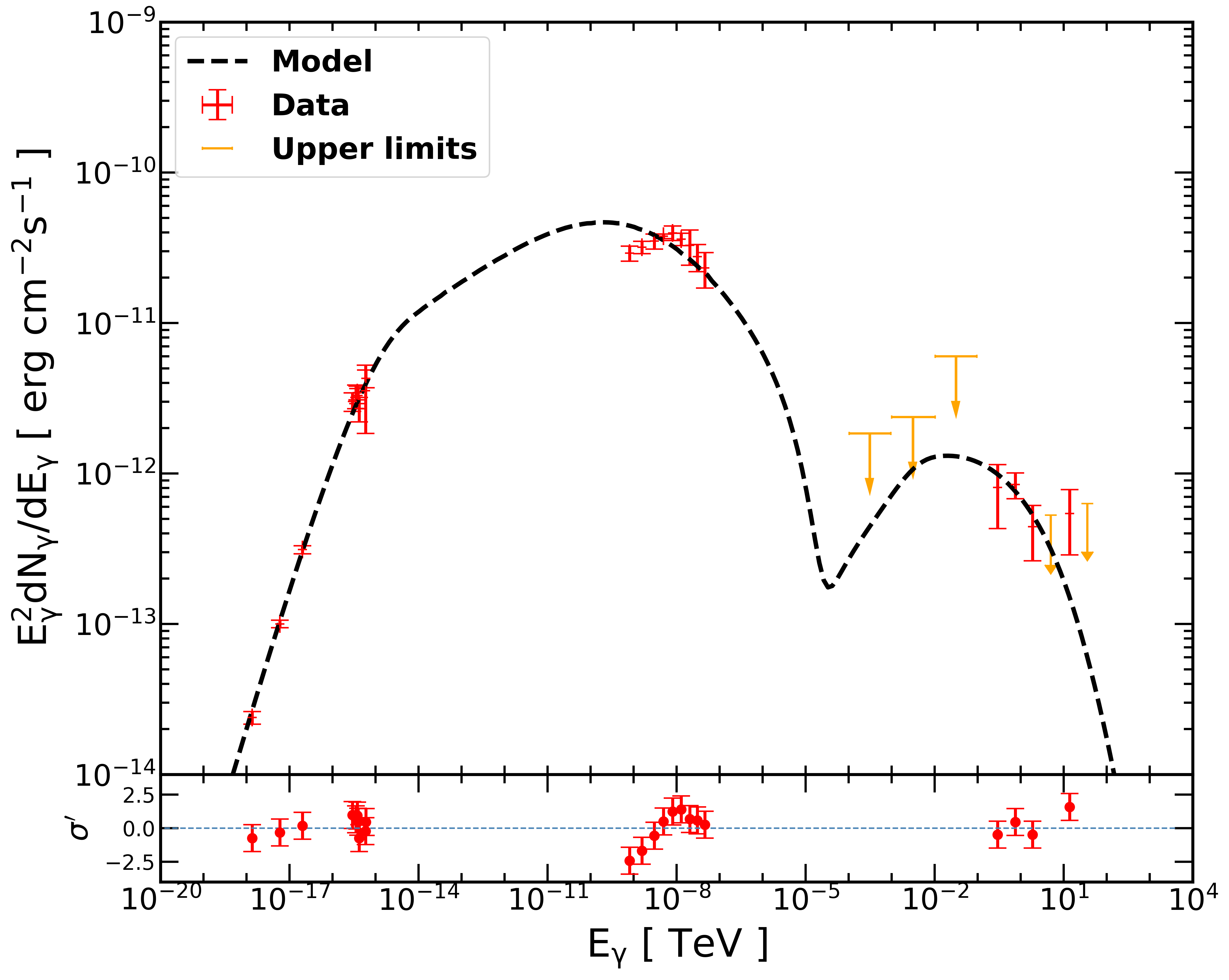}
   \caption{Best-fit SEDs for PWN Kes~75 (top) and G21.5 (bottom) from the 0D model using {\sc TIDE}. The data used in the fits are the same as the ones in Table~\ref{tbl:Data_energyrange_ref}. We also show \textit{Fermi} and H.E.S.S. upper limits for G21.5 (in orange). Note that they are shown only for representational purposes and were not part of the formal model fit.}
    \label{sed_0D}
\end{figure}

For comparison to the above results, we have  used {\sc TIDE}, a one-zone leptonic, time-dependent 0D model to describe the SED and dynamical evolution of a PWN (for a detailed description refer to \citealt{Martin2012, Torres2014, Martin2016, Martin2022}). We use here the latest and more complete incarnation of the model as 
described in the latter reference. This allows us to produce a fit of several parameters at once, and in particular, to note whether there are alternative sets of parameters 
producing good fits, which does not appear to be the case. One should note, however, that there is 
a difference in the magnetic field description compared to what is used in the 1D model. 
Differently to what has been commented above, {\sc TIDE} defines the $B$-field (independent of position) by solving a time-dependent equation that balances energy injection into the PWN and losses by expansion. This model difference may impact the comparison of the results of the 0D and 1D codes.

Considering the young ages of the pulsars, both PWNe we consider are believed to be in the free-expansion phase. Our 0D model could obtain a good fit to the SED data for these two PWNe. All model parameters, measured and assumed, along with fitted values and explored ranges, are presented in Table~\ref{pars_0D}. Additionally, Fig.~\ref{sed_0D} shows the best-fit SEDs for PWN Kes 75 and G21.5. In the case of the latter PWN, we approximated the piecewise X-ray power law fits \citep{2020ApJ90432H} to the data by points with errors to facilitate the automated fitting process. The deduced radii of these two PWNe are in good agreement with the measured values, and their magnetic fields are also typical compared to what is found in other PWNe. 

As outlined in Section~\ref{sec:ObsReviewKes75}, two X-ray outbursts were detected from PWN Kes~75 in 2006 and 2020 and they resulted in similar changes in the braking index values, decreasing from 2.65 (2.7) to 2.16 (2.19). 
The similarity between these changes seems to suggest the possibility of repeated outbursts. 
Typically, such variations in braking index values do not have a significant impact on SED and PWN properties when the true age of the pulsar, $t_{\rm age}$, is significantly smaller than the age that would lead to a negative initial spin-down age $\tau_0$ (e.g., \citealt{Martin2016,Sarkar2022}). 
By adopting $n=2.16$, the fitted value of $t_{\rm age}$ for Kes~75 is 1.05 kyr, similar to what was used above (i.e., 728 yr). When assuming $n=2.65$, any value of $t_{\rm age}$ larger than 0.883 kyr will result in a negative $\tau_0$. 
For $n=2.65$, if we use 0.882 kyr as an upper limit when fitting for $t_{\rm age}$, we find that $t_{\rm age}$ is very close to this upper limit, resulting in an unusually small $\tau_0$ and an unusually large $L_0$. By imposing a fixed $t_{\rm age}$ value of 0.7~kyr, our 0D model would yield distinct fitted parameters from those in Table~\ref{pars_0D} and an $R_{\rm PWN}$ of approximately 0.6 pc, which is notably lower than the observed value of 0.814 pc. This mismatch may be due to the influence of plausible repeated outbursts, or to the deviation of 2.65 from the value of $n$ that dominates along the entire evolution. Both frameworks infer (or assume) an age of $t_{\rm age}\sim1000$~yr for G21.5.

The preferred values of $\alpha_1$ and $\alpha_2$ for Kes~75 (see Table~\ref{tbl:Kes75}) are similar but outside the range suggested by the 0D code, and agree extremely well for
G21.5 (see Table~\ref{tbl:G21p5}). The difference in the case of Kes~75 could be explained as due to the degeneracy of the model parameters - while it may possible to find a good SED fit within the limits suggested by {\sc TIDE}, it might not describe the rest of the five spatio-temporal features we have considered in our preferred model fitting method. The inferred break energies are also quite similar between the two models. The preferred present-age magnetic field values for Kes~75 and G21.5 by the 1D model are $135$ and $140$ $\mu$G, while the 0D model suggests much smaller\footnote{The first value is the $B$-field at $r_0$; the volume-averaged value for the 1D model is of the order of $\sim10\mu$G for $R_{\rm PWN}\sim0.8$~pc, which is closer to the values preferred by the 0D model, although not exactly the same. The value varies depending on the choice of the $R_{\rm PWN}$.} values of $\sim 30$ and $\sim 47$ $\mu$G. These values should be compared to the equipartition magnetic fields of  $\sim 40$ $\mu$G for Kes 75 \citep{Ng08} and $\sim 300$ $\mu$G for G21.5 \citep{deJager2008Ferreira}. The difference between the preferred magnetic field values between the two models could be attributed to the difference in the magnetic field prescription between the two approaches. The considered magnetic energy fractions and soft-photon temperatures and energy densities are a bit different, indicating some freedom in these parameters. Finally, the 1D model can constrain model parameters connected with spatial aspects, such as the diffusion coefficient, bulk flow speed, and $B$-field profile, while the 0D model may infer dynamically-linked quantities such as the ejected mass and PWN radius that the 1D model does not consider. We thus see a broad concordance in the predicted SEDs of the two models, as well as in the values of the overlapping model parameters, despite there also being differences as discussed above.

\section{Discussion}\label{sec:disc}
We presented a novel, comprehensive parameter study for Kes~75 and G21.5 in the context of a 1D spatial PWN model, allowing us to  collectively study their spectral and spatio-temporal features. We studied the impact of various modelling components like the magnetic field, bulk flow speed, diffusion, and braking index on such features.

For Kes~$75$, we introduced an abrupt increase in bulk flow speed over the last $50$ years of its lifespan, justifying this via the energy deposition from its pulsar's outbursts as observed over past two decades. However, we did not see any considerable effect on the spectral features, even for a few percent increase in the flow speed at different stages of the PWN evolution. The expansion of the nebula in the X-ray energy band does seem to closer reproduce the data from \citet{Reynolds2018} if we increase the fractional bulk flow significantly (see Fig.~\ref{fig:Kes75ExpansionBulkFlowBurstPercent}). This strongly suggests that without a substantial sudden increase in the bulk flow (or enhanced diffusion; see below) to replicate the outburst behaviour, the observational features cannot be well explained with the model.

\cite{Hu2022} applied a pure diffusion for a number of sources and provided separate fits for the X-ray photon index and SB profiles. They suggested that diffusion is the dominating factor for particle transport in these sources. We performed a detailed diffusion variation study to understand its effect on the expansion and other spatio-temporal characteristics (see Fig.~\ref{fig:Kes75Diffusion} and \ref{fig:Kes75_kappa_scale_exp_new}). We showed that Bohm diffusion was favoured to explain the expansion, as compared to power-law-type diffusion. Bohm diffusion also sufficiently describes the SB profile. However, at least in the framework of other limitations of the model, it leads to unreasonable predictions when fitting the X-ray photon indices over several epochs in time and at varying radial distances, and hence, invoking a power-law-type diffusion provides an overall adequate representation to the four mentioned features. This highlights the importance of joint fits of various observational features instead of focusing on either the SED or a couple of spatial or temporal features only, as has usually been done in prior PWN modelling.

Our integrated flux for Kes~75 is more or less constant across the epochs spanning 16 years. \citet{Ng08} found that the total flux of the PWN did not change significantly between 2000 and 2006 in their \textit{Chandra} data, although some small-scale spatial features did exhibit flux variability. However,  \citet{Reynolds2018} found an unprecedented drop in flux of $\sim9\%\pm2\%$ from 2009 to 2016. It is not clear what is the cause of this relatively large drop in flux, mostly constrained to the northern half of the PWN, might be. Some effect beyond gradual evolutionary changes seem to be required. This may hint that future models need to incorporate inhomogeneities and more complex morphologies.

The model of \citet{StraalEtAl2023} invokes a power source of target photons for Kes~75 with a temperature $T\sim1\times10^5$~K and an energy density $u\sim3\times10^3$~eV\,cm$^{-3}$, and interpret this as the presence of a Wolf-Rayet star, thought to be the binary companion of the progenitor of Kes~75. Our results do not require such a bright source of target photons, which is a requirement in their model due to the inclusion of GeV data. We also note that we assume a harder injection spectrum ($\alpha_2=2.4$ as compared to their $\alpha_2 = 3.05$).

As can be seen from Fig.~\ref{fig:BestFit_Kes75}, our model can adequately represent the data, especially considering the attempt to explain all six observational features with one parameter set. The parameters are found to still be somewhat degenerate, but much more constrained when combining both spectral and spatial data. While we fit the data by eye in this paper, we are exploring a statistical approach that defines a workable combined statistical metric with which to characterise the goodness of fit for all data sets, given the disparate errors of each set. We also note  computational constraints that impact on the number of model parameters that we can free (Roy et al., in preparation). Including spatial data from more energy bands, and also modelling more sources should further break model degeneracies.

Using the multi-wavelength data of PWN G21.5, \citet{Tanaka_Takahara_2011,Torres2014} studied its physical properties, such as its age, PWN magnetic field, and the radiative processes occurring within the nebula. These time-dependent models explain very well the multi-wavelength emission properties of these objects. To understand the spatial as well radiative properties of these objects, \citet{van2020MNRAS3091V} recently developed a 1D spatio-temporal code. Such 1D models have also been tested by other groups, for example, \citet{2017ApJ...838..142I} used 1D models to investigate the simultaneous interpretation of the SED and radially-dependent SB. They found that the interpretation was difficult in the 1D model and emphasised that inclusion of a diffusion process would be useful. Later, they also included effects of diffusion (power-law-type) and found that their predicted X-ray SB profiles are consistent with observations \citep{2018ApJ...867..141I}. As shown in our earlier work for G21.5 \citep{van2020MNRAS3091V}, the Bohm-type diffusion could explain the SED and photon index profiles. In this work, we consider the power-law type diffusion scenario and whereas it can explain the SED and SB profiles well, below $\sim 20^{\prime\prime}$ the photon-index profile can not be accounted for.
A different approach applied to G21.5 made use of a hybrid hydrodynamic-radiative model \citep{Olmi2020}. Using power-law indices (assumed as free parameters) for specifying the particle distribution, they prescribed how to obtain the pressure in relativistic particles as a function of time using a hydrodynamical (HD) simulation - after considering radiation losses and magnetic energy input throughout the PWN lifetime. Recently, \citet{2020ApJ90432H} could explain the photon indices measured across the X-ray bands with their HD model, as well as the multi-wavelength spectrum of PWN G21.5.
We believe that the differences in the ability of the models to explain different aspects of the PWNe behaviour is most likely due to the differences in the magnetic field treatment and its radial dependence between the respective models and details of the HD approach. We note that \citet{
2020ApJ90432H} inferred a softer index $\alpha_1 \sim 2.9$ and harder index $\alpha_2 \sim 2.4$ of the particle injection spectrum, while we infer $\alpha_1 \sim 1.1$ and harder index $\alpha_2 \sim 2.5$, as is more typical. We cannot fit the spectral data when $\alpha_1$ is softer than $\alpha_2$. This difference may stem from the different types of model we apply to the data.

We lastly used a 0D code, {\sc TIDE}, which was run independently using a Nelder-Mead approach to find the best-fitting model parameter ranges for Kes~75 and G21.5, to compare with those of our 1D code \citep{Martin2022}. The present-day magnetic field value is substantially different between the two methods, which makes it apparent that the difference in the magnetic field prescription is also here a major factor influencing the values of model parameters. Overall, however,
we find a broad correspondence, taking note of the different approaches and the fact that each model is constraining somewhat different model parameter sets.

\section{Conclusions}\label{sec:con}
The availability of spectral and spatio-temporal data for Kes~75 and G21.5 inspired us to model these two young PWNe in some detail. With the inclusion of PWN features over spatial, spectral, and temporal scales, our model offers a novel approach involving the joint exploration and explanation of the SED, flux over different epochs, SB profile, expansion over several years, and X-ray photon index, utilising all currently available multi-wavelength data for these sources\footnote{The preprint of \citet{Sathyaprakash2024} has recently come to our attention. They provide SB profile observations from \textit{XMM-Newton} in different energy bands for the post-2020 outburst epoch. We verified that the profile we model in this paper for $0.5 - 8$~keV range is very similar in shape to the profiles for the sub-bands within their $0.3 - 8$~keV range. Given the similarity of these profiles, our conclusions should be largely unaffected upon including these data. We will explore the effect of including energy-dependent profiles in a future work.}. We found an overall reasonable representation of the data produced by the model for these sources, and compared this with the results from an independent 0D model ({\sc TIDE}). Our model parameters are broadly similar to what has been used by other authors, and additionally, we have now been able to constrain some parameters that are spatial in nature. Indeed, the major motivation for this paper was to apply a multi-zone model to both spectral and spatial PWN data to investigate whether we can obtain stronger constraints or new constraints on the modelled source properties. This first application of our multi-zone code to Kes~75 revealed that the the broad trends in all of the 6 data sets (Fig.~\ref{fig:BestFit_Kes75}) could be qualitatively captured, but less so in the case of the PWN expansion rate. In the case of G21.5, the code can qualitatively capture two out of three trends we aimed to explain simultaneously, with the X-ray index vs radius proving to be challenging to fit. We also produced plots for the integrated flux and X-ray photon index over different epochs for G21.5 using the previously-inferred model parameters, but we note that observation time span is extremely small, and future data taken over a longer duration of several years (like in the case of Kes~75) could be useful to consider these features in the exploration of the model parameters. Also, the variation of the X-ray photon index with respect to energy shows a harder to softer transition in the data, which the model captures reasonably well. The apparent incapability of the model to jointly fit all datasets implies that there is still some model development needed, and that invoking impulsive injection of energy or different diffusion scenarios are not adequate. Perhaps the PWN expansion may be driven by internal pressure, and not predominantly by diffusion. Different enhancements may have different impacts on the model outputs and will be the subject of a future study. A first approach may be to incorporate MHD or HD results for the $B$-field and $V$ profiles (also based on certain assumptions in those contexts) rather than using parametric forms to attempt to improve the fits.

Our comprehensive parameter study revealed the underlying degeneracy in the parameters, and the need of fine-tuning of independent parameters to find reasonable (compromise) joint fits for all the features. On the other hand, we also found that the effort to jointly explain spectral and spatial data aided in breaking some degeneracies. To improve our model in future, we can reconsider our magnetic field and flow speed profile prescriptions.

For the current work, we explored the effects of different parameter combinations, relying only on by-eye fitting. While our preferred model does explain the data to some extent, we see opportunities for model refinement, as well as using Markov chain Monte Carlo (MCMC) methods or a faster Nelder-Mead approach (as used in the {\sc TIDE} code above) to further explore the parameter space and determine model parameters that best align with the data more efficiently. The issue then becomes how to define a suitable figure of merit that captures the quality of the fit of diverse data sets (spectra, fluxes vs time, spatial profiles, etc.), as well as computational constraints.

The high-resolution \textit{Chandra} images of \citet{Ng08} indicate a strong axial symmetry of the core PWN on $10^{\prime\prime}$ scale, while our models are spherically symmetric and focus on a larger spatial scale. There is a clear need for the development of an axisymmetric 2D PWN model, given the ubiquity of jet-and-torus morphologies, also in Kes~75. Future work will include the development of 2D models that will allow us to more accurately model the PWN structure on smaller spatial scales, also taking into account results from HD or MHD models that can provide more realistic magnetic field and flow profiles.

For Kes~75, modelling outbursts over a relatively short duration of its lifetime did not produce significant changes, as is the case for instance when spin-down rate transitions happen (see, e.g., \citealt{ge2019}). In future, continuous observation of the glitching behaviour of J1846 and corresponding changes in braking index and flux could offer insights on how to more realistically model this behaviour.

Given the dependence of the results presented here on parameters pertaining to particle diffusion, such as the energy dependence of this coefficient as discussed in Section~\ref{sec-dif} and $\kappa_{\rm X}$ (see Equation~\ref{eqn:KolmogorovDiffusion}), as well as the theoretical dependence of these quantities on the turbulent properties of the PWN plasma, further insights into the exact nature of turbulence in these environments are needed. For example, from theory, the parameter $\kappa_{\rm X}$ would depend closely on quantities such as the magnetic variance, correlation scale, and turbulent anisotropy \cite[see, e.g.,][and references therein]{EngelbrechtEA22}. Although some observational evidence for turbulence in PWNe has been recently presented \citep[e.g.,][]{MizunoEA23,BucciantiniEA23}, said observations are not yet detailed enough to provide the required information. An alternative route would be to employ turbulence transport models, such as those employed in heliospheric cosmic-ray transport studies \citep[see, e.g.,][]{ZankEA18,OughtonEngelbrecht21,AdhikariEA21fluids,EngelbrechtEA22,KleimannEA23}. Although such an approach would provide the necessary inputs, these models would need to be modified to take into account the different (relativistic, low-$\sigma$) plasma conditions in PWNe, and would be difficult to constrain with observations. Nevertheless, this latter approach may be a useful future avenue for constraining diffusion-related parameters in PWNe.

We emphasise that all the spatial and temporal data available to us comes from a very narrow keV band (between $0.5-8$ keV), which restricts the exploration and constraints of the parameters specific only to that energy band. Availability of such features over multi-wavelength bands could be very useful to remove degeneracies in the model and finding strict constraints for various parameters. Future work will also involve studying more young sources as data become available.
\section*{Acknowledgements}
We thank the anonymous referee for their valuable comments that greatly improved the manuscript. AK and CV would like to thank Stephen P.\ Reynolds, Carlo van Rensburg, and Sunil Chandra for helpful discussions. This work is based on research supported wholly / in part by the National Research Foundation of South Africa (NRF; Grant Number 99072). The Grant holder acknowledges that opinions, findings and conclusions or recommendations expressed in any publication generated by the NRF-supported research is that of the author(s), and that the NRF accepts no liability whatsoever in this regard. This work was also partially supported by the National Scholarship Council (PhD fellowship from the China Scholarship Council (CSC) (No. 202107030003)) and by the grant PID2021-124581OB-I00 of MCIU/AEI/10.13039/501100011033 and 2021SGR00426 of the Generalitat de Catalunya, by the Spanish program Unidad de Excelencia María de Maeztu CEX2020-001058-M and by MCIU with funding from European Union NextGeneration EU (PRTR-C17.I1). SJT would like to thank JPJS Bilateral Program (No. JPJSBP120229940), Fostering Joint International Research (B) (No. JP20KK0064), the Sumitomo Foundation (No. 210629), and the Research Foundation For Opto-Science and Technology for support. We also acknowledge the use of the NASA Astrophysics Data Service (ADS).
\section*{Data Availability}
There are no new observational data in this paper.
Further theoretical details will be made available upon reasonable request to the authors.
\bibliographystyle{mnras}
\bibliography{0000_PWN_Kes75_G21p5}

\begin{thebibliography}{}
\makeatletter
\relax
\def\mn@urlcharsother{\let\do\@makeother \do\$\do\&\do\#\do\^\do\_\do\%\do\~}
\def\mn@doi{\begingroup\mn@urlcharsother \@ifnextchar [ {\mn@doi@} {\mn@doi@[]}}
\def\mn@doi@[#1]#2{\def\@tempa{#1}\ifx\@tempa\@empty \href {http://dx.doi.org/#2} {doi:#2}\else \href {http://dx.doi.org/#2} {#1}\fi \endgroup}
\def\mn@eprint#1#2{\mn@eprint@#1:#2::\@nil}
\def\mn@eprint@arXiv#1{\href {http://arxiv.org/abs/#1} {{\tt arXiv:#1}}}
\def\mn@eprint@dblp#1{\href {http://dblp.uni-trier.de/rec/bibtex/#1.xml} {dblp:#1}}
\def\mn@eprint@#1:#2:#3:#4\@nil{\def\@tempa {#1}\def\@tempb {#2}\def\@tempc {#3}\ifx \@tempc \@empty \let \@tempc \@tempb \let \@tempb \@tempa \fi \ifx \@tempb \@empty \def\@tempb {arXiv}\fi \@ifundefined {mn@eprint@\@tempb}{\@tempb:\@tempc}{\expandafter \expandafter \csname mn@eprint@\@tempb\endcsname \expandafter{\@tempc}}}

\bibitem[\protect\citeauthoryear{Abdollahi et~al.,}{Abdollahi et~al.}{2020}]{FermiDR2_2020}
Abdollahi S.,  et~al., 2020, \mn@doi [\apjs] {10.3847/1538-4365/ab6bcb}, 247, 33

\bibitem[\protect\citeauthoryear{{Abdollahi} et~al.,}{{Abdollahi} et~al.}{2022}]{2023Fermi4FGL_DR4}
{Abdollahi} S.,  et~al., 2022, \mn@doi [\apjs] {10.3847/1538-4365/ac6751}, \href {https://ui.adsabs.harvard.edu/abs/2022ApJS..260...53A} {260, 53}

\bibitem[\protect\citeauthoryear{{Abeysekara} et~al.,}{{Abeysekara} et~al.}{2017}]{Abe2017Sci911A}
{Abeysekara} A.~U.,  et~al., 2017, \mn@doi [Science] {10.1126/science.aan4880}, \href {https://ui.adsabs.harvard.edu/abs/2017Sci...358..911A} {358, 911}

\bibitem[\protect\citeauthoryear{{Ackermann} et~al.,}{{Ackermann} et~al.}{2011}]{GeV_l_2011ApJ35A}
{Ackermann} M.,  et~al., 2011, \mn@doi [\apj] {10.1088/0004-637X/726/1/35}, \href {https://ui.adsabs.harvard.edu/abs/2011ApJ...726...35A} {726, 35}

\bibitem[\protect\citeauthoryear{Adhikari, Zank  \& Zhao}{Adhikari et~al.}{2021}]{AdhikariEA21fluids}
Adhikari L.,  Zank G.~P.,   Zhao L.,  2021, \mn@doi [Fluids] {10.3390/fluids6100368}, 6, 368

\bibitem[\protect\citeauthoryear{{Aharonian} \& {Atoyan}}{{Aharonian} \& {Atoyan}}{1996}]{AharonianAtoyan1996}
{Aharonian} F.~A.,  {Atoyan} A.~M.,  1996, \aap, \href {https://ui.adsabs.harvard.edu/abs/1996A&A...309..917A} {309, 917}

\bibitem[\protect\citeauthoryear{{Altenhoff}, {Downes}, {Goad}, {Maxwell}  \& {Rinehart}}{{Altenhoff} et~al.}{1970}]{1970AnAS....1..319A}
{Altenhoff} W.~J.,  {Downes} D.,  {Goad} L.,  {Maxwell} A.,   {Rinehart} R.,  1970, A\&AS, \href {https://ui.adsabs.harvard.edu/abs/1970A&AS....1..319A} {1, 319}

\bibitem[\protect\citeauthoryear{{Amato}, {Salvati}, {Bandiera}, {Pacini}  \& {Woltjer}}{{Amato} et~al.}{2000}]{AmatoEtAl2000}
{Amato} E.,  {Salvati} M.,  {Bandiera} R.,  {Pacini} F.,   {Woltjer} L.,  2000, \mn@doi [\aap] {10.48550/arXiv.astro-ph/9911163}, \href {https://ui.adsabs.harvard.edu/abs/2000A&A...359.1107A} {359, 1107}

\bibitem[\protect\citeauthoryear{{Archibald}, {Kaspi}, {Livingstone}  \& {McLaughlin}}{{Archibald} et~al.}{2008}]{Archibald08}
{Archibald} A.~M.,  {Kaspi} V.~M.,  {Livingstone} M.~A.,   {McLaughlin} M.~A.,  2008, \mn@doi [\apj] {10.1086/591661}, \href {https://ui.adsabs.harvard.edu/abs/2008ApJ...688..550A} {688, 550}

\bibitem[\protect\citeauthoryear{{Archibald}, {Kaspi}, {Beardmore}, {Gehrels}  \& {Kennea}}{{Archibald} et~al.}{2015}]{Archibald15}
{Archibald} R.~F.,  {Kaspi} V.~M.,  {Beardmore} A.~P.,  {Gehrels} N.,   {Kennea} J.~A.,  2015, \mn@doi [\apj] {10.1088/0004-637X/810/1/67}, \href {https://ui.adsabs.harvard.edu/abs/2015ApJ...810...67A} {810, 67}

\bibitem[\protect\citeauthoryear{{Archibald} et~al.,}{{Archibald} et~al.}{2016}]{2016ArchibaldbrakingindexGT3}
{Archibald} R.~F.,  et~al., 2016, \mn@doi [\apjl] {10.3847/2041-8205/819/1/L16}, \href {https://ui.adsabs.harvard.edu/abs/2016ApJ...819L..16A} {819, L16}

\bibitem[\protect\citeauthoryear{{Ballet}, {Bruel}, {Burnett}, {Lott}  \& {The Fermi-LAT collaboration}}{{Ballet} et~al.}{2023}]{2023FermiDR4}
{Ballet} J.,  {Bruel} P.,  {Burnett} T.~H.,  {Lott} B.,   {The Fermi-LAT collaboration} 2023, \mn@doi [arXiv e-prints] {10.48550/arXiv.2307.12546}, \href {https://ui.adsabs.harvard.edu/abs/2023arXiv230712546B} {p. arXiv:2307.12546}

\bibitem[\protect\citeauthoryear{{Bandiera}, {Bucciantini}, {Mart{\'\i}n}, {Olmi}  \& {Torres}}{{Bandiera} et~al.}{2023a}]{Bandiera2023a}
{Bandiera} R.,  {Bucciantini} N.,  {Mart{\'\i}n} J.,  {Olmi} B.,   {Torres} D.~F.,  2023a, \mn@doi [\mnras] {10.1093/mnras/stad134}, \href {https://ui.adsabs.harvard.edu/abs/2023MNRAS.520.2451B} {520, 2451}

\bibitem[\protect\citeauthoryear{{Bandiera}, {Bucciantini}, {Olmi}  \& {Torres}}{{Bandiera} et~al.}{2023b}]{Bandiera2023}
{Bandiera} R.,  {Bucciantini} N.,  {Olmi} B.,   {Torres} D.~F.,  2023b, \mn@doi [\mnras] {10.1093/mnras/stad2387}, \href {https://ui.adsabs.harvard.edu/abs/2023MNRAS.525.2839B} {525, 2839}

\bibitem[\protect\citeauthoryear{{Becker} \& {Helfand}}{{Becker} \& {Helfand}}{1984}]{Becker1984}
{Becker} R.~H.,  {Helfand} D.~J.,  1984, \mn@doi [\apj] {10.1086/162285}, \href {https://ui.adsabs.harvard.edu/abs/1984ApJ...283..154B} {283, 154}

\bibitem[\protect\citeauthoryear{{Becker} \& {Szymkowiak}}{{Becker} \& {Szymkowiak}}{1981}]{becker1981ApJ23B}
{Becker} R.~H.,  {Szymkowiak} A.~E.,  1981, \mn@doi [\apjl] {10.1086/183615}, \href {https://ui.adsabs.harvard.edu/abs/1981ApJ...248L..23B} {248, L23}

\bibitem[\protect\citeauthoryear{{Bietenholz} \& {Bartel}}{{Bietenholz} \& {Bartel}}{2008}]{2008MNRAS861411B}
{Bietenholz} M.~F.,  {Bartel} N.,  2008, \mn@doi [\mnras] {10.1111/j.1365-2966.2008.13058.x}, \href {https://ui.adsabs.harvard.edu/abs/2008MNRAS.386.1411B} {386, 1411}

\bibitem[\protect\citeauthoryear{{Blondin}, {Chevalier}  \& {Frierson}}{{Blondin} et~al.}{2001}]{Blondin2001}
{Blondin} J.~M.,  {Chevalier} R.~A.,   {Frierson} D.~M.,  2001, \mn@doi [ApJ] {10.1086/324042}, \href {http://adsabs.harvard.edu/abs/2001ApJ...563..806B} {563, 806}

\bibitem[\protect\citeauthoryear{{Blumenthal} \& {Gould}}{{Blumenthal} \& {Gould}}{1970}]{1970RvMP237B}
{Blumenthal} G.~R.,  {Gould} R.~J.,  1970, \mn@doi [Reviews of Modern Physics] {10.1103/RevModPhys.42.237}, \href {https://ui.adsabs.harvard.edu/abs/1970RvMP...42..237B} {42, 237}

\bibitem[\protect\citeauthoryear{{Blumer}, {Safi-Harb}, {McLaughlin}  \& {Fiore}}{{Blumer} et~al.}{2021}]{Blumer2021}
{Blumer} H.,  {Safi-Harb} S.,  {McLaughlin} M.~A.,   {Fiore} W.,  2021, \mn@doi [\apjl] {10.3847/2041-8213/abf11d}, \href {https://ui.adsabs.harvard.edu/abs/2021ApJ...911L...6B} {911, L6}

\bibitem[\protect\citeauthoryear{{Bock} \& {Gaensler}}{{Bock} \& {Gaensler}}{2005}]{BockGaensler2005}
{Bock} D.~C.~J.,  {Gaensler} B.~M.,  2005, \mn@doi [ApJ] {10.1086/429789}, \href {https://ui.adsabs.harvard.edu/abs/2005ApJ...626..343B} {626, 343}

\bibitem[\protect\citeauthoryear{{Bucciantini}, {Blondin}, {Del Zanna}  \& {Amato}}{{Bucciantini} et~al.}{2003}]{Bucciantini2003}
{Bucciantini} N.,  {Blondin} J.~M.,  {Del Zanna} L.,   {Amato} E.,  2003, \mn@doi [\aap] {10.1051/0004-6361:20030624}, \href {https://ui.adsabs.harvard.edu/abs/2003A&A...405..617B} {405, 617}

\bibitem[\protect\citeauthoryear{{Bucciantini}, {Bandiera}, {Blondin}, {Amato}  \& {Del Zanna}}{{Bucciantini} et~al.}{2004}]{Bucciantini2004}
{Bucciantini} N.,  {Bandiera} R.,  {Blondin} J.~M.,  {Amato} E.,   {Del Zanna} L.,  2004, \mn@doi [A\&A] {10.1051/0004-6361:20034400}, \href {https://ui.adsabs.harvard.edu/abs/2004A&A...422..609B} {422, 609}

\bibitem[\protect\citeauthoryear{{Bucciantini}, {Arons}  \& {Amato}}{{Bucciantini} et~al.}{2011}]{Bucciantini2011}
{Bucciantini} N.,  {Arons} J.,   {Amato} E.,  2011, \mn@doi [\mnras] {10.1111/j.1365-2966.2010.17449.x}, \href {https://ui.adsabs.harvard.edu/abs/2011MNRAS.410..381B} {410, 381}

\bibitem[\protect\citeauthoryear{{Bucciantini} et~al.,}{{Bucciantini} et~al.}{2023}]{BucciantiniEA23}
{Bucciantini} N.,  et~al., 2023, \mn@doi [Nature Astronomy] {10.1038/s41550-023-01936-8}, \href {https://ui.adsabs.harvard.edu/abs/2023NatAs...7..602B} {7, 602}

\bibitem[\protect\citeauthoryear{{Caballero-Lopez}, {Engelbrecht}  \& {Richardson}}{{Caballero-Lopez} et~al.}{2019}]{CaballeroEA19}
{Caballero-Lopez} R.~A.,  {Engelbrecht} N.~E.,   {Richardson} J.~D.,  2019, \mn@doi [\apj] {10.3847/1538-4357/ab3c57}, \href {https://ui.adsabs.harvard.edu/abs/2019...883...73C} {883, 73}

\bibitem[\protect\citeauthoryear{{Camilo}, {Ransom}, {Gaensler}, {Slane}, {Lorimer}, {Reynolds}, {Manchester}  \& {Murray}}{{Camilo} et~al.}{2006}]{camilo2006ApJ637456C}
{Camilo} F.,  {Ransom} S.~M.,  {Gaensler} B.~M.,  {Slane} P.~O.,  {Lorimer} D.~R.,  {Reynolds} J.,  {Manchester} R.~N.,   {Murray} S.~S.,  2006, \mn@doi [\apj] {10.1086/498386}, \href {https://ui.adsabs.harvard.edu/abs/2006ApJ...637..456C} {637, 456}

\bibitem[\protect\citeauthoryear{{Cerutti} \& {Giacinti}}{{Cerutti} \& {Giacinti}}{2020}]{Cerutti20}
{Cerutti} B.,  {Giacinti} G.,  2020, \mn@doi [\aap] {10.1051/0004-6361/202038883}, \href {https://ui.adsabs.harvard.edu/abs/2020A&A...642A.123C} {642, A123}

\bibitem[\protect\citeauthoryear{{Chevalier} \& {Fransson}}{{Chevalier} \& {Fransson}}{1992}]{ChevalierFransson1992}
{Chevalier} R.~A.,  {Fransson} C.,  1992, \mn@doi [\apj] {10.1086/171674}, \href {https://ui.adsabs.harvard.edu/abs/1992ApJ...395..540C} {395, 540}

\bibitem[\protect\citeauthoryear{{De Sarkar}, {Zhang}, {Mart{\'\i}n}, {Torres}, {Li}  \& {Hou}}{{De Sarkar} et~al.}{2022}]{Sarkar2022}
{De Sarkar} A.,  {Zhang} W.,  {Mart{\'\i}n} J.,  {Torres} D.~F.,  {Li} J.,   {Hou} X.,  2022, \mn@doi [\aap] {10.1051/0004-6361/202244841}, \href {https://ui.adsabs.harvard.edu/abs/2022A&A...668A..23D} {668, A23}

\bibitem[\protect\citeauthoryear{{Del Zanna}, {Volpi}, {Amato}  \& {Bucciantini}}{{Del Zanna} et~al.}{2006}]{2006AnA621D}
{Del Zanna} L.,  {Volpi} D.,  {Amato} E.,   {Bucciantini} N.,  2006, \mn@doi [\aap] {10.1051/0004-6361:20064858}, \href {https://ui.adsabs.harvard.edu/abs/2006A&A...453..621D} {453, 621}

\bibitem[\protect\citeauthoryear{{Dempers} \& {Engelbrecht}}{{Dempers} \& {Engelbrecht}}{2020}]{DempersEngelbrecht20}
{Dempers} N.,  {Engelbrecht} N.~E.,  2020, \mn@doi [Advances in Space Research] {10.1016/j.asr.2020.01.040}, \href {https://ui.adsabs.harvard.edu/abs/2020AdSpR..65.2072D} {65, 2072}

\bibitem[\protect\citeauthoryear{{Di Mauro}, {Manconi}  \& {Donato}}{{Di Mauro} et~al.}{2020}]{dimar20203035D}
{Di Mauro} M.,  {Manconi} S.,   {Donato} F.,  2020, \mn@doi [\prd] {10.1103/PhysRevD.101.103035}, \href {https://ui.adsabs.harvard.edu/abs/2020PhRvD.101j3035D} {101, 103035}

\bibitem[\protect\citeauthoryear{{Djannati-Ata{\"\i}}, {deJager}, {Terrier}, {Gallant}  \& {Hoppe}}{{Djannati-Ata{\"\i}} et~al.}{2008a}]{2008ICRC823D}
{Djannati-Ata{\"\i}} A.,  {deJager} O.~C.,  {Terrier} R.,  {Gallant} Y.~A.,   {Hoppe} S.,  2008a, in International Cosmic Ray Conference. pp 823--826 (\mn@eprint {arXiv} {0710.2247}), \mn@doi{10.48550/arXiv.0710.2247}

\bibitem[\protect\citeauthoryear{{Djannati-Ata{\"i}}, {deJager}, {Terrier}, {Gallant}  \& {Hoppe}}{{Djannati-Ata{\"i}} et~al.}{2008b}]{Djannati2008}
{Djannati-Ata{\"i}} A.,  {deJager} O.~C.,  {Terrier} R.,  {Gallant} Y.~A.,   {Hoppe} S.,  2008b, International Cosmic Ray Conference, \href {http://adsabs.harvard.edu/abs/2008ICRC....2..823D} {2, 823}

\bibitem[\protect\citeauthoryear{{Ek{\c{s}}i}, {Anda{\c{c}}}, {{\c{C}}{\i}k{\i}nto{\u{g}}lu}, {G{\"u}gercino{\u{g}}lu}, {Vahdat Motlagh}  \& {K{\i}z{\i}ltan}}{{Ek{\c{s}}i} et~al.}{2016}]{2016Eksi}
{Ek{\c{s}}i} K.~Y.,  {Anda{\c{c}}} I.~C.,  {{\c{C}}{\i}k{\i}nto{\u{g}}lu} S.,  {G{\"u}gercino{\u{g}}lu} E.,  {Vahdat Motlagh} A.,   {K{\i}z{\i}ltan} B.,  2016, \mn@doi [\apj] {10.3847/0004-637X/823/1/34}, \href {https://ui.adsabs.harvard.edu/abs/2016ApJ...823...34E} {823, 34}

\bibitem[\protect\citeauthoryear{{Engelbrecht}}{{Engelbrecht}}{2019}]{Engelbrecht19}
{Engelbrecht} N.~E.,  2019, \mn@doi [ApJ] {10.3847/1538-4357/aafe7f}, \href {https://ui.adsabs.harvard.edu/abs/2019ApJ...872..124E} {872, 124}

\bibitem[\protect\citeauthoryear{{Engelbrecht} \& {Moloto}}{{Engelbrecht} \& {Moloto}}{2021}]{EngelbrechtMoloto21}
{Engelbrecht} N.~E.,  {Moloto} K.~D.,  2021, \mn@doi [ApJ] {10.3847/1538-4357/abd3a5}, \href {https://ui.adsabs.harvard.edu/abs/2021ApJ...908..167E} {908, 167}

\bibitem[\protect\citeauthoryear{{Engelbrecht} et~al.,}{{Engelbrecht} et~al.}{2022}]{EngelbrechtEA22}
{Engelbrecht} N.~E.,  et~al., 2022, \mn@doi [Space Sci. Rev.] {10.1007/s11214-022-00896-1}, \href {https://ui.adsabs.harvard.edu/abs/2022SSRv..218...33E} {218, 33}

\bibitem[\protect\citeauthoryear{{Ferreira} \& {de Jager}}{{Ferreira} \& {de Jager}}{2008}]{Ferreira_deJager2008}
{Ferreira} S.~E.~S.,  {de Jager} O.~C.,  2008, \mn@doi [A\&A] {10.1051/0004-6361:20077824}, \href {http://adsabs.harvard.edu/abs/2008A%26A...478...17F} {478, 17}

\bibitem[\protect\citeauthoryear{{Fleishman} \& {Bietenholz}}{{Fleishman} \& {Bietenholz}}{2007}]{2007MNRAS625F}
{Fleishman} G.~D.,  {Bietenholz} M.~F.,  2007, \mn@doi [\mnras] {10.1111/j.1365-2966.2007.11450.x}, \href {https://ui.adsabs.harvard.edu/abs/2007MNRAS.376..625F} {376, 625}

\bibitem[\protect\citeauthoryear{{Gaensler} \& {Slane}}{{Gaensler} \& {Slane}}{2006}]{Gaensler06}
{Gaensler} B.~M.,  {Slane} P.~O.,  2006, \mn@doi [ARA\&A] {10.1146/annurev.astro.44.051905.092528}, \href {http://adsabs.harvard.edu/abs/2006ARA%26A..44...17G} {44, 17}

\bibitem[\protect\citeauthoryear{{Gallant} \& {Tuffs}}{{Gallant} \& {Tuffs}}{1999}]{gallan1999ESASP313G}
{Gallant} Y.~A.,  {Tuffs} R.~J.,  1999, in {Cox} P.,  {Kessler} M.,  eds,  ESA Special Publication Vol. 427, The Universe as Seen by ISO. p.~313

\bibitem[\protect\citeauthoryear{{Gavriil}, {Gonzalez}, {Gotthelf}, {Kaspi}, {Livingstone}  \& {Woods}}{{Gavriil} et~al.}{2008}]{Gavriil2008}
{Gavriil} F.~P.,  {Gonzalez} M.~E.,  {Gotthelf} E.~V.,  {Kaspi} V.~M.,  {Livingstone} M.~A.,   {Woods} P.~M.,  2008, \mn@doi [Science] {10.1126/science.1153465}, \href {https://ui.adsabs.harvard.edu/abs/2008Sci...319.1802G} {319, 1802}

\bibitem[\protect\citeauthoryear{{Ge} et~al.,}{{Ge} et~al.}{2019}]{ge2019}
{Ge} M.~Y.,  et~al., 2019, \mn@doi [Nature Astronomy] {10.1038/s41550-019-0853-5}, \href {https://ui.adsabs.harvard.edu/abs/2019NatAs...3.1122G} {3, 1122}

\bibitem[\protect\citeauthoryear{{Gelfand}, {Slane}  \& {Temim}}{{Gelfand} et~al.}{2014}]{Gelfand2014}
{Gelfand} J.~D.,  {Slane} P.~O.,   {Temim} T.,  2014, \mn@doi [Astronomische Nachrichten] {10.1002/asna.201312039}, \href {https://ui.adsabs.harvard.edu/abs/2014AN....335..318G} {335, 318}

\bibitem[\protect\citeauthoryear{{Gotthelf}, {Vasisht}, {Boylan-Kolchin}  \& {Torii}}{{Gotthelf} et~al.}{2000}]{Gotthelf2000}
{Gotthelf} E.~V.,  {Vasisht} G.,  {Boylan-Kolchin} M.,   {Torii} K.,  2000, \mn@doi [\apjl] {10.1086/312923}, \href {https://ui.adsabs.harvard.edu/abs/2000ApJ...542L..37G} {542, L37}

\bibitem[\protect\citeauthoryear{{Gotthelf}, {Safi-Harb}, {Straal}  \& {Gelfand}}{{Gotthelf} et~al.}{2021}]{Gotthelf_2021}
{Gotthelf} E.~V.,  {Safi-Harb} S.,  {Straal} S.~M.,   {Gelfand} J.~D.,  2021, \mn@doi [\apj] {10.3847/1538-4357/abd32b}, \href {https://ui.adsabs.harvard.edu/abs/2021ApJ...908..212G} {908, 212}

\bibitem[\protect\citeauthoryear{{Gupta}, {Mitra}, {Green}  \& {Acharyya}}{{Gupta} et~al.}{2005}]{gupta2005CSc853G}
{Gupta} Y.,  {Mitra} D.,  {Green} D.~A.,   {Acharyya} A.,  2005, \mn@doi [Current Science] {10.48550/arXiv.astro-ph/0508257}, \href {https://ui.adsabs.harvard.edu/abs/2005CSci...89..853G} {89, 853}

\bibitem[\protect\citeauthoryear{{H.E.S.S. Collaboration} et~al.,}{{H.E.S.S. Collaboration} et~al.}{2018}]{HESSgps2018}
{H.E.S.S. Collaboration} et~al., 2018, \mn@doi [A\&A] {10.1051/0004-6361/201732098}, \href {https://ui.adsabs.harvard.edu/abs/2018A&A...612A...1H} {612, A1}

\bibitem[\protect\citeauthoryear{{Hattori}, {Straal}, {Zhang}, {Temim}, {Gelfand}  \& {Slane}}{{Hattori} et~al.}{2020}]{2020ApJ90432H}
{Hattori} S.,  {Straal} S.~M.,  {Zhang} E.,  {Temim} T.,  {Gelfand} J.~D.,   {Slane} P.~O.,  2020, \mn@doi [\apj] {10.3847/1538-4357/abba32}, \href {https://ui.adsabs.harvard.edu/abs/2020ApJ...904...32H} {904, 32}

\bibitem[\protect\citeauthoryear{{Herbst} et~al.,}{{Herbst} et~al.}{2022}]{HerbstEA22}
{Herbst} K.,  et~al., 2022, \mn@doi [Space Sci. Rev.] {10.1007/s11214-022-00894-3}, \href {https://ui.adsabs.harvard.edu/abs/2022SSRv..218...29H} {218, 29}

\bibitem[\protect\citeauthoryear{{Hitomi Collaboration} et~al.,}{{Hitomi Collaboration} et~al.}{2018}]{hitomi2018PASJ7038A}
{Hitomi Collaboration} et~al., 2018, \mn@doi [\pasj] {10.1093/pasj/psy027}, \href {https://ui.adsabs.harvard.edu/abs/2018PASJ...70...38A} {70, 38}

\bibitem[\protect\citeauthoryear{{Holler}, {Sch{\"o}ck}, {Eger}, {Kie{\ss}ling}, {Valerius}  \& {Stegmann}}{{Holler} et~al.}{2012}]{Holler12}
{Holler} M.,  {Sch{\"o}ck} F.~M.,  {Eger} P.,  {Kie{\ss}ling} D.,  {Valerius} K.,   {Stegmann} C.,  2012, \mn@doi [A\&A] {10.1051/0004-6361/201118121}, \href {https://ui.adsabs.harvard.edu/abs/2012A&A...539A..24H} {539, A24}

\bibitem[\protect\citeauthoryear{{Hu}, {Ishizaki}, {Ng}, {Tanaka}  \& {Mong}}{{Hu} et~al.}{2022}]{Hu2022}
{Hu} C.-P.,  {Ishizaki} W.,  {Ng} C.~Y.,  {Tanaka} S.~J.,   {Mong} Y.~L.,  2022, \mn@doi [\apj] {10.3847/1538-4357/ac4d2d}, \href {https://ui.adsabs.harvard.edu/abs/2022ApJ...927...87H} {927, 87}

\bibitem[\protect\citeauthoryear{{Hu} et~al.,}{{Hu} et~al.}{2023}]{HuEtal2023}
{Hu} C.-P.,  et~al., 2023, \mn@doi [arXiv e-prints] {10.48550/arXiv.2306.00902}, \href {https://ui.adsabs.harvard.edu/abs/2023arXiv230600902H} {p. arXiv:2306.00902}

\bibitem[\protect\citeauthoryear{{Ishizaki}, {Tanaka}, {Asano}  \& {Terasawa}}{{Ishizaki} et~al.}{2017}]{2017ApJ...838..142I}
{Ishizaki} W.,  {Tanaka} S.~J.,  {Asano} K.,   {Terasawa} T.,  2017, \mn@doi [\apj] {10.3847/1538-4357/aa679b}, \href {https://ui.adsabs.harvard.edu/abs/2017ApJ...838..142I} {838, 142}

\bibitem[\protect\citeauthoryear{{Ishizaki}, {Asano}  \& {Kawaguchi}}{{Ishizaki} et~al.}{2018}]{2018ApJ...867..141I}
{Ishizaki} W.,  {Asano} K.,   {Kawaguchi} K.,  2018, \mn@doi [\apj] {10.3847/1538-4357/aae389}, \href {https://ui.adsabs.harvard.edu/abs/2018ApJ...867..141I} {867, 141}

\bibitem[\protect\citeauthoryear{{Johnston} \& {Karastergiou}}{{Johnston} \& {Karastergiou}}{2017}]{Johnston2017}
{Johnston} S.,  {Karastergiou} A.,  2017, \mn@doi [\mnras] {10.1093/mnras/stx377}, \href {https://ui.adsabs.harvard.edu/abs/2017MNRAS.467.3493J} {467, 3493}

\bibitem[\protect\citeauthoryear{{Jokipii}}{{Jokipii}}{1966}]{Jokipii66}
{Jokipii} J.~R.,  1966, \mn@doi [ApJ] {10.1086/148912}, \href {https://ui.adsabs.harvard.edu/abs/1966ApJ...146..480J} {146, 480}

\bibitem[\protect\citeauthoryear{{Joshi}, {Tanaka}, {Miranda}  \& {Razzaque}}{{Joshi} et~al.}{2023}]{2023MNRAS5858J}
{Joshi} J.~C.,  {Tanaka} S.~J.,  {Miranda} L.~S.,   {Razzaque} S.,  2023, \mn@doi [\mnras] {10.1093/mnras/stad467}, \href {https://ui.adsabs.harvard.edu/abs/2023MNRAS.520.5858J} {520, 5858}

\bibitem[\protect\citeauthoryear{{Kargaltsev}, {Pavlov}, {Teter}  \& {Sanwal}}{{Kargaltsev} et~al.}{2003}]{Kargaltsev03}
{Kargaltsev} O.~Y.,  {Pavlov} G.~G.,  {Teter} M.~A.,   {Sanwal} D.,  2003, \mn@doi [\nar] {10.1016/S1387-6473(03)00077-0}, \href {https://ui.adsabs.harvard.edu/abs/2003NewAR..47..487K} {47, 487}

\bibitem[\protect\citeauthoryear{{Kennel} \& {Coroniti}}{{Kennel} \& {Coroniti}}{1984}]{Kennel1984a}
{Kennel} C.~F.,  {Coroniti} F.~V.,  1984, \mn@doi [ApJ] {10.1086/162356}, \href {http://adsabs.harvard.edu/abs/1984ApJ...283..694K} {283, 694}

\bibitem[\protect\citeauthoryear{{Kesteven}}{{Kesteven}}{1968}]{Kesteven1968}
{Kesteven} M.~J.~L.,  1968, \mn@doi [Australian Journal of Physics] {10.1071/PH680369}, \href {https://ui.adsabs.harvard.edu/abs/1968AuJPh..21..369K} {21, 369}

\bibitem[\protect\citeauthoryear{{Kirk}, {Lyubarsky}  \& {Petri}}{{Kirk} et~al.}{2009}]{KirkEA09}
{Kirk} J.~G.,  {Lyubarsky} Y.,   {Petri} J.,  2009, in {Becker} W.,  ed.,  Astrophysics and Space Science Library Vol. 357, Astrophysics and Space Science Library. p.~421 (\mn@eprint {arXiv} {astro-ph/0703116}), \mn@doi{10.1007/978-3-540-76965-1_16}

\bibitem[\protect\citeauthoryear{{Kleimann} et~al.,}{{Kleimann} et~al.}{2022}]{KleimannEA22}
{Kleimann} J.,  et~al., 2022, \mn@doi [\ssr] {10.1007/s11214-022-00902-6}, \href {https://ui.adsabs.harvard.edu/abs/2022SSRv..218...36K} {218, 36}

\bibitem[\protect\citeauthoryear{{Kleimann}, {Oughton}, {Fichtner}  \& {Scherer}}{{Kleimann} et~al.}{2023}]{KleimannEA23}
{Kleimann} J.,  {Oughton} S.,  {Fichtner} H.,   {Scherer} K.,  2023, \mn@doi [\apj] {10.3847/1538-4357/acd84e}, \href {https://ui.adsabs.harvard.edu/abs/2023ApJ...953..133K} {953, 133}

\bibitem[\protect\citeauthoryear{{Kopp}, {Venter}, {B{\"u}sching}  \& {de Jager}}{{Kopp} et~al.}{2013}]{Kopp2013}
{Kopp} A.,  {Venter} C.,  {B{\"u}sching} I.,   {de Jager} O.~C.,  2013, \mn@doi [ApJ] {10.1088/0004-637X/779/2/126}, \href {http://adsabs.harvard.edu/abs/2013ApJ...779..126K} {779, 126}

\bibitem[\protect\citeauthoryear{{Krimm}, {Lien}, {Page}, {Palmer}, {Tohuvavohu}  \& {Neil Gehrels Swift Observatory Team}}{{Krimm} et~al.}{2020}]{Krimm2020}
{Krimm} H.~A.,  {Lien} A.~Y.,  {Page} K.~L.,  {Palmer} D.~M.,  {Tohuvavohu} A.,   {Neil Gehrels Swift Observatory Team} 2020, GRB Coordinates Network, \href {https://ui.adsabs.harvard.edu/abs/2020GCN.28187....1K} {28187, 1}

\bibitem[\protect\citeauthoryear{{Kuiper}, {Hermsen}  \& {Dekker}}{{Kuiper} et~al.}{2018}]{Kuiper2018}
{Kuiper} L.,  {Hermsen} W.,   {Dekker} A.,  2018, \mn@doi [\mnras] {10.1093/mnras/stx3128}, \href {https://ui.adsabs.harvard.edu/abs/2018MNRAS.475.1238K} {475, 1238}

\bibitem[\protect\citeauthoryear{{Kumar} \& {Safi-Harb}}{{Kumar} \& {Safi-Harb}}{2008}]{Kumar2008}
{Kumar} H.~S.,  {Safi-Harb} S.,  2008, \mn@doi [ApJ] {10.1086/588284}, \href {https://ui.adsabs.harvard.edu/abs/2008ApJ...678L..43K} {678, L43}

\bibitem[\protect\citeauthoryear{{Li} \& {Gao}}{{Li} \& {Gao}}{2023}]{2023LiBiao}
{Li} B.-P.,  {Gao} Z.-F.,  2023, \mn@doi [Astronomische Nachrichten] {10.1002/asna.20220111}, \href {https://ui.adsabs.harvard.edu/abs/2023AN....34420111L} {344, e20220111}

\bibitem[\protect\citeauthoryear{{Liu} \& {Wang}}{{Liu} \& {Wang}}{2021}]{2021ApJ221L}
{Liu} R.-Y.,  {Wang} X.-Y.,  2021, \mn@doi [\apj] {10.3847/1538-4357/ac2ba0}, \href {https://ui.adsabs.harvard.edu/abs/2021ApJ...922..221L} {922, 221}

\bibitem[\protect\citeauthoryear{{Livingstone} \& {Kaspi}}{{Livingstone} \& {Kaspi}}{2006}]{Livingstone2006}
{Livingstone} M.~A.,  {Kaspi} V.~M.,  2006, in AAS High Energy Astrophysics Division \#9. p.~1.21

\bibitem[\protect\citeauthoryear{{Livingstone}, {Ng}, {Kaspi}, {Gavriil}  \& {Gotthelf}}{{Livingstone} et~al.}{2011}]{Livingstone2011}
{Livingstone} M.~A.,  {Ng} C.~Y.,  {Kaspi} V.~M.,  {Gavriil} F.~P.,   {Gotthelf} E.~V.,  2011, \mn@doi [\apj] {10.1088/0004-637X/730/2/66}, \href {https://ui.adsabs.harvard.edu/abs/2011ApJ...730...66L} {730, 66}

\bibitem[\protect\citeauthoryear{{Lower} et~al.,}{{Lower} et~al.}{2021}]{Lower2021}
{Lower} M.~E.,  et~al., 2021, \mn@doi [\mnras] {10.1093/mnras/stab2678}, \href {https://ui.adsabs.harvard.edu/abs/2021MNRAS.508.3251L} {508, 3251}

\bibitem[\protect\citeauthoryear{{Lu}, {Gao}, {Zhu}  \& {Zhang}}{{Lu} et~al.}{2017}]{Fang2017}
{Lu} F.-W.,  {Gao} Q.-G.,  {Zhu} B.-T.,   {Zhang} L.,  2017, \mn@doi [MNRAS] {10.1093/mnras/stx2223}, \href {https://ui.adsabs.harvard.edu/abs/2017MNRAS.472.2926L} {472, 2926}

\bibitem[\protect\citeauthoryear{{Lu}, {Gao}, {Zhu}  \& {Zhang}}{{Lu} et~al.}{2019}]{Fang2019}
{Lu} F.-W.,  {Gao} Q.-G.,  {Zhu} B.-T.,   {Zhang} L.,  2019, \mn@doi [\aap] {10.1051/0004-6361/201834320}, \href {https://ui.adsabs.harvard.edu/abs/2019A&A...624A.144L} {624, A144}

\bibitem[\protect\citeauthoryear{{Lu}, {Zhu}, {Hu}  \& {Zhang}}{{Lu} et~al.}{2023}]{LuEA23}
{Lu} F.-W.,  {Zhu} B.-T.,  {Hu} W.,   {Zhang} L.,  2023, \mn@doi [\apj] {10.3847/1538-4357/ace0c2}, \href {https://ui.adsabs.harvard.edu/abs/2023ApJ...953..116L} {953, 116}

\bibitem[\protect\citeauthoryear{{Lyubarsky}}{{Lyubarsky}}{2003}]{Lyub}
{Lyubarsky} Y.~E.,  2003, \mn@doi [\mnras] {10.1046/j.1365-8711.2003.06927.x}, \href {https://ui.adsabs.harvard.edu/abs/2003MNRAS.345..153L} {345, 153}

\bibitem[\protect\citeauthoryear{{Manchester}, {Hobbs}, {Teoh}  \& {Hobbs}}{{Manchester} et~al.}{2005}]{2005AJ....129.1993M}
{Manchester} R.~N.,  {Hobbs} G.~B.,  {Teoh} A.,   {Hobbs} M.,  2005, \mn@doi [\aj] {10.1086/428488}, \href {https://ui.adsabs.harvard.edu/abs/2005AJ....129.1993M} {129, 1993}

\bibitem[\protect\citeauthoryear{{Martin} \& {Torres}}{{Martin} \& {Torres}}{2022}]{Martin2022}
{Martin} J.,  {Torres} D.~F.,  2022, \mn@doi [J. High Energy Astrophys.] {10.1016/j.jheap.2022.09.003}, \href {https://ui.adsabs.harvard.edu/abs/2022JHEAp..36..128M} {36, 128}

\bibitem[\protect\citeauthoryear{{Mart{\'{\i}}n}, {Torres}  \& {Rea}}{{Mart{\'{\i}}n} et~al.}{2012}]{Martin2012}
{Mart{\'{\i}}n} J.,  {Torres} D.~F.,   {Rea} N.,  2012, \mn@doi [MNRAS] {10.1111/j.1365-2966.2012.22014.x}, \href {http://adsabs.harvard.edu/abs/2012MNRAS.427..415M} {427, 415}

\bibitem[\protect\citeauthoryear{{Mart{\'\i}n}, {Torres}  \& {Pedaletti}}{{Mart{\'\i}n} et~al.}{2016}]{Martin2016}
{Mart{\'\i}n} J.,  {Torres} D.~F.,   {Pedaletti} G.,  2016, \mn@doi [MNRAS] {10.1093/mnras/stw684}, \href {https://ui.adsabs.harvard.edu/abs/2016MNRAS.459.3868M} {459, 3868}

\bibitem[\protect\citeauthoryear{{Martin}, {Torres}  \& {Zhang}}{{Martin} et~al.}{2020}]{Martin20}
{Martin} J.,  {Torres} D.~F.,   {Zhang} B.,  2020, \mn@doi [J. High Energy Astrophys.] {10.1016/j.jheap.2020.09.001}, \href {https://ui.adsabs.harvard.edu/abs/2020JHEAp..28...10M} {28, 10}

\bibitem[\protect\citeauthoryear{{Matheson} \& {Safi-Harb}}{{Matheson} \& {Safi-Harb}}{2005}]{Matheson2005}
{Matheson} H.,  {Safi-Harb} S.,  2005, \mn@doi [Advances in Space Research] {10.1016/j.asr.2005.04.050}, \href {http://adsabs.harvard.edu/abs/2005AdSpR..35.1099M} {35, 1099}

\bibitem[\protect\citeauthoryear{{Matthaeus}, {Qin}, {Bieber}  \& {Zank}}{{Matthaeus} et~al.}{2003}]{MattEA03}
{Matthaeus} W.~H.,  {Qin} G.,  {Bieber} J.~W.,   {Zank} G.~P.,  2003, \mn@doi [\apjl] {10.1086/376613}, \href {https://ui.adsabs.harvard.edu/abs/2003ApJ...590L..53M} {590, L53}

\bibitem[\protect\citeauthoryear{{McBride} et~al.,}{{McBride} et~al.}{2008}]{McBride08}
{McBride} V.~A.,  et~al., 2008, \mn@doi [\aap] {10.1051/0004-6361:20078432}, \href {https://ui.adsabs.harvard.edu/abs/2008A&A...477..249M} {477, 249}

\bibitem[\protect\citeauthoryear{{McKee}}{{McKee}}{1974}]{McKee1974}
{McKee} C.~F.,  1974, \mn@doi [ApJ] {10.1086/152721}, \href {http://adsabs.harvard.edu/abs/1974ApJ...188..335M} {188, 335}

\bibitem[\protect\citeauthoryear{{Minnie}, {Bieber}, {Matthaeus}  \& {Burger}}{{Minnie} et~al.}{2007}]{MinnieEA07}
{Minnie} J.,  {Bieber} J.~W.,  {Matthaeus} W.~H.,   {Burger} R.~A.,  2007, \mn@doi [\apj] {10.1086/518765}, \href {https://ui.adsabs.harvard.edu/abs/2007...663.1049M} {663, 1049}

\bibitem[\protect\citeauthoryear{{Mizuno}, {Ohno}, {Watanabe}, {Bucciantini}, {Gunji}, {Shibata}, {Slane}  \& {Weisskopf}}{{Mizuno} et~al.}{2023}]{MizunoEA23}
{Mizuno} T.,  {Ohno} H.,  {Watanabe} E.,  {Bucciantini} N.,  {Gunji} S.,  {Shibata} S.,  {Slane} P.,   {Weisskopf} M.~C.,  2023, \mn@doi [\pasj] {10.1093/pasj/psad070}, \href {https://ui.adsabs.harvard.edu/abs/2023PASJ...75.1298M} {75, 1298}

\bibitem[\protect\citeauthoryear{{Ng}, {Slane}, {Gaensler}  \& {Hughes}}{{Ng} et~al.}{2008}]{Ng08}
{Ng} C.~Y.,  {Slane} P.~O.,  {Gaensler} B.~M.,   {Hughes} J.~P.,  2008, \mn@doi [\apj] {10.1086/591146}, \href {https://ui.adsabs.harvard.edu/abs/2008ApJ...686..508N} {686, 508}

\bibitem[\protect\citeauthoryear{{Nynka} et~al.,}{{Nynka} et~al.}{2014}]{nynka2014ApJ_78972N}
{Nynka} M.,  et~al., 2014, \mn@doi [\apj] {10.1088/0004-637X/789/1/72}, \href {https://ui.adsabs.harvard.edu/abs/2014ApJ...789...72N} {789, 72}

\bibitem[\protect\citeauthoryear{{Olmi} \& {Bucciantini}}{{Olmi} \& {Bucciantini}}{2019}]{Olmi2019}
{Olmi} B.,  {Bucciantini} N.,  2019, \mn@doi [MNRAS] {10.1093/mnras/stz382}, \href {https://ui.adsabs.harvard.edu/abs/2019MNRAS.484.5755O} {484, 5755}

\bibitem[\protect\citeauthoryear{{Olmi} \& {Torres}}{{Olmi} \& {Torres}}{2020}]{Olmi2020}
{Olmi} B.,  {Torres} D.~F.,  2020, \mn@doi [\mnras] {10.1093/mnras/staa1052}, \href {https://ui.adsabs.harvard.edu/abs/2020MNRAS.494.4357O} {494, 4357}

\bibitem[\protect\citeauthoryear{{Olmi}, {Del Zanna}, {Amato}, {Bucciantini}  \& {Mignone}}{{Olmi} et~al.}{2016}]{OlmiEA16}
{Olmi} B.,  {Del Zanna} L.,  {Amato} E.,  {Bucciantini} N.,   {Mignone} A.,  2016, \mn@doi [Journal of Plasma Physics] {10.1017/S0022377816000957}, \href {https://ui.adsabs.harvard.edu/abs/2016JPlPh..82f6301O} {82, 635820601}

\bibitem[\protect\citeauthoryear{{Oughton} \& {Engelbrecht}}{{Oughton} \& {Engelbrecht}}{2021}]{OughtonEngelbrecht21}
{Oughton} S.,  {Engelbrecht} N.~E.,  2021, \mn@doi [New Astronomy] {10.1016/j.newast.2020.101507}, \href {https://ui.adsabs.harvard.edu/abs/2021NewA...8301507O} {83, 101507}

\bibitem[\protect\citeauthoryear{{Pacini} \& {Salvati}}{{Pacini} \& {Salvati}}{1973}]{Pacini1973}
{Pacini} F.,  {Salvati} M.,  1973, \mn@doi [ApJ] {10.1086/152495}, \href {http://adsabs.harvard.edu/abs/1973ApJ...186..249P} {186, 249}

\bibitem[\protect\citeauthoryear{Parker}{Parker}{1958}]{Parker1958}
Parker E.~N.,  1958, ApJ, 128, 664

\bibitem[\protect\citeauthoryear{{Parthasarathy} et~al.,}{{Parthasarathy} et~al.}{2020}]{Parthasarathy2020MNRAS}
{Parthasarathy} A.,  et~al., 2020, \mn@doi [\mnras] {10.1093/mnras/staa882}, \href {https://ui.adsabs.harvard.edu/abs/2020MNRAS.494.2012P} {494, 2012}

\bibitem[\protect\citeauthoryear{{Peng}, {Bao}, {Lu}  \& {Zhang}}{{Peng} et~al.}{2022}]{2022ApJ7P}
{Peng} Q.-Y.,  {Bao} B.-W.,  {Lu} F.-W.,   {Zhang} L.,  2022, \mn@doi [\apj] {10.3847/1538-4357/ac4161}, \href {https://ui.adsabs.harvard.edu/abs/2022ApJ...926....7P} {926, 7}

\bibitem[\protect\citeauthoryear{{Porth}, {Vorster}, {Lyutikov}  \& {Engelbrecht}}{{Porth} et~al.}{2016}]{PorthEA16}
{Porth} O.,  {Vorster} M.~J.,  {Lyutikov} M.,   {Engelbrecht} N.~E.,  2016, \mn@doi [\mnras] {10.1093/mnras/stw1152}, \href {https://ui.adsabs.harvard.edu/abs/2016MNRAS.460.4135P} {460, 4135}

\bibitem[\protect\citeauthoryear{{Ranasinghe} \& {Leahy}}{{Ranasinghe} \& {Leahy}}{2018}]{2018AJ204R}
{Ranasinghe} S.,  {Leahy} D.~A.,  2018, \mn@doi [\aj] {10.3847/1538-3881/aab9be}, \href {https://ui.adsabs.harvard.edu/abs/2018AJ....155..204R} {155, 204}

\bibitem[\protect\citeauthoryear{{Reynolds} \& {Chevalier}}{{Reynolds} \& {Chevalier}}{1984}]{Reynolds1984}
{Reynolds} S.~P.,  {Chevalier} R.~A.,  1984, \mn@doi [ApJ] {10.1086/161831}, \href {http://adsabs.harvard.edu/abs/1984ApJ...278..630R} {278, 630}

\bibitem[\protect\citeauthoryear{{Reynolds}, {Borkowski}  \& {Gwynne}}{{Reynolds} et~al.}{2018}]{Reynolds2018}
{Reynolds} S.~P.,  {Borkowski} K.~J.,   {Gwynne} P.~H.,  2018, \mn@doi [ApJ] {10.3847/1538-4357/aab3d3}, \href {https://ui.adsabs.harvard.edu/abs/2018ApJ...856..133R} {856, 133}

\bibitem[\protect\citeauthoryear{{Roy}, {Gupta}  \& {Lewandowski}}{{Roy} et~al.}{2012}]{roy2012MNRAS213R}
{Roy} J.,  {Gupta} Y.,   {Lewandowski} W.,  2012, \mn@doi [\mnras] {10.1111/j.1365-2966.2012.21380.x}, \href {https://ui.adsabs.harvard.edu/abs/2012MNRAS.424.2213R} {424, 2213}

\bibitem[\protect\citeauthoryear{{Rybicki} \& {Lightman}}{{Rybicki} \& {Lightman}}{1979}]{RybickiLightman1979}
{Rybicki} G.~B.,  {Lightman} A.~P.,  1979, {Radiative processes in astrophysics}.
New York, Wiley-Interscience, 1979.~393 p.

\bibitem[\protect\citeauthoryear{{Salter}, {Reynolds}, {Hogg}, {Payne}  \& {Rhodes}}{{Salter} et~al.}{1989}]{Salter1989}
{Salter} C.~J.,  {Reynolds} S.~P.,  {Hogg} D.~E.,  {Payne} J.~M.,   {Rhodes} P.~J.,  1989, \mn@doi [ApJ] {10.1086/167191}, \href {http://adsabs.harvard.edu/abs/1989ApJ...338..171S} {338, 171}

\bibitem[\protect\citeauthoryear{{Sathyaprakash} et~al.,}{{Sathyaprakash} et~al.}{2024}]{Sathyaprakash2024}
{Sathyaprakash} R.,  et~al., 2024, \mn@doi [arXiv e-prints] {10.48550/arXiv.2401.08010}, \href {https://ui.adsabs.harvard.edu/abs/2024arXiv240108010S} {p. arXiv:2401.08010}

\bibitem[\protect\citeauthoryear{{Schlickeiser}}{{Schlickeiser}}{2002}]{S02}
{Schlickeiser} R.,  2002, {Cosmic Ray Astrophysics}.
Astronomy and Astrophysics Library; Physics and Astronomy Online Library. Berlin: Springer.

\bibitem[\protect\citeauthoryear{{Sch{\"o}ck}, {B{\"u}sching}, {de Jager}, {Eger}  \& {Vorster}}{{Sch{\"o}ck} et~al.}{2010}]{Schock2010}
{Sch{\"o}ck} F.~M.,  {B{\"u}sching} I.,  {de Jager} O.~C.,  {Eger} P.,   {Vorster} M.~J.,  2010, \mn@doi [A\&A] {10.1051/0004-6361/201014151}, \href {http://adsabs.harvard.edu/abs/2010A%26A...515A.109S} {515, A109}

\bibitem[\protect\citeauthoryear{Shalchi}{Shalchi}{2009}]{Shalchi09}
Shalchi A.,  2009, Nonlinear Cosmic Ray Diffusion Theories.
 Astrophysics and Space Science Library Vol. 362, Springer, Berlin, \mn@doi{10.1007/978-3-642-00309-7}

\bibitem[\protect\citeauthoryear{{Shalchi}}{{Shalchi}}{2020}]{Shalchi20}
{Shalchi} A.,  2020, \mn@doi [Space Sci. Rev.] {10.1007/s11214-020-0644-4}, \href {https://ui.adsabs.harvard.edu/abs/2020SSRv..216...23S} {216, 23}

\bibitem[\protect\citeauthoryear{{Shebalin}, {Matthaeus}  \& {Montgomery}}{{Shebalin} et~al.}{1983}]{Shebalin83}
{Shebalin} J.~V.,  {Matthaeus} W.~H.,   {Montgomery} D.,  1983, \mn@doi [Journal of Plasma Physics] {10.1017/S0022377800000933}, \href {http://adsabs.harvard.edu/abs/1983JPlPh..29..525S} {29, 525}

\bibitem[\protect\citeauthoryear{Slane}{Slane}{2016}]{Slane2016}
Slane P.,  2016, Pulsar Wind Nebulae.
Springer International Publishing, Cham, pp 1--21, \mn@doi{10.1007/978-3-319-20794-0_95-1}, \url {https://doi.org/10.1007/978-3-319-20794-0_95-1}

\bibitem[\protect\citeauthoryear{{Slane}}{{Slane}}{2017}]{SlanePWN2017}
{Slane} P.,  2017, {Pulsar Wind Nebulae}, \mn@doi{10.1007/978-3-319-21846-5_95.
}

\bibitem[\protect\citeauthoryear{{Slane}, {Chen}, {Schulz}, {Seward}, {Hughes}  \& {Gaensler}}{{Slane} et~al.}{2000}]{slane2000ApJ533L29S}
{Slane} P.,  {Chen} Y.,  {Schulz} N.~S.,  {Seward} F.~D.,  {Hughes} J.~P.,   {Gaensler} B.~M.,  2000, \mn@doi [\apjl] {10.1086/312589}, \href {https://ui.adsabs.harvard.edu/abs/2000ApJ...533L..29S} {533, L29}

\bibitem[\protect\citeauthoryear{{Straal}, {Gelfand}  \& {Eagle}}{{Straal} et~al.}{2023}]{StraalEtAl2023}
{Straal} S.~M.,  {Gelfand} J.~D.,   {Eagle} J.~L.,  2023, \mn@doi [\apj] {10.3847/1538-4357/aca1a9}, \href {https://ui.adsabs.harvard.edu/abs/2023ApJ...942..103S} {942, 103}

\bibitem[\protect\citeauthoryear{{Tanaka} \& {Asano}}{{Tanaka} \& {Asano}}{2017}]{Tanaka2017}
{Tanaka} S.~J.,  {Asano} K.,  2017, \mn@doi [\apj] {10.3847/1538-4357/aa6f13}, \href {https://ui.adsabs.harvard.edu/abs/2017ApJ...841...78T} {841, 78}

\bibitem[\protect\citeauthoryear{{Tanaka} \& {Ishizaki}}{{Tanaka} \& {Ishizaki}}{2024}]{ShutaWataru2024}
{Tanaka} S.~J.,  {Ishizaki} W.,  2024, \mn@doi [Progress of Theoretical and Experimental Physics] {10.1093/ptep/ptae069}, \href {https://ui.adsabs.harvard.edu/abs/2024PTEP.2024e3E03T} {2024, 053E03}

\bibitem[\protect\citeauthoryear{{Tanaka} \& {Kashiyama}}{{Tanaka} \& {Kashiyama}}{2023}]{Tanaka2023}
{Tanaka} S.~J.,  {Kashiyama} K.,  2023, \mn@doi [\mnras] {10.1093/mnras/stad2504}, \href {https://ui.adsabs.harvard.edu/abs/2023MNRAS.525.2750T} {525, 2750}

\bibitem[\protect\citeauthoryear{{Tanaka} \& {Takahara}}{{Tanaka} \& {Takahara}}{2010}]{2010ApJ1248T}
{Tanaka} S.~J.,  {Takahara} F.,  2010, \mn@doi [\apj] {10.1088/0004-637X/715/2/1248}, \href {https://ui.adsabs.harvard.edu/abs/2010ApJ...715.1248T} {715, 1248}

\bibitem[\protect\citeauthoryear{{Tanaka} \& {Takahara}}{{Tanaka} \& {Takahara}}{2011}]{Tanaka_Takahara_2011}
{Tanaka} S.~J.,  {Takahara} F.,  2011, \mn@doi [ApJ] {10.1088/0004-637X/741/1/40}, \href {http://adsabs.harvard.edu/abs/2011ApJ...741...40T} {741, 40}

\bibitem[\protect\citeauthoryear{{Tang} \& {Chevalier}}{{Tang} \& {Chevalier}}{2012}]{Tang2012}
{Tang} X.,  {Chevalier} R.~A.,  2012, \mn@doi [\apj] {10.1088/0004-637X/752/2/83}, \href {https://ui.adsabs.harvard.edu/abs/2012ApJ...752...83T} {752, 83}

\bibitem[\protect\citeauthoryear{{Terrier}, {Djannati-Atai}, {Hoppe}, {Marand on}, {Renaud}  \& {de Jager}}{{Terrier} et~al.}{2008}]{Terrier2008}
{Terrier} R.,  {Djannati-Atai} A.,  {Hoppe} S.,  {Marand on} V.,  {Renaud} M.,   {de Jager} O.,  2008, in {Aharonian} F.~A.,  {Hofmann} W.,   {Rieger} F.,  eds,  American Institute of Physics Conference Series Vol. 1085, American Institute of Physics Conference Series. pp 316--319, \mn@doi{10.1063/1.3076670}

\bibitem[\protect\citeauthoryear{{Teufel} \& {Schlickeiser}}{{Teufel} \& {Schlickeiser}}{2003}]{TeufelSchlickeiser03}
{Teufel} A.,  {Schlickeiser} R.,  2003, \mn@doi [\aap] {10.1051/0004-6361:20021471}, \href {https://ui.adsabs.harvard.edu/abs/2003A&A...397...15T} {397, 15}

\bibitem[\protect\citeauthoryear{{Tian} \& {Leahy}}{{Tian} \& {Leahy}}{2008}]{tian2008MNRAS391L4T}
{Tian} W.~W.,  {Leahy} D.~A.,  2008, \mn@doi [\mnras] {10.1111/j.1745-3933.2008.00557.x}, \href {https://ui.adsabs.harvard.edu/abs/2008MNRAS.391L..54T} {391, L54}

\bibitem[\protect\citeauthoryear{{Tong} \& {Kou}}{{Tong} \& {Kou}}{2017}]{2017TongKau}
{Tong} H.,  {Kou} F.~F.,  2017, \mn@doi [\apj] {10.3847/1538-4357/aa60c6}, \href {https://ui.adsabs.harvard.edu/abs/2017ApJ...837..117T} {837, 117}

\bibitem[\protect\citeauthoryear{{Torres} \& {Lin}}{{Torres} \& {Lin}}{2018}]{Torres2018}
{Torres} D.~F.,  {Lin} T.,  2018, \mn@doi [\apjl] {10.3847/2041-8213/aad6e1}, \href {https://ui.adsabs.harvard.edu/abs/2018ApJ...864L...2T} {864, L2}

\bibitem[\protect\citeauthoryear{{Torres}, {Cillis}  \& {Mart{\'\i}n Rodriguez}}{{Torres} et~al.}{2013}]{2013Ap7634T}
{Torres} D.~F.,  {Cillis} A.~N.,   {Mart{\'\i}n Rodriguez} J.,  2013, \mn@doi [\apjl] {10.1088/2041-8205/763/1/L4}, \href {https://ui.adsabs.harvard.edu/abs/2013ApJ...763L...4T} {763, L4}

\bibitem[\protect\citeauthoryear{{Torres}, {Cillis}, {Mart{\'{\i}}n}  \& {de O{\~n}a Wilhelmi}}{{Torres} et~al.}{2014}]{Torres2014}
{Torres} D.~F.,  {Cillis} A.,  {Mart{\'{\i}}n} J.,   {de O{\~n}a Wilhelmi} E.,  2014, \mn@doi [J. High Energy Astrophys.] {10.1016/j.jheap.2014.02.001}, \href {http://adsabs.harvard.edu/abs/2014JHEAp...1...31T} {1, 31}

\bibitem[\protect\citeauthoryear{{Tsujimoto} et~al.,}{{Tsujimoto} et~al.}{2011}]{tsuj2011AA25A25T}
{Tsujimoto} M.,  et~al., 2011, \mn@doi [\aap] {10.1051/0004-6361/201015597}, \href {https://ui.adsabs.harvard.edu/abs/2011A&A...525A..25T} {525, A25}

\bibitem[\protect\citeauthoryear{{Verbiest}, {Weisberg}, {Chael}, {Lee}  \& {Lorimer}}{{Verbiest} et~al.}{2012}]{Verbiest12}
{Verbiest} J.~P.~W.,  {Weisberg} J.~M.,  {Chael} A.~A.,  {Lee} K.~J.,   {Lorimer} D.~R.,  2012, \mn@doi [\apj] {10.1088/0004-637X/755/1/39}, \href {https://ui.adsabs.harvard.edu/abs/2012ApJ...755...39V} {755, 39}

\bibitem[\protect\citeauthoryear{{Vladimirov}, {J{\'o}hannesson}, {Moskalenko}  \& {Porter}}{{Vladimirov} et~al.}{2012}]{Vla2012ApJ68V}
{Vladimirov} A.~E.,  {J{\'o}hannesson} G.,  {Moskalenko} I.~V.,   {Porter} T.~A.,  2012, \mn@doi [\apj] {10.1088/0004-637X/752/1/68}, \href {https://ui.adsabs.harvard.edu/abs/2012ApJ...752...68V} {752, 68}

\bibitem[\protect\citeauthoryear{{Vorster}, {Tibolla}, {Ferreira}  \& {Kaufmann}}{{Vorster} et~al.}{2013a}]{Vorster2013}
{Vorster} M.~J.,  {Tibolla} O.,  {Ferreira} S.~E.~S.,   {Kaufmann} S.,  2013a, preprint, \href {http://adsabs.harvard.edu/abs/2013arXiv1308.1626V} {} (\mn@eprint {arXiv} {1308.1626})

\bibitem[\protect\citeauthoryear{{Vorster}, {Ferreira}, {de Jager}  \& {Djannati-Ata{\"i}}}{{Vorster} et~al.}{2013b}]{Vorster2013_3}
{Vorster} M.~J.,  {Ferreira} S.~E.~S.,  {de Jager} O.~C.,   {Djannati-Ata{\"i}} A.,  2013b, \mn@doi [\aap] {10.1051/0004-6361/201220276}, \href {http://adsabs.harvard.edu/abs/2013A%26A...551A.127V} {551, A127}

\bibitem[\protect\citeauthoryear{{Vorster}, {Tibolla}, {Ferreira}  \& {Kaufmann}}{{Vorster} et~al.}{2013c}]{2013ApJ773139V}
{Vorster} M.~J.,  {Tibolla} O.,  {Ferreira} S.~E.~S.,   {Kaufmann} S.,  2013c, \mn@doi [\apj] {10.1088/0004-637X/773/2/139}, \href {https://ui.adsabs.harvard.edu/abs/2013ApJ...773..139V} {773, 139}

\bibitem[\protect\citeauthoryear{{Wilkin}}{{Wilkin}}{1996}]{Wilkin1996}
{Wilkin} F.~P.,  1996, \mn@doi [\apjl] {10.1086/309939}, \href {https://ui.adsabs.harvard.edu/abs/1996ApJ...459L..31W} {459, L31}

\bibitem[\protect\citeauthoryear{{Zajczyk} et~al.,}{{Zajczyk} et~al.}{2012}]{2012AnA12Z}
{Zajczyk} A.,  et~al., 2012, \mn@doi [\aap] {10.1051/0004-6361/201117194}, \href {https://ui.adsabs.harvard.edu/abs/2012A&A...542A..12Z} {542, A12}

\bibitem[\protect\citeauthoryear{Zank, Adhikari, Zhao, Mostafavi, Zirnstein  \& Mc{Comas}}{Zank et~al.}{2018}]{ZankEA18}
Zank G.~P.,  Adhikari L.,  Zhao L.-L.,  Mostafavi P.,  Zirnstein E.~J.,   Mc{Comas} D.~J.,  2018, \mn@doi [\apj] {10.3847/1538-4357/aaebfe}, \href {http://adsabs.harvard.edu/abs/2018ApJ...869...23Z} {869, 23}

\bibitem[\protect\citeauthoryear{{Zhu}, {Zhang}  \& {Fang}}{{Zhu} et~al.}{2018}]{2018AnA609A110Z}
{Zhu} B.-T.,  {Zhang} L.,   {Fang} J.,  2018, \mn@doi [\aap] {10.1051/0004-6361/201629108}, \href {https://ui.adsabs.harvard.edu/abs/2018A&A...609A.110Z} {609, A110}

\bibitem[\protect\citeauthoryear{{Zhu}, {Lu}  \& {Zhang}}{{Zhu} et~al.}{2023}]{Zhu2023}
{Zhu} B.-T.,  {Lu} F.-W.,   {Zhang} L.,  2023, \mn@doi [\apj] {10.3847/1538-4357/acaaa0}, \href {https://ui.adsabs.harvard.edu/abs/2023ApJ...943...89Z} {943, 89}

\bibitem[\protect\citeauthoryear{{de Jager}, {Harding}, {Michelson}, {Nel}, {Nolan}, {Sreekumar}  \& {Thompson}}{{de Jager} et~al.}{1996}]{1996ApJ253D}
{de Jager} O.~C.,  {Harding} A.~K.,  {Michelson} P.~F.,  {Nel} H.~I.,  {Nolan} P.~L.,  {Sreekumar} P.,   {Thompson} D.~J.,  1996, \mn@doi [\apj] {10.1086/176726}, \href {https://ui.adsabs.harvard.edu/abs/1996ApJ...457..253D} {457, 253}

\bibitem[\protect\citeauthoryear{{de Jager}, {Ferreira}  \& {Djannati-Ata{\"\i}}}{{de Jager} et~al.}{2008}]{deJager2008Ferreira}
{de Jager} O.~C.,  {Ferreira} S.~E.~S.,   {Djannati-Ata{\"\i}} A.,  2008, in {Aharonian} F.~A.,  {Hofmann} W.,   {Rieger} F.,  eds,  American Institute of Physics Conference Series Vol. 1085, American Institute of Physics Conference Series. pp 199--202, \mn@doi{10.1063/1.3076638}

\bibitem[\protect\citeauthoryear{{de Rosa}, {Ubertini}, {Campana}, {Bazzano}, {Dean}  \& {Bassani}}{{de Rosa} et~al.}{2009}]{de2009MNRAS93527D}
{de Rosa} A.,  {Ubertini} P.,  {Campana} R.,  {Bazzano} A.,  {Dean} A.~J.,   {Bassani} L.,  2009, \mn@doi [\mnras] {10.1111/j.1365-2966.2008.14160.x}, \href {https://ui.adsabs.harvard.edu/abs/2009MNRAS.393..527D} {393, 527}

\bibitem[\protect\citeauthoryear{{van Der Swaluw}, {Achterberg}  \& {Gallant}}{{van Der Swaluw} et~al.}{1998}]{vanderSwaluw1998}
{van Der Swaluw} E.,  {Achterberg} A.,   {Gallant} Y.~A.,  1998, Memorie della Società Astronomia Italiana, \href {https://ui.adsabs.harvard.edu/abs/1998MmSAI..69.1017V} {69, 1017}

\bibitem[\protect\citeauthoryear{{van Rensburg}}{{van Rensburg}}{2020}]{CarloPhD2020}
{van Rensburg} C.,  2020, PhD thesis, North-West University, South Africa, \url {https://repository.nwu.ac.za/handle/10394/34852}

\bibitem[\protect\citeauthoryear{{van Rensburg}, {Venter}  \& {Kruger}}{{van Rensburg} et~al.}{2018a}]{2018arXiv180200216V}
{van Rensburg} C.,  {Venter} C.,   {Kruger} P.,  2018a, \mn@doi [arXiv e-prints] {10.48550/arXiv.1802.00216}, \href {https://ui.adsabs.harvard.edu/abs/2018arXiv180200216V} {p. arXiv:1802.00216}

\bibitem[\protect\citeauthoryear{{van Rensburg}, {Kr{\"u}ger}  \& {Venter}}{{van Rensburg} et~al.}{2018b}]{CvR2018MNRAS_G09}
{van Rensburg} C.,  {Kr{\"u}ger} P.~P.,   {Venter} C.,  2018b, \mn@doi [\mnras] {10.1093/mnras/sty826}, \href {https://ui.adsabs.harvard.edu/abs/2018MNRAS.477.3853V} {477, 3853}

\bibitem[\protect\citeauthoryear{{van Rensburg}, {Venter}, {Seyffert}  \& {Harding}}{{van Rensburg} et~al.}{2020}]{van2020MNRAS3091V}
{van Rensburg} C.,  {Venter} C.,  {Seyffert} A.~S.,   {Harding} A.~K.,  2020, \mn@doi [\mnras] {10.1093/mnras/staa016}, \href {https://ui.adsabs.harvard.edu/abs/2020MNRAS.492.3091V} {492, 3091}

\bibitem[\protect\citeauthoryear{{van der Swaluw}, {Achterberg}, {Gallant}  \& {T{\'o}th}}{{van der Swaluw} et~al.}{2001}]{vanderSwaluw2001}
{van der Swaluw} E.,  {Achterberg} A.,  {Gallant} Y.~A.,   {T{\'o}th} G.,  2001, \mn@doi [\aap] {10.1051/0004-6361:20011437}, \href {https://ui.adsabs.harvard.edu/abs/2001A&A...380..309V} {380, 309}

\makeatother
\end{thebibliography}
\appendix
\section{Burst-like injection of Energy into the PWN}\label{app:BurstLikeEnergyInjection}
If we consider a change in pulsar spin-down $dL = d\dot{E}$ over some period $\tau$, this may lead to particle acceleration and thus a change in bulk flow speed. To estimate the magnitude of this effect, let us assume that we can model the spin-down of a glitching pulsar using the usual braking model, but with an effective braking index $n_2$, compared to a non-glitching pulsar with braking index $n_1$. This ignores the spiky behaviour of $L$, and rather interpolates the curve using a different braking index than what was measured after a glitch (smaller, but presumably close to the pre-burst value). 
The energy deposited by the pulsar into the environment is given by
\begin{equation}
E(n_i,t) \equiv \int_0^t L(t)\,dt = \frac{E_0}{1-\beta_i}\left(x_i^{1-\beta_i} - 1\right),
\end{equation}
where $\beta_i \equiv (n_i + 1)/(n_i - 1)$, $i=1,2$, $x_i(t) \equiv 1 + t/\tau_i$, $\tau_{i}$ the typical spin-down time-scale $\tau_{i} = P/{(n_i-1)\dot{P}}$, and with $E_0\equiv L(0)\tau$. 
The injected energy over some short period $dt=t_b-t_a$ is given by
\begin{equation}
\Delta E(n_i,t_b) \equiv E(n_i,t_b) - E(n_i,t_a) = \frac{E_0}{1-\beta_i}\left(x_{b,i}^{1-\beta_i} - x_{a,i}^{1-\beta_i}\right).
\end{equation}
The fractional injection of energy due to an abrupt change in $L(t)$ over and above the usual cumulative energy injection via spin-down is
\begin{equation}
\frac{dE}{E}\equiv\frac{\Delta E(n_2,t_b) - \Delta E(n_1,t_b)}{E(n_1,t_b)} =\frac{\left(x_{b,2}^{1-\beta_2} - x_{a,2}^{1-\beta_2}\right) - \left(x_{b,1}^{1-\beta_1} - x_{a,1}^{1-\beta_1}\right)}{x_{b,1}^{1-\beta_1}-1}.
\end{equation}
Let us consider a division of rotational power into particle acceleration and electromagnetic fields:
\begin{equation}
    L = L_{\rm part}+L_{\rm EM} = \frac{1}{1+\sigma}L + \frac{\sigma}{1+\sigma}L,
\end{equation}
with $\sigma\equiv L_{\rm EM}/L_{\rm part}$. Close to the pulsar, $\sigma\gg 1$, while beyond the shock radius, $\sigma\ll1$. If we assume that the injected energy $E(t)$ is converted to non-relativistic kinetic energy $K$ of the bulk flow beyond the termination shock, the first term dominates over changes in the field strength due to this energy injection, and we have
\begin{equation}
    \frac{dE}{E}\sim\frac{dK}{K}\sim2\frac{dV_0}{V_0}.
\end{equation}
For the case of Kes~75, taking $n_1=2$ and $n_2=2.65$, $t_b = 720$~yr and $t_a=670$~yr, we find $dE/E\sim dK/K\sim 0.02$ for small $\sigma$, so that $dV_0/V_0\sim 0.01.$ This is a crude upper bound, but limits the fractional change in the bulk flow to a few percent, weakly depending on $n_i$ and the choices for $t_a$ and $t_b$. This confirms the idea that since $L$ did not change much after some recent glitches, the injected energy may lead to a small increase in bulk flow, limited by the applicability of the approximation of the effective braking model described here.

\bsp
\label{lastpage}
\end{document}